\documentclass[12pt,a4paper,titlepage]{report}
\usepackage{setspace}
\usepackage{amsmath, amssymb}
\usepackage{graphicx}
\usepackage[top=30truemm,bottom=30truemm,left=25truemm,right=25truemm]{geometry}
\usepackage[compress, numbers]{natbib}
\usepackage{theorem}
\theorembodyfont{\normalfont}
\newtheorem{theo}{Theorem}
\newtheorem{defi}{Definition}
\newtheorem{lemm}{Lemma}
\newenvironment{proof}{\noindent{\it Proof.}}{\hfill$\square$\\}

\title{Master Thesis\\
\ \\
Absolute Irreversibility\\in Information Thermodynamics}
\author{Y\^uto Murashita}
\date{February 2, 2015}

\newcommand{\eqn}[1]{\begin{eqnarray} #1 \end{eqnarray}}
\newcommand{\ave}[1]{\langle #1 \rangle}
\newcommand{\mP}[0]{\mathcal P}
\newcommand{\mD}[0]{\mathcal D}
\newcommand{\mN}[0]{\mathcal N}
\newcommand{\mA}[0]{\mathcal A}
\newcommand{\mM}[0]{\mathcal M}
\newcommand{\mZ}[0]{\mathcal Z}
\newcommand{\mY}[0]{\mathcal Y}
\newcommand{\mX}[0]{\mathcal X}
\newcommand{\mB}[0]{\mathcal B}
\newcommand{\mF}[0]{\mathcal F}
\newcommand{\dl}[0]{\partial}
\newcommand{\kb}[0]{k_{\rm B}}

\begin{document}
\maketitle
\begin{abstract}
Nonequilibrium equalities have attracted considerable interest in the context of statistical mechanics and information thermodynamics.
What is remarkable about nonequilibrium equalities is that they apply to rather general nonequilibrium situations beyond the linear response regime.
However, nonequilibrium equalities are known to be inapplicable to some important situations.
In this thesis, we introduce a concept of absolute irreversibility as a new class of irreversibility that encompasses the entire range of those irreversible situations to which the conventional nonequilibrium equalities are inapplicable.
In mathematical terms, absolute irreversibility corresponds to the singular part of probability measure and can be separated from the ordinary irreversible part by Lebesgue's decomposition theorem in measure theory.
This theorem guarantees the uniqueness of the decomposition of probability measure into singular and nonsingular parts, which enables us to give a well-defined mathematical and physical meaning to absolute irreversibility.
Consequently, we derive a new type of nonequilibrium equalities in the presence of absolute irreversibility.
Inequalities derived from our nonequilibrium equalities give stronger restrictions on the entropy production during nonequilibrium processes than the conventional second-law like inequalities.
Moreover, we present a new resolution of Gibbs' paradox from the viewpoint of absolute irreversibility.
This resolution applies to a classical mesoscopic regime, where two prevailing resolutions of Gibbs' paradox break down.
\end{abstract}
\tableofcontents
\chapter{Introduction}
In the mid-1990s, a significant breakthrough was achieved in the field of nonequilibrium statistical physics.
Evans, Cohen and Morris numerically found a new symmetry of the probability distribution function of the entropy production rate in a steady-state shear-driven flow \cite{ECM93}.
This symmetry, later formulated in the form of fluctuation theorems, was proven in chaotic systems by Gallavotti and Cohen \cite{GC95}, and later in various types of systems \cite{Kur98, LS99, SE00, ESM01}.
In this way, the fluctuation theorems present a ubiquitous and universal structure residing in nonequilibrium systems.
What is important about the fluctuation theorems is that they apply to systems far from equilibrium.
Moreover, they can be regarded as a generalized formulation of the linear response theory to rather general nonequilibrium situations \cite{Gal96}.
In this respect, the fluctuation theorems have attracted considerable attention.

In the course of the development of the fluctuation theorems, an important relation was found by Jarzynski \cite{Jar97Apr, Jar97Nov}.
The Jarzynski equality or the integral nonequilibrium equality relates the equilibrium free-energy difference between two configurations to an ensemble property of the work performed on the system during a rather general nonequilibrium process that starts from one of the two configurations.
Thus, the Jarzynski equality enables us to estimate the free-energy difference by finite-time measurements in a nonequilibrium process outside of the linear response regime.
Moreover, the Jarzynski equality concisely reproduces the second law of thermodynamics and the fluctuation-dissipation relation in the weak fluctuation limit.
In this way, the integral nonequilibrium equality is a fundamental relation with experimental applications.

Another significant relation is the Crooks fluctuation theorem or the detailed nonequilibrium equality \cite{Cro99, Cro00}.
The Crooks fluctuation theorem compares the realization probability of a trajectory in phase space under a given dynamics with that of the time-reversed trajectory under the time-reversed dynamics.
The ratio of these two probabilities is exactly quantified by the exponentiated entropy production.
Therefore, irreversibility of a path under time reversal is quantitatively characterized by the entropy production of the path itself.
Additionally, the Crooks fluctuation theorem succinctly reproduces the Jarzynski equality.

Recently, the subject of nonequilibrium equalities marks a new development in the field of feedback control.
The theory of feedback control dates back to Maxwell \cite{Max1871}.
In his textbook of thermodynamics, Maxwell pointed out that the second law can be violated if we have access to microscopic degrees of freedom of the system and illustrated the idea in his renowned gedankenexperiment later christened by Lord Kelvin as Maxwell's demon.
Maxwell's demon is able to reduce the entropy of an isolated many-particle gas, by measuring the velocity of the particles and manipulating them based on the information of the measurement outcomes without expenditure of work. 
Maxwell's demon triggered subsequent discussions on the relations between thermodynamics and information processing \cite{Szi29,Bri51,Lan61}.
Although Maxwell's demon has been a purely imaginary object for about one and a half century, thanks to technological advances, the realization of Maxwell's demon in real experiments is now within our hands \cite{TSUe10, KMSP14}.
Hence, the theory of feedback control has attracted considerable interest in these days,
and quantitative relations between thermodynamic quantities and information were established \cite{SU08, SU09}, forming a new field of information thermodynamics.
One of the quantitative relation is known as the second law of information thermodynamics, which states that the entropy reduction in the feedback process is restricted by the amount of mutual information obtained in the measurement process.
As the Jarzynski equality is a generalization of the second law of thermodynamics, the second law of information thermodynamics is generalized in the form of the integral nonequilibrium equality \cite{SU10}.
Thus, nonequilibrium equalities are an actively developing field in the context of information thermodynamics.

Despite of the wide applicability and interest of nonequilibrium equalities, they are known to be inapplicable to  free expansion \cite{Gro05Aug, Jar05}.
Moreover, information-thermodynamic nonequilibrium equalities cannot apply to situations that involve such high-accuracy measurements as error-free measurements \cite{SU12}.
This inapplicability roots from divergence of the exponentiated entropy production, and can be circumvented at the level of the detailed nonequilibrium equalities \cite{HV10}.
However, to obtain the corresponding integral nonequilibrium equalities, situation-specific modifications are needed, and moreover the form of the obtained equalities is rather unusual.
There has been no unified strategy to derive these exceptional integral nonequilibrium equalities in the situations to which the conventional integral nonequilibrium equalities cannot apply.

In this thesis, we propose a new concept of absolute irreversibility that constitutes those irreversible situations to which the conventional integral nonequilibrium equalities cannot apply.
In physical terms, absolute irreversibility refers to those irreversible situations in which paths in the time-reversed dynamics do not have the corresponding paths in the original forward dynamics, which makes stark contrast to ordinary irreversible situations, in which every time-reversed path has the corresponding original path.
Therefore, in the context of the detailed nonequilibrium equalities, irreversibility is so strong that the entropy production diverges, whereas, in ordinary irreversible situations, irreversibility is quantitatively characterized by a finite entropy production.
In mathematical terms, absolute irreversibility is characterized as the singular part of the probability measure in the time-reversed dynamics with respect to the probability measure in the original dynamics.
Therefore, based on Lebesgue's decomposition theorem in measure theory \cite{Hal74,Bar95}, the absolutely irreversible part is uniquely separated from the ordinary irreversible part.
As a result, we obtain nonequilibrium integral equalities that are applicable to absolutely irreversible situations \cite{MFU14}.
Furthermore, our nonequilibrium equalities give tighter restrictions on the entropy production than the conventional second-law like inequalities \cite{MFU14,AFMU14}.

As an illustrative application of our integral nonequilibrium equalities and the notion of absolute irreversibility, we consider the problem of gas mixing, which is what Gibbs' paradox deals with \cite{Gibbs1875}.
Gibbs' paradox is qualitatively resolved once we recognize the equivocal nature of the thermodynamic entropy \cite{Grad61,Kampen84,Jaynes92}.
Although the standard quantitative resolution of Gibbs' paradox in many textbooks is based on quantum statistical mechanics, this resolution is indeed irrelevant to Gibbs' paradox \cite{Kampen84,Jaynes92}.
Pauli gave a correct quantitative analysis of Gibbs' paradox based on the extensivity of the thermodynamic entropy \cite{Pauli73,Jaynes92}.
However, this resolution holds only in the thermodynamic limit, and ignores any sub-leading effects.
Based on our nonequilibrium equalities in the presence of absolute irreversibility, we give a quantitative resolution of Gibbs' paradox that is applicable even to a classical mesoscopic regime, where sub-leading effects play an important role.

This thesis is organized as follows.
In Chap. 2, we briefly review history of nonequilibrium equalities and a unified approach to derive them.
In Chap. 3, we review a part of historical discussions on Maxwell's demon and some fundamental relations under measurements and feedback control including the second law of information thermodynamics.
Then, we review and derive information-thermodynamic nonequilibrium equalities.
In Chaps. 4-6, we describe the main results of this study.
In Chap. 4, we introduce a concept of absolute irreversibility in an example of free expansion, to which the conventional integral nonequilibrium equalities do not apply, and define absolute irreversibility in mathematical terms.
Then, in situations without measurements and feedback control, we derive nonequilibrium equalities in the presence of absolute irreversibility based on Lebesgue's decomposition theorem, and verify them in several illustrative examples.
In Chap. 5, we generalize the results in Chap. 4 to obtain information-thermodynamic nonequilibrium equalities in the presence of absolute irreversibility and verify them analytically in a few simple examples.
In Chap. 6, we briefly review discussions and conventional resolutions on Gibbs' paradox and quantitatively resolve Gibbs' paradox based on our nonequilibrium equalities with absolute irreversibility.
In Chap. 7, we summarize this thesis and discuss some future prospects.

\chapter{Review of Nonequilibrium Equalities}
In this chapter, we review nonequilibrium equalities in classical statistical mechanics.
Nonequilibrium equalities are exact equalities applicable to quite general nonequilibrium systems, and are generalizations of the second law of thermodynamics and well-known relations in linear response theory.
Moreover, nonequilibrium equalities give one solution of Loschmidt's paradox.

In the former half of this chapter, we review various nonequilibrium equalities in the chronological order.
In the latter half, we review a unified method to derive the nonequilibrium equalities introduced in the former half.

\section{History}
First of all, we briefly review history of nonequilibrium equalities.

\subsection{Discovery of fluctuation theorem}
\begin{figure}
\begin{center}
\includegraphics[width=0.9\columnwidth]{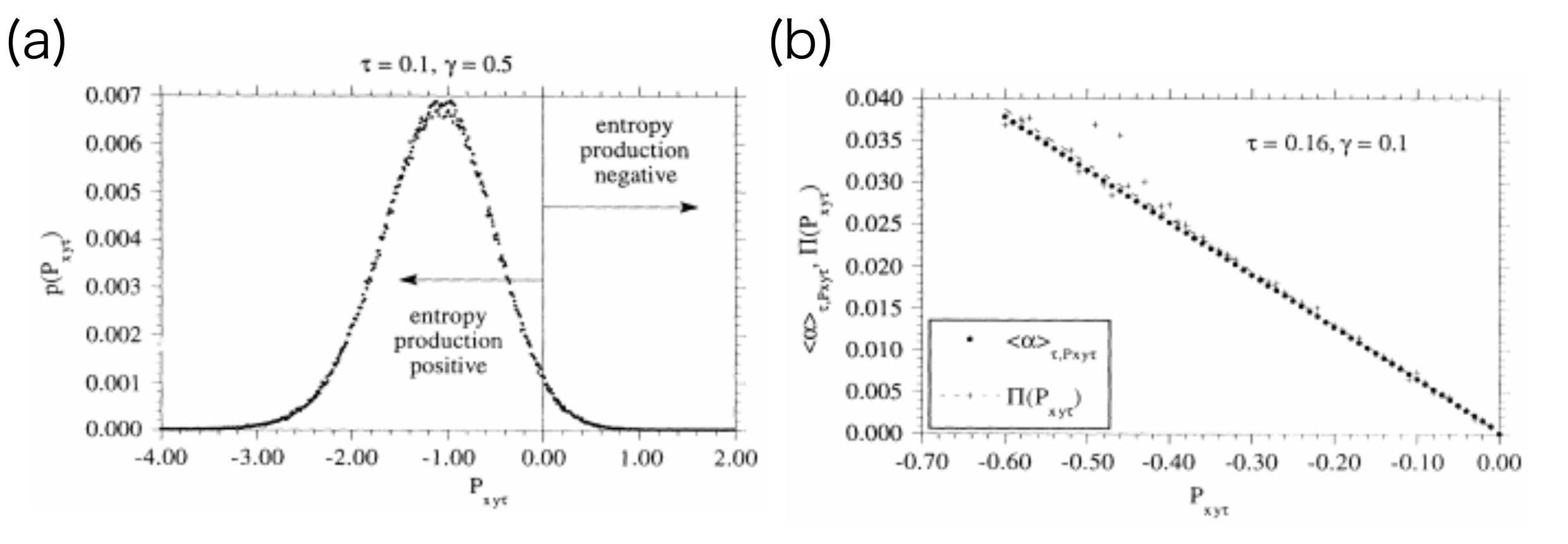}
\caption{\label{fig:ESM}
(a)
Probability distribution of the entropy production rate obtained by numerical simulations in shear-driven flow.
The abscissa shows the negative of the entropy production rate $\Sigma$ in arbitrary units. 
We can observe that the probability of events violating the second law does not vanish.
(b)
Demonstration of the steady-state fluctuation theorem (\ref{SSFT}).
The abscissa is the same as (a).
The markers represent the logarithm of the ratio (log-ratio) of the probability of a positive entropy production to that of the reversed sign (namely the log-ratio in the left-hand side of Eq. (\ref{SSFT})), and the dotted line represents the values of the log-ratio predicted by Eq.~(\ref{SSFT}).
We can confirm that the log-ratio is proportional to the negative of entropy production rate.
Reproduced from Figs. 1 and 2 in Ref. \cite{ECM93}.
Copyright 1993 by the American Physical Society.
}
\end{center}
\end{figure}

The field of nonequilibrium equalities was initiated by Evans, Cohen, and Morris in 1993 \cite{ECM93}.
In a steady state of thermostatted shear-driven flow, they numerically discovered a novel symmetry in the probability distribution of the entropy production rate, which is nowadays called the steady-state fluctuation theorem.
The theorem reads
\eqn{\label{SSFT}
	\lim_{t\to\infty}\frac{1}{t}\ln\frac{P(\Sigma)}{P(-\Sigma)} = \Sigma,
}
where $t$ is the time interval and $P(\Sigma)$ is the probability distribution function for the time-averaged entropy production rate $\Sigma$ in units of $k_{\rm B}T$, where $k_{\rm B}$ is the Boltzmann constant and $T$ is the absolute temperature of the system (see Fig. \ref{fig:ESM}).
They justified this symmetry by assuming a certain statistical ensemble of the nonequilibrium steady state.
Subsequently, in the same system, a similar symmetry is disclosed in a transient situation from the equilibrium state to the steady state \cite{ES94}.
It is known as the transient fluctuation theorem and written as
\eqn{\label{TFT}
	\frac{P(\Sigma)}{P(-\Sigma)} = e^{\Sigma t}.
}
This relation was proved under the same assumption as the steady-state fluctuation theorem (\ref{SSFT}),
which suggests that the fluctuation theorem is not a property particular to steady states, but applicable to wider classes of nonequilibrium situations.
Equations (\ref{SSFT}) and (\ref{TFT}) demonstrate that the probability of a positive entropy production is exponentially greater than that of the reversed sign.
Moreover, these fluctuation theorems reproduce the Green-Kubo relation \cite{Kubo57, KYN57} and Onsager's reciprocity relation \cite{Ons31, Ons31-2} in the limit of weak external fields \cite{Gal96}.
Therefore, the fluctuation theorems can be regarded as extensions of the well-established relations in the linear-response regime to a more general nonequilibrium regime.

These fluctuation theorems~(\ref{SSFT}) and (\ref{TFT}) resolve Loschmidt's paradox in the following sense.
Loschmidt's paradox originates from his criticism to the $H$-theorem proposed by Boltzmann, which demonstrates that the Shannon entropy of the probability distribution function of phase space increases with time in a system obeying the Boltzmann equation, although this equation is symmetric under time reversal.
It was claimed that the $H$-theorem is a derivation of the second law of thermodynamics, and the irreversible macroscopic law can be derived from the reversible microscopic dynamics.
However, Loschmidt criticized this argument by the observation that it should be impossible to deduce irreversible properties from the time-reversal symmetric dynamics.
If we have a process with a positive entropy production and reverse the velocity of all the particle in the system at once, we can generate a process with a negative entropy production since the dynamics is time-reversal symmetric.
Therefore, the entropy of the system should not always decrease, which is a defect of the $H$-theorem.
Fluctuation theorems~(\ref{SSFT}) and (\ref{TFT}) demonstrate that, as Loschmidt pointed out, paths with a negative entropy production have nonzero probability.
However, an important implication of the fluctuation theorems is that the probability of a negative entropy production is exponentially suppressed in large systems or in the long-time limit (namely when $\Sigma t\gg 1$).
Therefore, paths violating the second law cannot be observed in macroscopic systems in practice.
In this way, the fluctuation theorems reconcile Loschmidt's paradox with the second law originating from the reversible dynamics.

Soon after the discovery of the fluctuation theorems, the steady-state fluctuation theorem (\ref{SSFT}) was proved under the chaotic hypothesis, which is an extension of the ergodic hypothesis, in dissipative reversible systems \cite{GC95, GC95-2}.
Later, a proof free from the chaotic hypothesis was proposed in Langevin systems since the Langevin dynamics is ergodic in the sense that the system relaxes to the thermal equilibrium distribution in the long-time limit \cite{Kur98}, and then generalized to general Markov processes \cite{LS99}.
Moreover, both the steady-state fluctuation theorem (\ref{SSFT}) and the transient fluctuation theorem (\ref{TFT}) were shown in general thermostatted systems in a unified manner \cite{SE00}.
It was pointed out that this proof of the fluctuation theorems is applicable even to Hamiltonian systems, although it had been believed that the thermostatting mechanism is needed for the fluctuation theorems \cite{ESM01}.
Thus, the fluctuation theorems are known to apply to wide classes of nonequilibrium systems.

\begin{figure}
\begin{center}
\includegraphics[width=0.9\columnwidth]{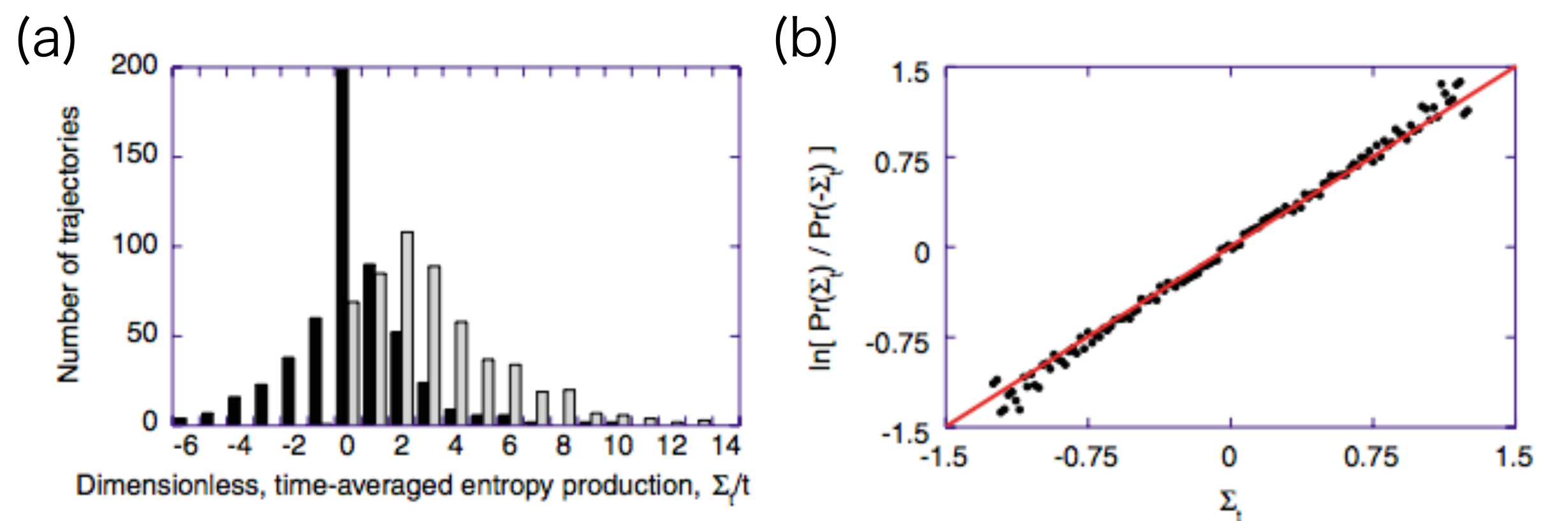}
\caption{\label{fig:WSMe}
(a)
Histograms of time-averaged entropy production of a dragged colloidal particle in units of $k_{\rm B}T$, where $k_{\rm B}$ is the Boltzmann constant.
Two different types of the bars represent two different measurement-time intervals.
We can observe that the probability of paths violating the second law does not vanish.
(b)
Log-ratio of the probability of entropy production to that of its negative.
The experimental results are consistent with the transient fluctuation theorem (\ref{TFT}).
Reproduced from Figs.\,\,1 and 4 in Ref. \cite{WSMe02}.
Copyright 2002 by the American Physical Society.
}
\end{center}
\end{figure}

The transient fluctuation theorem (\ref{TFT}) was experimentally demonstrated in Ref.~\cite{WSMe02}.
They prepared a colloidal particle in an optical trap at rest, and then translated the trap relative to the surrounding water.
In this transient situation, they obtained the probability distribution of entropy production, and observed trajectories of the particle violating the second law at the level of individual paths (see Fig.~\ref{fig:WSMe}).
The amount of this violation was confirmed to be consistent with Eq.~(\ref{TFT}).
Later, the steady-state fluctuation theorem (\ref{SSFT}) was also verified in the same setup \cite{WRCe05}.

\subsection{Jarzynski equality}
In 1997, Jarzynski discovered a remarkable exact nonequilibrium equality in a Hamiltonian system \cite{Jar97Apr}.
Let $H(\lambda, x)$ denote the Hamiltonian of the system, where $\lambda$ is an external parameter that we control to manipulate the system and $x$ represents internal degrees of freedom of the system.
The system is initially in equilibrium with the inverse temperature $\beta$, and we subject the system to a nonequilibrium process by our manipulation of $\lambda$ from $\lambda_{\rm i}$ to $\lambda_{\rm f}$.
Let $F(\lambda)$ denote the equilibrium free energy of the system under a given external parameter $\lambda$, i.e.,
\eqn{
	e^{-\beta F(\lambda)}= \int dx\ e^{-\beta H(\lambda, x)}.
}
The Jarzynski equality relates the free-energy difference $\Delta F:=F(\lambda_{\rm f}) - F(\lambda_{\rm i})$ to the probability distribution of work $W$ performed during the nonequilibrium process as
\eqn{\label{Jar}
	\ave{e^{-\beta(W-\Delta F)}}=1,
}
where the angular brackets mean the statistical average under the initial equilibrium state and the given nonequilibrium protocol.
Soon after the discovery, the same equality is proved in stochastic systems based on the master equation formalism \cite{Jar97Nov}.
It is noteworthy that we assume nothing about how fast we change the external parameter $\lambda$, and therefore the Jarzynski equality~(\ref{Jar}) remains valid under a rapid change of the parameter, which means that the Jarzynski equality applies to processes beyond the linear response regime.

Moreover, the Jarzynski equality~(\ref{Jar}) is an extension of conventional thermodynamic relations to the case of a rather general nonequilibrium regime \cite{Jar97Apr}.
First, it leads to a second-law-like inequality in isothermal processes.
Using Jensen's inequality, we obtain
\eqn{
	\ave{e^{-\beta(W-\Delta F)}} \ge e^{-\beta\ave{W-\Delta F}}.
}
Combining this inequality with the Jarzynski equality~(\ref{Jar}), we conclude
\eqn{\label{2ndlaw}
	\langle W \rangle\ge \Delta F,
}
which is the second law of thermodynamics in isothermal processes.
The equality condition is that $W$ has a single definite value, i.e., $W$ does not fluctuate.
We can regard $W-\Delta F$ as the total entropy production of the system.
Let $\Delta U$ and $Q$ denote the internal-energy difference and dissipated heat from the system to the heat bath, respectively.
Then, the first law of thermodynamics is $\Delta U = W-Q$.
We rewrite Eq.~(\ref{2ndlaw}) as
\eqn{
	\frac{\Delta U -\Delta F}{T} + \frac{Q}{T} \ge 0,
}
where $T$ is the temperature of the heat bath (and therefore the initial temperature of the system).
The first term represents the entropy production of the system $\Delta S$ because $\Delta F = \Delta U - T \Delta S$, and the second term is the entropy production of the heat bath.
Therefore, Eq.~(\ref{2ndlaw}) means that the entropy production of the total system must be positive.

Secondly, the Jarzynski equality~(\ref{Jar}) reproduces the fluctuation-dissipation relation in the linear response theory \cite{Jar97Apr}.
Let us denote $\sigma=\beta(W-\Delta F)$, and assume that the dissipated work $W-\Delta F$ is much smaller than the thermal energy $k_{\rm B}T$, namely, $|\sigma|\ll 1$.
Expanding $\ln\ave{e^{-\sigma}}$ up to the second order in $\sigma$, we obtain
\eqn{
	\ln\ave{e^{-\sigma}} \simeq -\ave{\sigma} + \frac{1}{2}(\ave{\sigma^2}-\ave{\sigma}^2).
}
Substituting the Jarzynski equality~(\ref{Jar}), we obtain
\eqn{\label{FDRforW}
	\ave{W}-\Delta F = \frac{1}{2k_{\rm B}T}(\ave{W^2}-\ave{W}^2).
}
The left-hand side is dissipation of the total system, and the right-hand side represents fluctuations of the work during the process.
This is one form of the fluctuation-dissipation relation.

Not only does the Jarzynski equality reproduce the second law of thermodynamics, but also it gives a stringent restriction on the probability of events violating the second law \cite{Jar99}.
Let $\sigma_0$ be a positive number and let us calculate the probability that the entropy production is smaller than $-\sigma_0$:
\eqn{
	{\rm Prob}[\sigma\le -\sigma_0]
	&=& \int_{-\infty}^{-\sigma_0} d\sigma\ P(\sigma)\nonumber\\
	&\le& \int_{-\infty}^{-\sigma_0} d\sigma\ P(\sigma) e^{-\sigma-\sigma_0}\nonumber\\
	&=& \ave{e^{-\sigma}} e^{-\sigma_0}\nonumber\\
	&=& e^{-\sigma_0},
}
where we use the Jarzynski equality~(\ref{Jar}) to obtain the last line.
Therefore, the probability of negative entropy production is exponentially suppressed.
While a negative entropy production ($\sigma<0$) may occasionally occur, a greatly negative entropy production ($|\sigma|\gg 1$) is effectively prohibited by the Jarzynski equality.

In addition to the above-described properties, the Jarzynski equality~(\ref{Jar}) enables us to determine the free-energy difference from our observation of how the system evolves under a  nonequilibrium process because
\eqn{\label{detF}
	\Delta F = -\frac{1}{\beta}\ln \ave{e^{-\beta W}},
}
although the free-energy difference is an equilibrium property of the system \cite{Jar97Nov, PS04}.
A naive method to determine the free-energy difference in experiments or numerical simulations is to conduct reversible measurements of work and use the fact $W=\Delta F$.
However, this method is not realistic in general because an extremely long time is needed to realize even approximately reversible processes.
A more sophisticated method is to use the linear response relation (\ref{FDRforW}) to determine the free energy difference through the work distribution obtained in the measurements.
This method may still be time-consuming because the manipulation must be slow enough for the system to remain in the linear response regime.
Equation (\ref{detF}) enables us to reduce the time of experiments or simulations when we calculate the free-energy difference, because Eq.~(\ref{detF}) is valid even for rapid nonequilibrium processes.
\footnote{The exact equality (\ref{detF}) requires more samples for convergence than the approximate equality (\ref{FDRforW}) does \cite{PS04}. Therefore, the total time required for convergence, which is the time of the process multiplied by the number of samples, can be longer when we use Eq. (\ref{detF}) than we use (\ref{FDRforW}).}

\begin{figure}
\begin{center}
\includegraphics[width=0.9\columnwidth]{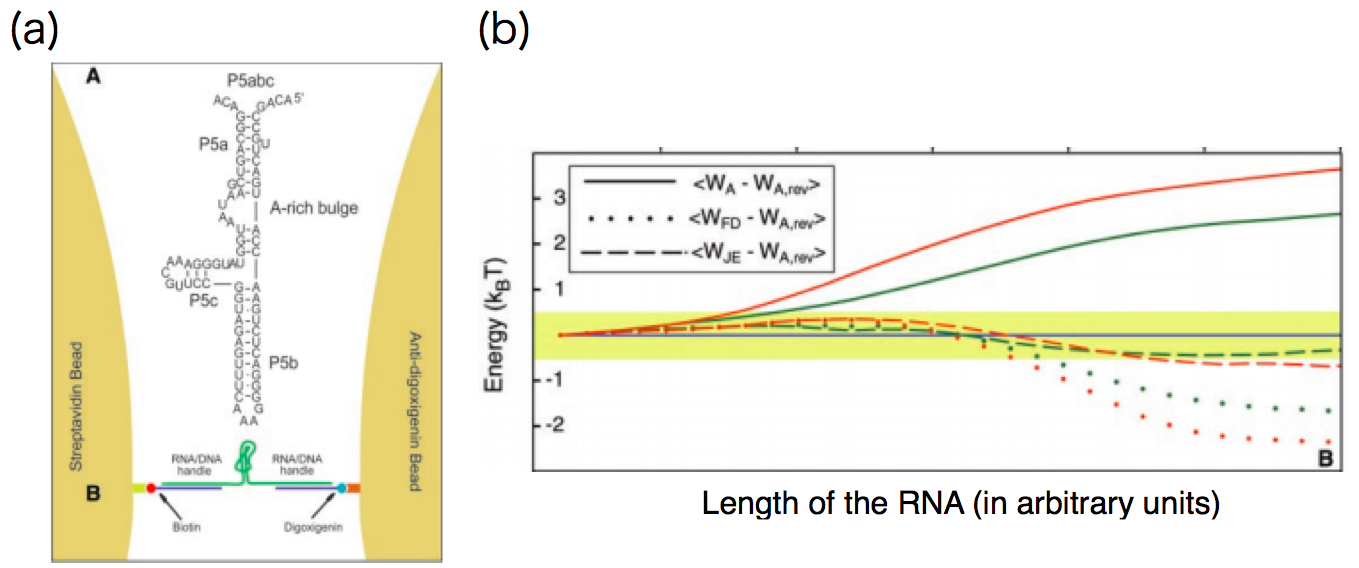}
\caption{\label{fig:LDSe}
(a) Experimental setup of the stretching experiment. An RNA molecule is attached to two beads.
One bead is in an optical trap to measure force and the other is linked to a piezoelectric actuator.
(b) Difference between the estimated free energy and its actual value. The solid curves represent the reversible estimation ($\Delta F=\ave{W}$).
The dotted curves represent the linear response estimation by Eq.~(\ref{FDRforW}).
The dashed curves represent the estimation based on the Jarzynski equality, namely, Eq. (\ref{detF}).
We see that the Jarzynski equality gives the best estimation.
Reproduced from Figs. 1 and 3 in Ref. \cite{LDSe02}.
Copyright 2002 by the American Association for the Advancement of Science.
}
\end{center}
\end{figure}

Hummer and Szabo found an experimentally useful variant of the Jarzynski equality \cite{HS01}.
The Hummer-Szabo equality can be utilized to rigorously reconstruct the free-energy landscape of a molecule from repeated measurements based, for example, on an atomic force microscope or an optical tweezer.
Shortly thereafter, in a setup with an optical tweezer shown in Fig. \ref{fig:LDSe} (a), the free-energy profile of a single molecule of RNA was reconstructed by mechanically stretching the RNA in an irreversible manner \cite{LDSe02} (see Fig. \ref{fig:LDSe} (b)).
This experiment demonstrated that the Jarzynski equality is useful in practice to determine free-energy differences of systems.
In a similar manner, the free-energy profile of a protein was estimated by stretching the protein by an atomic force microscope \cite{HSK07}.

\subsection{Crooks fluctuation theorem}
In 1998, Crooks offered a new proof of the Jarzynski equality in stochastic systems \cite{Cro98}.
What is remarkable about this proof is the method comparing an original process with the time-reversed process to derive nonequilibrium relations.
One year later, this idea led him to propose a novel relation now known as the Crooks fluctuation theorem \cite{Cro99}, which reads
\eqn{\label{CFT}
	\frac{P(\sigma)}{P^\dag(-\sigma)} = e^\sigma,
}
where $P(\sigma)$ is the probability distribution function of entropy production $\sigma$, and $P^\dag(\sigma)$ is that in the time-reversed process.
Later, this theorem was generalized to Hamiltonian systems with multiple heat baths by Jarzynski \cite{Jar00}.

The Crooks fluctuation theorem (\ref{CFT}) can be regarded as a generalized version of the steady-state fluctuation theorem (\ref{SSFT}) in systems symmetric with respect to reversal of the perturbation that drives the steady flow; in this case, we have $P^\dag(-\sigma)=P(-\sigma)$, which reduces Eq.~(\ref{CFT}) to Eq.~(\ref{SSFT}), because there is no difference between the original process and the time-reversed one.
What is more, the Crooks fluctuation theorem easily reproduces the Jarzynski equality as follows:
\eqn{\label{JRE}
	\ave{e^{-\sigma}}
	&=& \int_{-\infty}^\infty d\sigma\ e^{-\sigma}P(\sigma)
	\nonumber\\
	&=& \int_{-\infty}^\infty d\sigma\ P^\dag(\sigma)
	\nonumber\\
	&=& 1,
}
where we have used the Crooks fluctuation theorem (\ref{CFT}) to obtain the second line, and the normalization of probability to obtain the last line.
Moreover, 
the Crooks fluctuation theorem implies that 
\eqn{\label{Pcross}
	\sigma = 0\ \Leftrightarrow P(\sigma) = P^\dag(-\sigma).
}

Soon after, Crooks found a significant generalization of his own theorem \cite{Cro00}.
He related irreversibility of an individual path $\Gamma$ in phase space to its own entropy production $\sigma[\Gamma]$,
\footnote{
We use box brackets to indicate that $f[\Gamma]$ is a functional of $\Gamma$ instead of a function.
}
that is,
\eqn{\label{PCFT}
	\frac{\mP[\Gamma]}{\mP^\dag[\Gamma^\dag]} = e^{\sigma[\Gamma]},
}
where $\Gamma^\dag$ represents the time-reversed path of $\Gamma$; 
$\mP$ is the path-probability functional under a given dynamics, and $\mP^\dag$ is that under the time-reversed dynamics.
To reproduce Eq.~(\ref{CFT}) from Eq.~(\ref{PCFT}), we note
\eqn{
	P(\sigma)
	&=& \int \mD\Gamma\mP[\Gamma] \delta (\sigma[\Gamma]-\sigma)
	\nonumber \\
	&=& \int \mD\Gamma\mP^\dag[\Gamma^\dag] e^{\sigma[\Gamma]}
	\delta (\sigma[\Gamma]-\sigma)
	\nonumber \\
	&=& e^\sigma \int \mD\Gamma^\dag\mP^\dag[\Gamma^\dag] \delta(\sigma[\Gamma^\dag]+\sigma)
	\nonumber \\
	&=& e^\sigma P^\dag(-\sigma),
}
where $\mD\Gamma$ denotes the natural measure on the set of all paths, and we use Eq.~(\ref{PCFT}) to obtain the second line, and we assume entropy production is odd under time reversal, namely $\sigma[\Gamma]=-\sigma[\Gamma^\dag]$, to obtain the third line.
A more general form of the nonequilibrium integral equality can be derived based on Eq.~(\ref{PCFT}).
Let $\mF[\Gamma]$ be an arbitrary functional.
Then, we obtain
\eqn{\label{MFT}
	\langle \mF e^{-\sigma} \rangle &=&
	\int\mD\Gamma\mP[\Gamma]e^{-\sigma[\Gamma]}\mF[\Gamma]\nonumber\\
	&=&
	\int\mD\Gamma^\dag\mP^\dag[\Gamma^\dag]\mF[\Gamma]\nonumber\\
	&=& \langle \mF \rangle^\dag,
}
where $\langle\cdots\rangle^\dag$ denotes the average over the time-reversed probability $\mP^\dag[\Gamma^\dag]$.
When we set $\mF$ to unity, Eq.~(\ref{MFT}) reduces to Eq.~(\ref{JRE}).
The idea to consider the path-probability of an individual path is a crucial element to treat nonequilibrium equalities in a unified manner as described in Sec. 2.2.

\begin{figure}
\begin{center}
\includegraphics[width = 0.9\columnwidth]{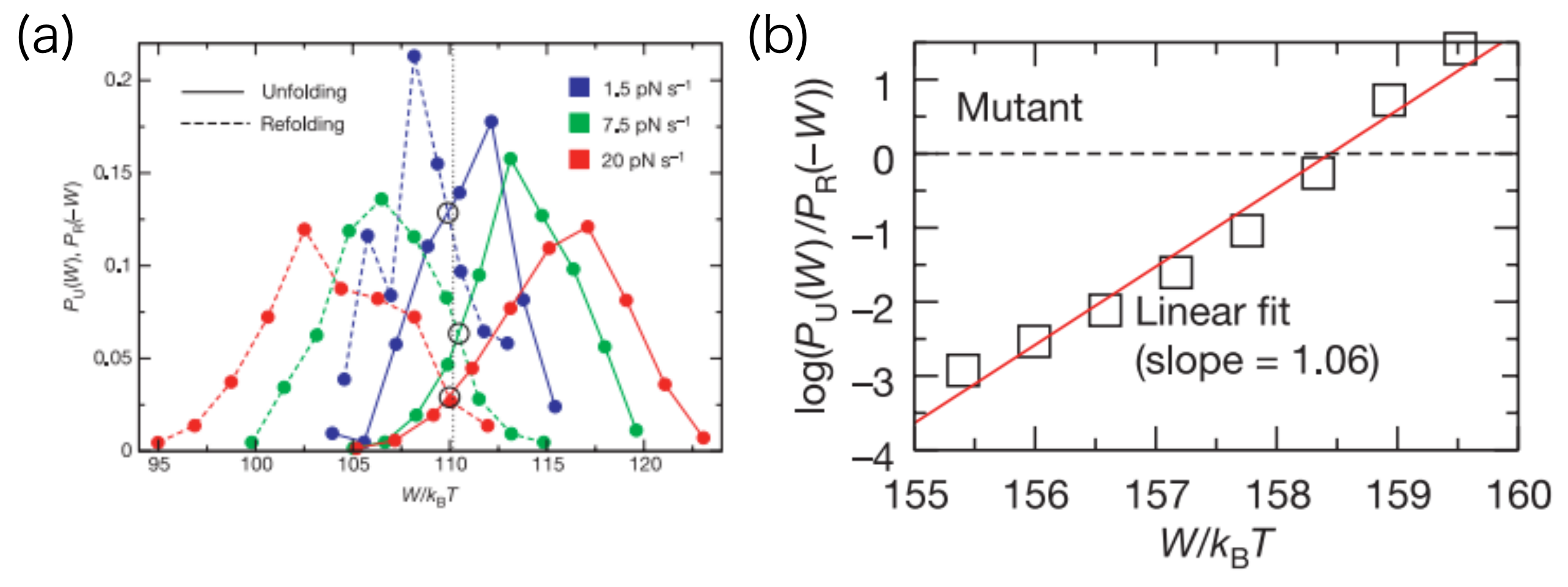}
\caption{\label{fig:CLJe}
(a)
Work distribution of the unfolding process (solid lines) and the refolding process (dashed lines).
Three different colors correspond to three different speed of unfolding and refolding.
We find that the work value where two distributions at the same speed cross each other is independent of the speed (as shown by the vertical dotted line).
Equation (\ref{Pcross}) implies that this value corresponds to zero entropy production, namely, the work value is equal to the free-energy difference.
(b)
Verification of the Crooks fluctuation theorem (\ref{CFT}).
In the setup of this experiment, the entropy production $\sigma$ reduces to work $W$ minus the free-energy difference.
It is illustrated that the log-ratio of the probability distributions of work is proportional to work itself.
Reproduced from Fig. 2 and the inset of Fig. 3 in Ref. \cite{CRJe05}.
Copyright 2005 by the Nature Publishing Group.
}
\end{center}
\end{figure}

The Crooks fluctuation theorem (\ref{CFT}) was experimentally verified \cite{CRJe05} in a similar setup in Ref. \cite{LDSe02}, which was used to verify the Jarzynski equality.
The work extracted when an RNA is unfolded and refolded was measured, and the work distribution of the unfolding process and that of the refolding process, which is the time reversed process of unfolding, were obtained (see Fig. \ref{fig:CLJe} (a)).
It was verified that the two distributions are consistent with the Crooks fluctuation theorem (\ref{CFT}) (see Fig. \ref{fig:CLJe} (b)).
Moreover, the free-energy difference of a folded RNA and an unfolded one was obtained using Eq.~(\ref{Pcross}), and the obtained value is consistent with a numerically obtained one.

\subsection{Further equalities}
In this section, we will briefly review further nonequilibrium equalities.
To this end, let us introduce some kinds of entropy production.
Mathematical definitions of these kinds of entropy production are presented in Sec. 2.2.

The total entropy production $\Delta s_{\rm tot}$ is the sum of the Shannon entropy production of the system $\Delta s$ and the entropy production of the heat bath $\Delta s_{\rm bath}$, that is,
\eqn{
	\Delta s_{\rm tot} = \Delta s + \Delta s_{\rm bath}.
}
The entropy production of the bath is related to the heat $q$ dissipated from the system to the bath
\eqn{
	\Delta s_{\rm bath} = q/T,
}
where $T$ is the absolute temperature of the bath.
In a steady-state situation, the heat $q$ is split into two parts as
\eqn{
	q = q_{\rm hk} + q_{\rm ex},
}
where $q_{\rm hk}$ is called the housekeeping heat, which is the inevitable heat dissipation to maintain the corresponding nonequilibrium steady state, and $q_{\rm ex}$ is called the excess heat, which arises due to a non-adiabatic change of the external control parameter.
Along with these definitions of heat, we define two kinds of entropy production as
\eqn{
	\Delta s_{\rm hk} = q_{\rm hk}/T,\ \Delta s_{\rm ex} = q_{\rm ex}/T.
}

\subsubsection*{Hatano-Sasa relation}
The Hatano-Sasa relation is a generalization of the Jarzynski equality \cite{HSa01}.
The Jarzynski equality relates the free-energy difference of two equilibrium states to the nonequilibrium average starting from one of the equilibrium states.
In a similar manner, the Hatano-Sasa relation relates the difference of nonequilibrium potentials $\phi$ of two nonequilibrium steady states, which is a generalization of $U-F$ in equilibrium situations, to an average starting from one of the steady state.
The Hatano-Sasa relation reads
\eqn{\label{HatanoSasa}
	\ave{e^{-\Delta \phi - \Delta s_{\rm ex}/k_{\rm B}}}= 1.
}
From Jensen's inequality, we obtain
\eqn{\label{GCI}
	\ave{\Delta \phi} + \frac{\ave{\Delta s_{\rm ex}}}{k_{\rm B}} \ge 0.
}
This inequality can be interpreted as a nonequilibrium version of the Clausius inequality ($\Delta s \ge -q/T$).
In fact, the inequality (\ref{GCI}) can be rewritten as
\eqn{
	k_{\rm B}\ave{\Delta \phi} \ge - \frac{\ave{q_{\rm ex}}}{T}.
}
The equality can be achieved in quasi-static transitions between the two nonequilibrium steady states, which is also analogous to the Clausius inequality, whose equality is also achieved in quasi-static transitions.

\subsubsection*{Seifert relation}
The Seifert relation applies to an arbitrary nonequilibrium process starting from an arbitrary initial state \cite{Sei05}.
The integral Seifert relation is given by
\eqn{\label{SR}
	\ave{e^{-\Delta s_{\rm tot}/k_{\rm B}}}=1.
}
The corresponding inequality is
\eqn{
	\ave{\Delta s_{\rm tot}} \ge 0,
}
which can be considered as a second law in a nonequilibrium process.
The detailed version, which is analogous to the Crooks fluctuation theorem, is given by
\eqn{\label{DSR}
	\frac{P(-\Delta s_{\rm tot})}{P(\Delta s_{\rm tot})} = e^{-\Delta s_{\rm tot}/k_{\rm B}}.
}
It is noteworthy that these relations hold for an arbitrary time interval.
Equation (\ref{DSR}) can be regarded as a refinement of the steady state fluctuation theorem (\ref{SSFT}).
The steady state fluctuation theorem (\ref{SSFT}) holds only in the long-time limit.
This is because $\Sigma$ in Eq.~(\ref{SSFT}) is in fact the entropy production rate of the bath and does not include that of the system.
Therefore, in Eq.~(\ref{SSFT}), the time interval should be long enough to ignore the entropy production of the system compared with that of the bath.

\subsubsection*{Relation for housekeeping entropy production}
In Ref. \cite{SS05}, a relation for housekeeping entropy production is also obtained as
\eqn{\label{hkr}
	\ave{e^{-\Delta s_{\rm hk}/k_{\rm B}}} = 1,
}
which leads to
\eqn{
	\ave{\Delta s_{\rm hk}} \ge 0.
}
\section{Unified formulation based on reference probabilities}
\begin{table}[b]
\caption{\label{tab}
Choices of the reference dynamics and the associated entropy productions.
Specific choices of the reference dynamics lead to specific entropy productions with specific physical meanings.
The meaning of the ``boundary term'' is explained later.
}
\begin{center}
\begin{tabular}{|c|c|}
\hline
Reference dynamics & Entropy production $\sigma$\\
\hline\hline
time reversal & $\Delta s_{\rm bath}/k_{\rm B}+({\rm boundary\ term})$\\
steady-flow reversal & $\Delta s_{\rm hk}/k_{\rm B}+({\rm boundary\ term})$\\
time reversal + steady-flow reversal & $\Delta s_{\rm ex}/k_{\rm B}+({\rm boundary\ term})$\\
\hline
\end{tabular}
\end{center}
\end{table}

In this section, we review a unified strategy to derive the nonequilibrium equalities introduced in the previous section.

In Ref. \cite{Cro00}, Crooks revealed that a unified approach to derive nonequilibrium equalities is to compare the nonequilibrium process with the time-reversed process and obtain a detailed fluctuation theorem, namely, the Crooks fluctuation theorem given by
\eqn{\label{CFTRNE}
	\frac{\mP^\dag[\Gamma^\dag]}{\mP[\Gamma]} = e^{-\beta(W[\Gamma]-\Delta F)},
} 
where quantities with a superscript $\dag$ are the ones in the time-reversed process.
Later, Hatano and Sasa derived the nonequilibrium equality (\ref{HatanoSasa}) in steady states by comparing the original dynamics with its ``time-reversed dual'' dynamics in a sense to be specified later.
In their derivation, they essentially used a detailed fluctuation theorem given by
\eqn{\label{HS2}
	\frac{\mP^\dag[\Gamma^\dag]}{\mP[\Gamma]} = e^{-\Delta \phi[\Gamma] - \Delta s_{\rm ex}[\Gamma]/k_{\rm B}},
}
where quantities accompanied by $\dag$ are, in this equality, the ones in the time-reversed dual process.
Their work indicates that a further unification is possible, namely, a detailed fluctuation theorem will be obtained when we compare the original dynamics with a properly chosen reference dynamics.
With this speculation, let us generalize the detailed fluctuation theorems (\ref{CFTRNE}) and (\ref{HS2}) to
\eqn{\label{DFT}
	\frac{\mP^{\rm r}[\Gamma]}{\mP[\Gamma]} = e^{-\sigma[\Gamma]},
}
where $\mP^{\rm r}$ represents a reference probability of a reference path, and $\sigma[\Gamma]$ is a formal entropy production.
In the case of the Crooks fluctuation theorem (\ref{CFT}), the reference is the time reversal, and $\sigma[\Gamma]$ reduces to $W[\Gamma]-\Delta F$.
In the case considered by Hatano and Sasa, in a similar way, the reference is the time-reversed dual, and $\sigma[\Gamma]$ reduces to $\Delta \phi[\Gamma] + \Delta s_{\rm ex}[\Gamma]/k_{\rm B}$.
Once we obtain the detailed fluctuation theorem (\ref{DFT}), we succinctly derive an integral nonequilibrium equality given by
\eqn{
	\ave{e^{-\sigma}} = 1,
}
because
\eqn{
	\ave{e^{-\sigma}}
	&=&
	\int \mD\Gamma e^{-\sigma}\mP[\Gamma]
	\nonumber\\
	&=&
	\int \mD\Gamma \mP^{\rm r}[\Gamma]
	\nonumber\\
	&=&
	1,
}
where we use Eq.~(\ref{DFT}) to obtain the second line, and we use the normalization condition for the reference probability to obtain the last line.
In the same way, we obtain
\eqn{\label{RMFT}
	\langle \mF e^{-\sigma} \rangle = \langle \mF \rangle^{\rm r},
}
where $\mF[\Gamma]$ is an arbitrary functional, and $\langle\cdots\rangle^{\rm r}$ represents the average over the reference probability $\mP^{\rm r}[\Gamma]$.

In the rest of this section, we validate Eq.~(\ref{DFT}) in specific systems.
To be precise, we show that appropriate choices of the reference probability make the formal entropy production $\sigma$ reduce to physically meaningful entropy productions.
The results are summarized in Table \ref{tab}.

\subsection{Langevin system}
First of all, we consider a one-dimensional overdamped Langevin system.
One reason why we deal with the Langevin system as a paradigm is that steady states can be simply achieved by applying an external driving force.
Moreover, the Langevin system is thermodynamically sound in that it relaxes to the thermal equilibrium state, i.e., the Gibbs state after a sufficiently long time without the external driving force.

This part is mainly based on a review article by Seifert, i.e., Ref. \cite{Sei12}.

\subsubsection*{Basic properties and definition of heat}
Let $x(t)$ denote the position of a particle at time $t$ in a thermal environment with temperature $T$.
We consider a nonequilibrium process from time $t=0$ to $\tau$ controlled by an external parameter $\lambda(t)$.
The overdamped Langevin equation is given by
\eqn{\label{Laneq}
	\dot x(t) = \mu F(x(t),\lambda(t)) + \zeta(t),
}
where $\mu$ is the mobility, and $F(x, \lambda)$ is a systematic force applied to the particle with the position $x$ when the external control parameter is $\lambda$,
and $\zeta(t)$ represents a random force.
We assume that $\zeta(t)$ is a white Gaussian noise satisfying 
\eqn{
	\ave{\zeta(t)\zeta(t')} = 2D \delta (t-t'),
}
where $D$ is the diffusion constant.
The systematic force $F(x, \lambda)$ consists of two parts, that is,
\eqn{
	F(x,\lambda) = -\dl_x V(x, \lambda) + f(x, \lambda).
}
The first part is due to the conservative potential $V(x,\lambda)$ and $f(x,\lambda)$ is the external driving force.
Under this Langevin dynamics, the probability to generate an entire trajectory $\{x\}$ starting from $x_0=x(0)$ under a given entire protocol $\{\lambda\}$ is calculated as
\eqn{\label{PIform}
	P[\{x\}|x_0, \{\lambda\}] = \mathcal{N} e^{-\mA[\{x\}, \{\lambda\}]},
}
where the action $\mA[\{x\},\{\lambda\}]$ of the trajectory $\{x\}$ is
\eqn{\label{action}
	\mA[\{x\},\{\lambda\}]
	=
	\int_0^\tau dt \left[
		\frac{(\dot x -\mu F(x, \lambda))^2}{4D} + \mu\frac{\dl_x F(x, \lambda)}{2}
	\right]
}
(see Appendix A for a derivation).
Another strategy to describe the system is to trace the probability $p(x,t)$ to find the particle at $x$ at time $t$.
We can show $p(x,t)$ to obey the Fokker-Planck equation given by
\eqn{\label{FPeq}
	\dl_t p(x,t) = -\dl_x j(x,t),
}
where the probability current $j(x,t)$ is defined as
\eqn{
	j(x,t) = \mu F(x,\lambda(t))p(x,t) - D\dl_x p(x,t)
}
(see Appendix A for a derivation).
With this probability, the entropy of the system is defined as the stochastic Shannon entropy of the system, i.e., 
\eqn{
	s(t) = -k_{\rm B}\ln p(x(t), t).
}
The mean local velocity is defined as
\eqn{
	v(x,t) = \frac{j(x,t)}{p(x,t)}.
}

When the external driving force is not applied, i.e., $f(x, \lambda)=0$, the system relaxes to the thermal equilibrium state given by
\eqn{\label{eqstate}
	p_{\rm eq}(x,\lambda) = e^{-\beta (V(x,\lambda)-F(\lambda))},
}
where $\beta$ is the inverse temperature and free energy $F(\lambda)$ is defined as
\eqn{
	e^{-\beta F(\lambda)} = \int dx e^{-\beta V(x, \lambda)},
}
because the kinetic energy contributes nothing due to the assumption of overdamping.

Next, we consider how we should define heat in this Langevin system \cite{Sek97}.
The dissipated heat is the energy flow from the system to the bath.
This energy transfer is done by the viscous friction force $-\gamma \dot x$ and the thermal noise $\gamma\zeta(t)$, where $\gamma=1/\mu$ is the friction coefficient.
Therefore, the ``work'' done by these force should be identified with the heat flowing into the system.
Thus we define the dissipated heat as
\eqn{
	q[\{x\}] = - \int_0^\tau dt \dot x (-\gamma\dot x + \gamma\zeta).
}
Using the Langevin equation~(\ref{Laneq}), we obtain
\eqn{\label{defq}
	q[\{x\}] = \int_0^\tau dt \dot x F(x,\lambda).	
}
Now we can define the entropy production of the heat bath as
\eqn{\label{defentbath}
	\Delta s_{\rm bath} = \frac{q}{T}.
}
In this way, the concepts of the heat and entropy are generalized to the level of an individual stochastic trajectory.

\subsubsection*{Steady-state properties and definitions of thermodynamic quantities}
When the external driving force $f(x,\lambda)$ is applied at a fixed $\lambda$, the system relaxes to a nonequilibrium steady state $p_{\rm ss} (x, \lambda)$.
In the analogy of the equilibrium state~(\ref{eqstate}), let us define a nonequilibrium potential $\phi(x,\lambda)$ by
\eqn{\label{sssta}
	p_{\rm ss} (x,\lambda) = e^{-\phi(x,\lambda)}.
}
The steady current is defined as
\eqn{\label{sscur}
	j_{\rm ss}(x,\lambda) = \mu F(x,\lambda)p_{\rm ss}(x, \lambda)
	-D\dl_x p_{\rm ss}(x, \lambda),
}
and the mean velocity in the steady state is given by
\eqn{\label{ssvel}
	v_{\rm ss}(x,\lambda) &=& \frac{j_{\rm ss}(x,\lambda)}{p_{\rm ss}(x,\lambda)}
	\nonumber\\
	&=& \mu F(x, \lambda) - D\dl_x \ln p_{\rm ss}(x,\lambda)
	\nonumber\\
	&=& \mu F(x, \lambda) + D\dl_x \phi(x,\lambda).
}
Using this expression, we rewrite Eq.~(\ref{defq}) as
\eqn{
	q[\{x\}]
	=
	\frac{1}{\mu}\int_0^\tau dt \dot x v_{\rm ss}(x,\lambda)
	-
	k_{\rm B}T \int_0^\tau dt \dot x \dl_x \phi(x,\lambda),
}
where the Einstein relation $D=\mu k_{\rm B} T$ is assumed.
The first term is the inevitable dissipation proportional to the steady mean velocity $v_{\rm ss}$ and is therefore identified as the housekeeping heat, namely,
\eqn{
	q_{\rm hk}[\{x\}] = \frac{1}{\mu}\int_0^\tau dt \dot x v_{\rm ss}(x,\lambda),
}
and the second term is the excess contribution after the subtraction of the housekeeping part from the total heat, and defined as
\eqn{
	q_{\rm ex}[\{x\}] = 
	-k_{\rm B}T \int_0^\tau dt \dot x \dl_x \phi(x,\lambda).
}
Therefore, following Ref.~\cite{HSa01}, we define two kinds of entropy production as
\eqn{\label{defenthk}
	\Delta s_{\rm hk}[\{x\}] = \frac{k_{\rm B}}{D}\int_0^\tau dt \dot x v_{\rm ss}(x,\lambda),
}
and
\eqn{\label{defentex}
	\Delta s_{\rm ex}[\{x\}] &=& 
	-k_{\rm B} \int_0^\tau dt \dot x \dl_x \phi(x,\lambda)
	\nonumber\\
	&=&
	k_{\rm B}\left[
	 -\Delta \phi + \int_0^\tau dt \dot \lambda \dl_\lambda \phi(x,\lambda)
	 \right],
}
because $d\phi = (\dl_x\phi) dx + (\dl_\lambda\phi) d\lambda$.
Note that the ensemble average of the excess entropy production vanishes in steady states because $\lambda$ is independent of time, which justifies that the excess entropy production is indeed an excess part due to non-adiabatic changes of the control parameter $\lambda$.

\subsubsection*{Time reversal and $\Delta s_{\rm bath}$}
We consider a process starting from an initial probability distribution $p_0(x_0)$ under a protocol $\{\lambda\}$, and the time-reversed process starting from an initial probability distribution $p^\dag_0(x_0^\dag)$ under the time-reversed protocol $\{\lambda^\dag\}$ defined by $\lambda^\dag(t):=\lambda(\tau -t)$.
Here, we do not assume any relations between $p_0(x_0)$ and $p_0^\dag(x_0^\dag)$.
Now we compare the realization probability of the original path $\{x\}$ and that of the time-reversed path $\{x^\dag\}$ defined by $x^\dag(t)=x(\tau-t)$.
The original probability is
\eqn{
	\mP[\{x\}|\{\lambda\}]
	&=&
	\mP[\{x\}|x_0,\{\lambda\}]p_0(x_0)
	\nonumber\\
	&=&
	\mN p_0(x_0) e^{-\mA[\{x\},\{\lambda\}]},
}
and the time-reversed probability is
\eqn{
	\mP[\{x^\dag\}|\{\lambda^\dag\}]
	&=&
	\mP[\{x^\dag\}|x_0^\dag,\{\lambda^\dag\}]p_0^\dag(x_0^\dag)
	\nonumber\\
	&=&
	\mN p_0^\dag(x_0^\dag) e^{-\mA[\{x^\dag\},\{\lambda^\dag\}]}.
}
Therefore, the formal entropy production defined in Eq.~(\ref{DFT}) reduces to
\eqn{
	\sigma[\{x\}]
	&=&
	-\ln \frac{\mP[\{x^\dag\}|\{\lambda^\dag\}]}{\mP[\{x\}|\{\lambda\}]}
	\nonumber\\
	&=& (\mA[\{x^\dag\},\{\lambda^\dag\}] - \mA[\{x\}, \{\lambda\}])
	-\ln \frac{p_0^\dag(x_0^\dag)}{p_0(x_0)}.
}
The first term is called the bulk term, and the second term is called the boundary term, because the second one arises from the boundary conditions $p_0$ and $p_0^\dag$.
Using Eq.~(\ref{action}), we obtain
\eqn{
	\mA[\{x^\dag\},\{\lambda^\dag\}]
	&=&
	\int_0^\tau dt \left[
		\frac{(\dot x^\dag(t) -\mu F(x^\dag(t), \lambda^\dag(t)))^2}{4D} + \mu\frac{\dl_x F(x^\dag(t), \lambda^\dag(t))}{2}
	\right]
	\nonumber\\
	&=&
	\int_0^\tau dt \left[
		\frac{(-\dot x(\tau-t) -\mu F(x(\tau-t), \lambda(\tau-t)))^2}{4D} + \mu\frac{\dl_x F(x(\tau-t), \lambda(\tau-t))}{2}
	\right]
	\nonumber\\
	&=&
	\int_0^\tau dt \left[
		\frac{(\dot x(t) +\mu F(x(t), \lambda(t)))^2}{4D} + \mu\frac{\dl_x F(x(t), \lambda(t))}{2}
	\right],
}
where we change the integration variable from $t$ to $\tau-t$ to obtain the last line.
Therefore, the bulk term is
\eqn{\label{AAq}
	\mA[\{x^\dag\},\{\lambda^\dag\}]-\mA[\{x\},\{\lambda\}]
	=
	\frac{1}{k_{\rm B}T} \int_0^\tau dt \dot x F(x,\lambda)
	=\frac{q[\{x\}]}{\kb T},
}
where we use the definition~(\ref{defq}) to obtain the last equality.
By Eq.~(\ref{defentbath}), we obtain
\eqn{\label{sigmabath}
	\sigma[\{x\}] = \frac{\Delta s_{\rm bath}}{k_{\rm B}}
	-\ln \frac{p_0^\dag(x_0^\dag)}{p_0(x_0)}.
}

Here, we assume that the initial state is in equilibrium
\eqn{
	p_0(x_0)=p_{\rm eq}(x_0, \lambda_0)=e^{-\beta(U(x_0, \lambda_0)-F(\lambda_0))}.
}
Although the initial state of the reversed dynamics can be set to an arbitrary probability distribution, we set it also to the canonical ensemble as
\eqn{
	p_0^\dag(x_0^\dag) = p_{\rm eq}(x_0^\dag, \lambda_0^\dag)&=&e^{-\beta(U(x_0^\dag, \lambda_0^\dag)-F(\lambda_0^\dag))}\nonumber\\
	&=&e^{-\beta(U(x_\tau, \lambda_\tau)-F(\lambda_\tau))}.
}
In this case, we obtain
\eqn{
	\sigma[\{x\}]
	&=&
	\frac{\Delta s_{\rm bath}}{k_{\rm B}} + \beta(\Delta U -\Delta F)
	\nonumber\\
	&=&
	\beta(q+\Delta U -\Delta F)
	\nonumber\\
	&=&
	\beta(W-\Delta F),
}
which means that we obtain the Crooks fluctuation theorem~(\ref{CFT}).
Therefore, we also obtain the Jarzynski equality~(\ref{Jar}) succinctly.

Next, we assume nothing about $p_0$, and set the initial probability of the reversed dynamics to the final probability of the original dynamics, namely,
\eqn{
	p_0^\dag(x_0^\dag) = p(x_0^\dag, \tau)=p(x_\tau, \tau).
}
In this case, the boundary term reduces to the Shannon entropy production of the system as
\eqn{
	\sigma[\{x\}] &=& \frac{\Delta s_{\rm bath}}{k_{\rm B}} - \ln p(x_\tau, \tau) + \ln p(x_0, 0)
	\nonumber\\
	&=&
	\frac{\Delta s_{\rm bath}+s(\tau) -s(0)}{k_{\rm B}}
	\nonumber\\
	&=&
	\frac{\Delta s_{\rm bath}+\Delta s}{k_{\rm B}}.
}
Therefore, we obtain
\eqn{
	\sigma[\{x\}] = \frac{\Delta s_{\rm tot}}{k_{\rm B}},
}
and the detailed Seifert relation~(\ref{DSR}), which automatically derives the integral Seifert relation~(\ref{SR}).

Finally, we assume nothing about $p_0$, and set the initial probability of the reversed dynamics to $p_0$.
Then, we obtain
\eqn{
	\sigma[\{x\}] = \Omega[\{x\}],
}
where
\eqn{
	\Omega[\{x\}]=\frac{\Delta s_{\rm bath}}{k_{\rm B}} - \ln p_0(x_\tau) + \ln p_0(x_0)
}
is the dissipation functional defined in Ref.~\cite{SE00} and used in the transient fluctuation theorem~(\ref{TFT}).

\subsubsection{Dual (steady-flow reversal) and $\Delta s_{\rm hk}$}
Next, we consider the steady-flow-reversed dynamics.
To this aim, we define the dual of the external force $F(x,\lambda)$ as
\eqn{\label{defdual}
	F^\dag(x,\lambda) = F(x, \lambda) -\frac{2v_{{\rm ss}, F}(x,\lambda)}{\mu},
}
where we use $v_{{\rm ss},F}$ to represent the steady velocity under force $F$.
Let us demonstrate that the steady state under force $F$ is also the steady state under force $F^\dag$.
Comparing Eq.~(\ref{ssvel}), we obtain
\eqn{\label{defdual2}
	\mu F^\dag(x, \lambda) = -\mu F(x, \lambda) -2D\dl_x \phi_F(x, \lambda),
}
or
\eqn{\label{rvs}
	\mu F^\dag(x, \lambda) + D\dl_x \phi_F(x, \lambda)
	= -(\mu F(x, \lambda) + D\dl_x \phi_F(x, \lambda)).
}
Since $p_{{\rm ss}, F}(x, \lambda)= e^{-\phi_F(x, \lambda)}$ is the steady solution of the Fokker-Planck equation~(\ref{FPeq}), we have
\eqn{
	0 = \dl_x j_{{\rm ss}, F}(x, \lambda)
	&=& \dl_x (\mu F(x, \lambda) p_{{\rm ss}, F}(x, \lambda)-D\dl_x p_{{\rm ss}, F}(x, \lambda))
	\nonumber\\
	&=&
	\dl_x[(\mu F(x, \lambda)+ D\dl_x\phi_F(x,\lambda))p_{{\rm ss}, F}(x, \lambda)].
}
Using Eq.~(\ref{rvs}), we obtain
\eqn{
	0 &=& \dl_x[(\mu F^\dag(x, \lambda)+ D\dl_x\phi_F(x,\lambda))p_{{\rm ss}, F}(x, \lambda)]
	\nonumber\\
	&=&
	\dl_x (\mu F^\dag(x, \lambda) p_{{\rm ss}, F}(x, \lambda)-D\dl_x p_{{\rm ss}, F}(x, \lambda)),
}
which means $p_{{\rm ss}, F}$ is also the steady state solution under force $F^\dag$, that is,
\eqn{
	p_{{\rm ss}, F}(x, \lambda) &=& p_{{\rm ss}, F^\dag}(x, \lambda)
	\ (=:p_{\rm ss}(x, \lambda)),\\
	\phi_F(x,\lambda) &=& \phi_{F^\dag}(x,\lambda)
	\ (=:\phi(x,\lambda)).
}
Comparing Eqs.~(\ref{sscur}) and (\ref{rvs}), we have
\eqn{
	j_{{\rm ss}, F}(x, \lambda) =
	-j_{{\rm ss}, F^\dag}(x, \lambda)
	\ (=:j_{{\rm ss}}(x, \lambda)).
}
Therefore, the dual dynamics has the same steady state as the original dynamics and the negative of the steady current in the original dynamics.

Now, we consider a process starting from an initial probability distribution $p_0(x_0)$ and the dual process starting from an initial probability distribution $p^\dag_0(x_0)$.
The original probability is
\eqn{
	\mP_F[\{x\}|\{\lambda\}] &=& \mP_F[\{x\}|x_0, \{\lambda\}]p_0(x_0)
	\nonumber\\
	&=& \mN p_0(x_0) e^{-\mA_F[\{x\}, \{\lambda\}]}
}
and the dual probability is
\eqn{
	\mP_{F^\dag}[\{x\}|\{\lambda\}] &=& \mP_{F^\dag}[\{x\}|x_0, \{\lambda\}]p_0(x_0)
	\nonumber\\
	&=& \mN p_0^\dag(x_0) e^{-\mA_{F^\dag}[\{x\}, \{\lambda\}]}.
}
Therefore, the formal entropy production reduces to
\eqn{
	\sigma[\{x\}]
	=
	(\mA_{F^\dag}[\{x\}, \{\lambda\}]-\mA_F[\{x\}, \{\lambda\}])
	-\ln\frac{p_0^\dag(x_0)}{p_0(x_0)}.
}
The bulk term can be calculated as
\eqn{
	&&\mA_{F^\dag}[\{x\}, \{\lambda\}]
	-\mA_{F}[\{x\}, \{\lambda\}]
	\nonumber\\
	&&=
	\int_0^\tau dt \left[
		\frac{(\dot x-\mu F^\dag)^2}{4D} + \mu \frac{\dl_x F^\dag}{2}
	\right]
	-
	\int_0^\tau dt \left[
		\frac{(\dot x-\mu F)^2}{4D} + \mu \frac{\dl_x F}{2}
	\right]
	\nonumber\\
	&&=
	\int_0^\tau dt \left[
		\frac{\mu}{4D}(2\dot x -\mu F-\mu F^\dag)(F-F^\dag)
		+ \frac{\mu}{2}\dl_x(F^\dag-F)
	\right].
}
Using Eqs.~(\ref{defdual}) and (\ref{defdual2}), we obtain
\eqn{
	\mA_{F^\dag}[\{x\}, \{\lambda\}]
	-\mA_{F}[\{x\}, \{\lambda\}]
	&=&
	\int_0^\tau dt \left[
		\frac{1}{D}(\dot x +D\dl_x \phi)v_{\rm ss}
		-\dl_x v_{\rm ss}
	\right]
	\nonumber \\
	&=&
	\frac{1}{D}\int_0^\tau dt \dot x v_{\rm ss}
	-\int_0^\tau dt e^{\phi}\dl_x j_{\rm ss}
	\nonumber \\
	&=&
	\frac{1}{D}\int_0^\tau dt \dot x v_{\rm ss},
}
where we use $j_{\rm ss} = v_{\rm ss}e^{-\phi}$ to obtain the second line and the fact that the divergence of the current vanishes in the steady state to obtain the last line.
By the definition~(\ref{defenthk}), we obtain
\eqn{
	\sigma[\{x\}]
	= \frac{\Delta s_{\rm hk}}{k_{\rm B}} - \ln \frac{p_0^\dag(x_0)}{p_0(x_0)}.
}
When we set $p_0^\dag$ to $p_0$, the formal entropy production reduces to
\eqn{
	\sigma[\{x\}] =  \frac{\Delta s_{\rm hk}}{k_{\rm B}},
}
and therefore we obtain the integral fluctuation theorem for the housekeeping entropy production~(\ref{hkr}).

\subsubsection*{Time-reversed dual and $\Delta s_{\rm ex}$}
We consider a process starting from an initial probability distribution $p_0(x_0)$ under a protocol $\{\lambda\}$, and compare it with the dual process starting from an initial probability distribution $p_0^\dag(x_0^\dag)$ under the time-reversed protocol $\{\lambda^\dag\}$.
The original probability is
\eqn{
	\mP_F[\{x\}|\{\lambda\}]
	&=&
	\mP_F[\{x\}|x_0, \{\lambda\}] p_0(x_0)
	\nonumber\\
	&=&
	\mN p_0(x_0) e^{-\mA_F[\{x\}, \{\lambda\}]}
}
and the time-reversed dual probability is
\eqn{
	\mP_{F^\dag}[\{x^\dag\}|\{\lambda^\dag\}]
	&=&
	\mP_{F^\dag}[\{x^\dag\}|x_0^\dag, \{\lambda^\dag\}] p_0^\dag(x_0^\dag)
	\nonumber\\
	&=&
	\mN p_0^\dag(x_0^\dag) e^{-\mA_{F^\dag}[\{x^\dag\}, \{\lambda^\dag\}]}
}
Therefore, we obtain
\eqn{
	\sigma[\{x\}]
	=
	(\mA_{F^\dag}[\{x^\dag\}, \{\lambda^\dag\}]-\mA_F[\{x\}, \{\lambda\}])
	-\ln \frac{p_0^\dag(x_0^\dag)}{p_0(x_0)}.
}
The bulk term is calculated as
\eqn{
	&&\mA_{F^\dag}[\{x^\dag\}, \{\lambda^\dag\}]
	-\mA_{F}[\{x\}, \{\lambda\}]
	\nonumber\\
	&&=
	\int_0^\tau dt \left[
		\frac{(-\dot x-\mu F^\dag)^2}{4D} + \mu \frac{\dl_x F^\dag}{2}
	\right]
	-
	\int_0^\tau dt \left[
		\frac{(\dot x-\mu F)^2}{4D} + \mu \frac{\dl_x F}{2}
	\right]
	\nonumber\\
	&&=
	\int_0^\tau dt \left[
		-\frac{\mu}{4D}(F+F^\dag)(-2\dot x + \mu (F-F^\dag)) +
		\frac{\mu}{2}\dl_x(F^\dag-F)
	\right].
}
Using Eqs.~(\ref{defdual}) and (\ref{defdual2}), we obtain
\eqn{
	\mA_{F^\dag}[\{x^\dag\}, \{\lambda^\dag\}]
	-\mA_{F}[\{x\}, \{\lambda\}]
	&=&
	\int_0^\tau dt \left[
		(\dl_x \phi)(- \dot x + v_{\rm ss}) -\dl_x v_{\rm ss}
	\right]	
	\nonumber\\
	&=&
	-\int_0^\tau dt \dot x \dl_x\phi - \int_0^\tau dt e^\phi \dl_x j_{\rm ss}
	\nonumber\\
	&=&
	-\int_0^\tau dt \dot x \dl_x\phi.
}
By the definition~(\ref{defentex}), we obtain
\eqn{
	\sigma[\{x\}] = \frac{\Delta s_{\rm ex}}{k_{\rm B}} - \ln\frac{p_0^\dag(x_0^\dag)}{p_0(x_0)}.
}

We assume that the initial state is the nonequilibrium steady state given by
\eqn{
	p_0(x_0) = e^{-\phi(x_0, \lambda_0)}
}
and set the initial state of the time-reversed dual dynamics to the nonequilibrium steady state as
\eqn{
	p_0^\dag(x_0^\dag) &=& e^{-\phi(x_0^\dag, \lambda_0^\dag)}\nonumber\\
	&=& e^{-\phi(x_\tau, \lambda_\tau)}.
}
Then, we obtain
\eqn{
	\sigma[\{x\}] &=& \frac{\Delta s_{\rm ex}}{k_{\rm B}} + \phi(x_\tau,\lambda_\tau) -\phi(x_0, \lambda_0)\nonumber\\
	&=&
	 \frac{\Delta s_{\rm ex}}{k_{\rm B}} + \Delta \phi.
}
Therefore, we reproduce the Hatano-Sasa relation~(\ref{HatanoSasa}) in a simple manner.

\subsection{Hamiltonian system}
Next, we consider a Hamiltonian system consisting of a system and a heat bath with inverse temperature $\beta$.
We consider only time reversal as the reference dynamics because it is difficult to define steady flow in a Hamiltonian system in general.

This part is partly based on Ref. \cite{Jar04}.

\subsubsection{Setup}
Let $z$ denote the position in the phase space of the total system.
We separate the degrees of freedom $z$ into two parts as $z=(x, y)$, where $x$ denotes the degrees of freedom of the system, and $y$ is the degrees of freedom of the bath.
We assume that the Hamiltonian of the total system can be decomposed into
\eqn{\label{Hdecom}
	H_{\rm tot}(z, \lambda) = H(x, \lambda) + H_{\rm int}(x, y, \lambda) + H_{\rm bath}(y),
}
where $\lambda$ is an external control parameter.
We also assume that the Hamiltonian is invariant under time reversal.
We vary $\lambda$ from time $t=0$ to $\tau$, and the system is subject to a nonequilibrium process.
The free energy of the bath $F_{\rm bath}$ is defined by
\eqn{
	e^{-\beta F_{\rm bath}} = \int dy e^{-\beta H_{\rm bath}(y)},
}
and is invariant during the process.
Moreover, we define an effective Hamiltonian of the system $H_{\rm ef}(x,\lambda)$
by tracing out the degrees of freedom of the bath as
\eqn{\label{defeff}
	e^{-\beta H_{\rm ef}(x, \lambda)}e^{-\beta F_{\rm bath}}
	=
	e^{-\beta H(x, \lambda)}\int dy e^{-\beta(H_{\rm int}(x, y)+H_{\rm bath}(y))}.
}
The free energy of the total system $F_{\rm tot}(\lambda)$ defined by
\eqn{
	e^{-\beta F_{\rm tot}(\lambda)} = \int dz e^{-\beta H_{\rm tot}(z, \lambda)}
}
and the free energy $F(\lambda)$ based on $H_{\rm ef}(x, \lambda)$ defined by
\eqn{
	e^{-\beta F(\lambda)} = \int dx e^{-\beta H_{\rm ef}(x, \lambda)}
}
are related by
\eqn{\label{fff}
	F_{\rm tot}(\lambda) = F(\lambda) + F_{\rm bath}.
}

\subsubsection{Time reversal}
We compare the original process $\{z\}$ starting from an initial probability distribution of the total system $p_0(z_0)$ with the time-reversed process $\{z^\dag\}$ starting from an initial probability distribution of the total system $p_0^\dag(z_0^\dag)$, where $z^\dag(t) = z^*(\tau -t)$ and the superscript $*$ represents the sign reversal of momenta.
We note that the probability to realize a path $\{z\}$ is the same as the probability to have $z_0$ in the initial state because the Hamiltonian system is deterministic as a whole.
Therefore, we obtain
\eqn{
	\sigma[\{z\}]
	&=&
	-\ln \frac{\mP[\{z^\dag\}|\{\lambda^\dag\}]}{\mP[\{z\}|\{\lambda\}]}
	\nonumber\\
	&=&
	-\ln \frac{p_0^\dag(z_0^\dag)}{p_0(z_0)}.
}

Now, we assume that the initial state is the equilibrium state of the total Hamiltonian
\eqn{
	p_0(z_0) = e^{-\beta(H_{\rm tot}(z_0, \lambda_0)-F_{\rm tot}(\lambda_0))}.
}
Moreover, we set the initial state of the reversed process to the equilibrium state of the total Hamiltonian
\eqn{
	p_0^\dag(z_0^\dag) = e^{-\beta(H_{\rm tot}(z_t, \lambda_t)-F_{\rm tot}(\lambda_t))}.
}
Thus, we obtain
\eqn{
	\sigma[\{x\}] = \beta(\Delta H_{\rm tot} - \Delta F_{\rm tot}).
}
Since the total system is isolated, the difference of the total Hamiltonian is due to the work done by the external controller, that is,
\eqn{
	\Delta H_{\rm tot} = W.
}
Therefore, noting that $F_{\rm bath}$ is independent of $\lambda$, we obtain
\eqn{
	\sigma = \beta(W-\Delta F),
}
which gives the Crooks fluctuation theorem~(\ref{CFT}) and the Jarzynski equality~(\ref{Jar}).

Next, we assume that the initial state of the system is not correlated with the initial state of the bath and that the initial state of the bath is the canonical ensemble, namely,
\eqn{
	p_0(z_0) = p_0(x_0) e^{-\beta(H_{\rm bath}(y_0)-F_{\rm bath})}.
}
In addition, we set the initial state of the time-reversed dynamics 
to the product state of the final probability distribution of the system in the original process and the canonical ensemble of the bath
\eqn{
	p_0^\dag(z_0^\dag)
	&=& p_\tau(x_0^\dag)e^{-\beta(H_{\rm bath}(y_0^\dag) -F_{\rm bath})}
	\nonumber\\
	&=& p_\tau(x_\tau)e^{-\beta(H_{\rm bath}(y_\tau) -F_{\rm bath})}.
}
Therefore, we obtain
\eqn{
	\sigma[\{x\}] = \beta \Delta H_{\rm bath} - \ln p_\tau(x_\tau) + \ln p_0 (x_0).
}
We define an unaveraged Shannon entropy of the system by $s(t)= -k_{\rm B}\ln p_t(x_t)$, and we have
\eqn{
	\sigma = \beta \Delta H_{\rm bath}+\frac{\Delta s}{k_{\rm B}}.
}
The energy change of the bath $\Delta H_{\rm bath}$ can be regarded as the heat $Q$ dissipated from the system to the bath.
Moreover, the heat $Q$ is related to the entropy production of the bath $\Delta s_{\rm bath}$ as $Q/T=\Delta s_{\rm bath}$.
Thus, we obtain
\eqn{
	\sigma &=& \frac{\Delta s_{\rm bath} + \Delta s}{k_{\rm B}}\nonumber\\
	&=& \frac{\Delta s_{\rm tot}}{k_{\rm B}},
}
where $\Delta s_{\rm tot}=\Delta s_{\rm bath} + \Delta s$ is the total entropy production.
Therefore, we automatically reproduce the detailed Seifert relation~(\ref{DSR}) and the integral Seifert relation~(\ref{SR}).
\\
\\
The results obtained in this section are summarized in Table~\ref{tab:RNEsum}.

\begin{table}[h]
\begin{center}
\caption{\label{tab:RNEsum}
	Summary of choices of the reference probability and specific meanings of the entropy production.
}
{\renewcommand\arraystretch{1.1}
\begin{tabular}{|c|c|c|}
\hline
Reference dynamics & Reference initial state $p^\dag_0(x_0^{(\dag)})$ & Entropy production $\sigma$\\
\hline\hline
time reversal & canonical $p_{\rm eq}(x_\tau,\lambda_\tau)$ & dissipated work $\beta(W-\Delta F)$\\
\hline
time reversal & final state $p_\tau(x_\tau)$ & total $\Delta s_{\rm tot}/\kb$\\
\hline
time reversal & initial state $p_0(x_\tau)$ & dissipation functional $\Omega$\\
\hline
dual & initial state $p_0(x_0)$ & housekeeping $\Delta s_{\rm hk}/\kb$\\
\hline
time-reversed dual & steady state $p_{\rm ss}(x_\tau,\lambda_\tau)$ & excess $\Delta \phi + \Delta s_{\rm ex}/\kb$\\
\hline
\end{tabular}}
\end{center}
\end{table}

\include{RNE2}
\chapter{Review of Information Thermodynamics}
In this chapter, we review thermodynamics with measurements and feedback control.
Historically, Maxwell pointed out that thermodynamics, specifically the second law of thermodynamics, should break down when an intelligent being, known as Maxwell's demon, controls the system by utilizing information obtained by measurements.
Since then, numerous researches have been done on the foundation of the second law of thermodynamics \cite{MNV09}, and thermodynamics of information processing is established \cite{Sag12}.
Since the nonequilibrium equalities reviewed in the previous chapter are generalizations of the second law of thermodynamics,
the second-law-like inequality of information processing can be extended to nonequilibrium equalities.

First, we trace historical discussions on Maxwell's demon.
Then, we formulate the second law of information thermodynamics from a modern point of view.
Next, nonequilibrium equalities of information thermodynamics are reviewed.
Finally, we review experimental demonstrations of Maxwell's demon.

\section{Maxwell's demon}
In this section, we review historical discussions on Maxwell's demon.

\subsection{Original Maxwell's demon}
Maxwell's demon was proposed in his book titled ``Theory of Heat'' published in 1871 \cite{Max1871}.
In the second last section of the book, Maxwell discussed the ``limitation of the second law of thermodynamics'' in an example with an intelligent being, which was later christened Maxwell's demon by Lord Kelvin.

\begin{figure}
\begin{center}
\includegraphics[width = 0.7\columnwidth]{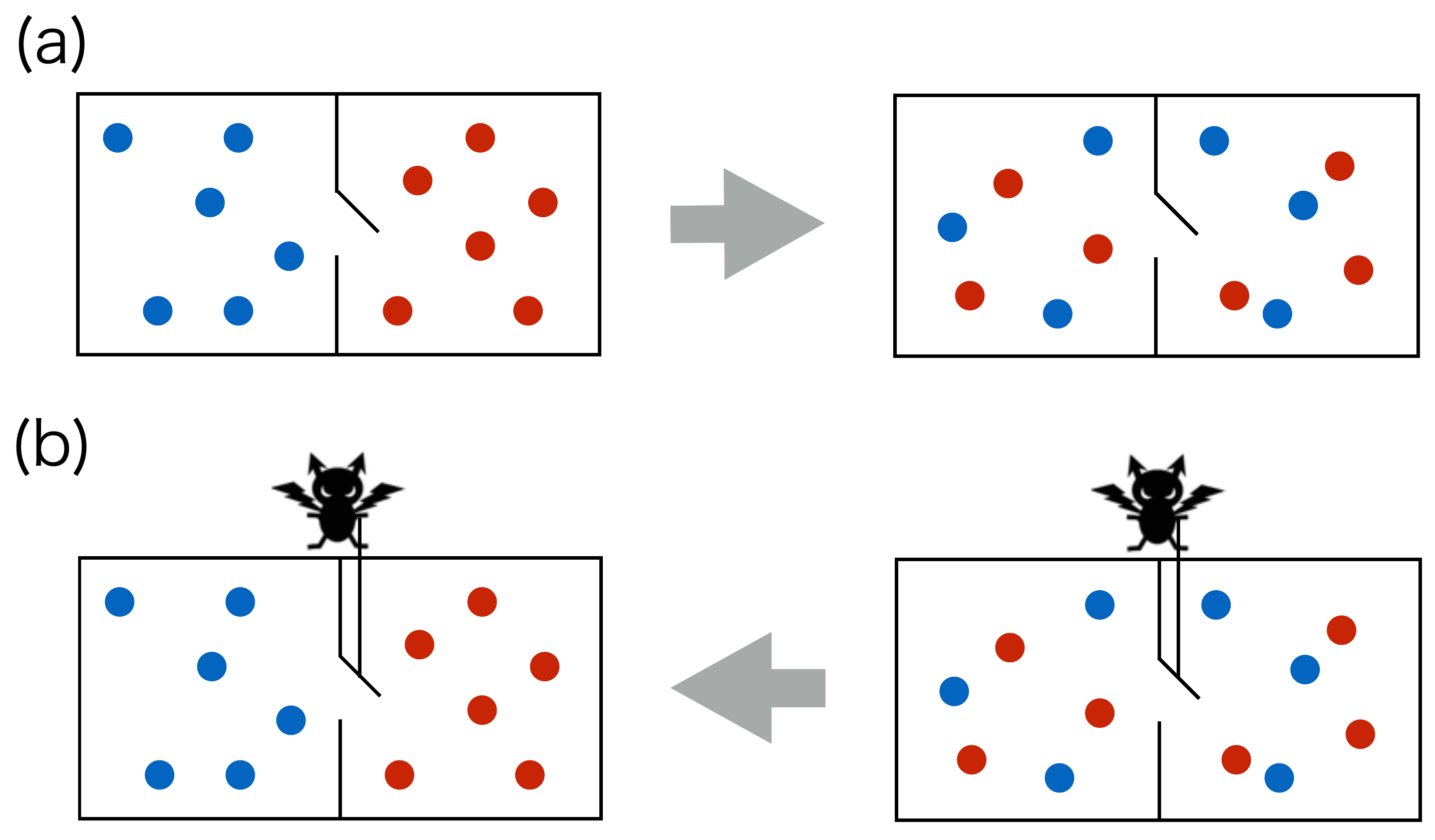}
\caption{\label{fig:demon}
Schematic illustration of the gedankenexperiment of Maxwell's demon.
(a) Second law.
Initially, the left box is filled with blue (slower and colder) molecules, and the right box is filled with red (faster and hotter) molecules.
When the window of the division is open, the entire system becomes uniform in temperature.
(b) Function of Maxwell's demon.
Initially, the temperature is uniform.
The demon measures the velocity of molecules and let the red particles go through the window from the left box to the right box and let the blue particles go from the right to the left.
Nonuniformity of temperature is then achieved without work or heat exchange.
}
\end{center}
\end{figure}

Let us consider a vessel filled with gas molecules.
The vessel is divided into two parts, and the division has a small window, through which a molecule passes from one side to the other when the window is open.
When the window opens and the temperature of one side is different from that of the other side, the second law of thermodynamics states that the temperature becomes uniform (see Fig.~\ref{fig:demon} (a)).
Maxwell's demon achieves the reverse process of this phenomenon (see Fig.~\ref{fig:demon} (b)).
At an initial time, the temperature is uniform throughout the vessel.
The demon observes molecules in the vessel.
Some molecules are faster than the mean velocity and others are slower because of thermal fluctuations.
The demon opens and closes the window to allow only the faster-than-average molecules to pass from the left side to the right side, and only the slower-than-average molecules to pass from the right to the left.
After some time, the demon succeeds in raising the temperature of the right side and lowering the temperature of the left without the expenditure of work.
This means that the entropy is reduced in an isolated system, which apparently contradicts the second law of thermodynamics.
In summary, Maxwell demonstrated that the control of the system based on the outcomes of the measurement can reduce the entropy of the system beyond the restriction from the second law of thermodynamics.

\subsection{Szilard engine}
In 1929, a simplest model of Maxwell's demon, now known as the Szilard engine, was proposed \cite{Szi29}.
Although the Szilard engine is apparently different from the original Maxwell's gedankenexperiment, it captures the essential features of the demon of utilizing measurement and feedback control to reduce the entropy of a system.
Moreover, the Szilard engine enables us to quantitatively analyze the role of the information. 

\begin{figure}
\begin{center}
\includegraphics[width = 0.3\columnwidth]{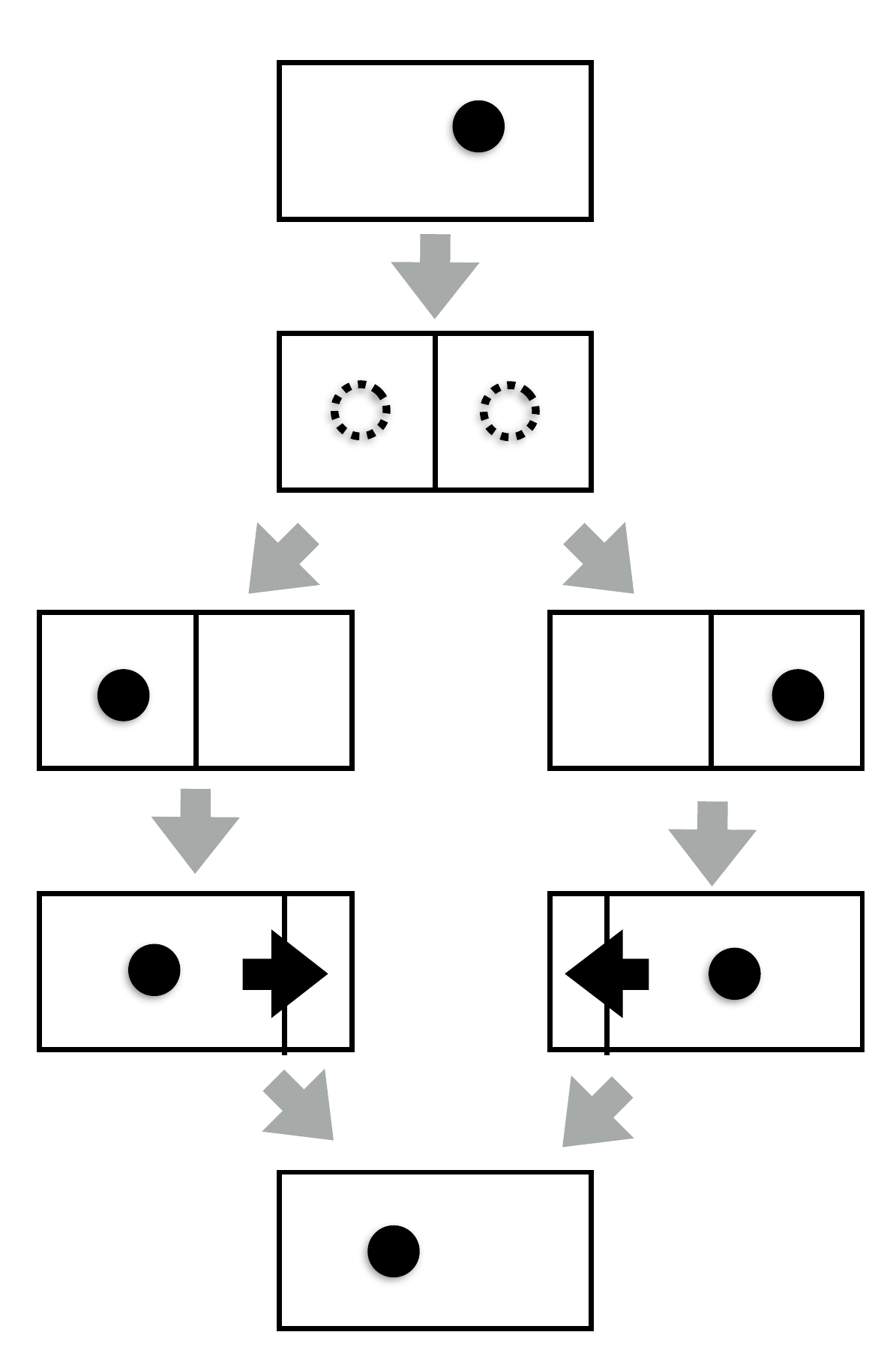}
\caption{\label{fig:SZE}
Protocol of the Szilard engine.
A single-particle gas particle is enclosed in the box, which is surrounded by a heat bath at temperature $T$.
A partition is inserted in the middle, and the position of the particle is measured.
Based on the measurement outcome, we decide in which direction we shift the division.
We isothermally expand the division to the end and remove the division.
}
\end{center}
\end{figure}

We elaborate on the protocol of the Szilard engine (see Fig.~\ref{fig:SZE}).
An ideal classical gas molecule is confined in a box with volume $V$, and the box is surrounded by an isothermal environment with temperature $T$.
We insert a division in the middle of the box and separate the box into two parts with the same volume $V/2$.
Then, we measure the position of the particle to determine whether the particle is in the left or right part.
We assume this measurement is error-free.
When we find the particle is in the left, we isothermally shift the division to the right end.
On the other hand, when we find the particle is in the right, we isothermally shift the division to the left end.
In both cases, we can extract a positive work of $k_{\rm B}T\ln 2$ (see the discussion below for the derivation of this result) from the particle in these processes.
We remove the division and the system returns to its initial state.
Therefore, we can repeatedly extract work from this isothermal cycle.

Szilard pointed out that the correlation made by the measurement is the resource for the entropy reduction and work extraction.
The measurement process creates a correlation between the position of the particle $x$ and the measurement outcome $y$.
Let us set the origin of $x$ at the middle.
When $x>0\ (x<0)$, we obtain $y=R\ (y=L)$, where $R$ ($L$) means the right (left) .
After the process of feedback control, namely isothermal shifting, this correlation vanishes because the particle can now be present in the entire system, so $x$ can be positive or negative regardless of the value of $y$.
Therefore, in the feedback process, we extract a positive work at the cost of eliminating the correlation between $x$ and $y$.

Let us quantitatively analyze this protocol from a modern point of view.
By the position measurement, we obtain the Shannon information
\eqn{
	I = \ln 2.
}
In the process of isothermal expansion, we extract work $W_{\rm ext}$.
Because we assume that the gas is ideal, the equation of state reads $pV = \kb T$.
Therefore, the work is calculated as
\eqn{
	W_{\rm ext} = \int_{V/2}^V pdV = \kb T \int_{V/2}^V \frac{dV}{V}=\kb T \ln 2 = \kb T I.
}
Therefore, we conjecture that the information obtained by the measurement can be quantitatively converted to the work during the feedback process.

As explained above, Szilard revealed that we can utilize the correlation established by the measurement to reduce the entropy of the system.
He also pointed out that the entropy reduction achieved in the feedback process must be compensated by a positive entropy production during the process to establish the correlation to be consistent with the second law of thermodynamics of the entire process.
However, it remained unexplained why the measurement process should be accompanied by a positive entropy production.

\subsection{Brillouin's argument}
An answer to the question was presented by Brillouin in 1951 \cite{Bri51}.
He argued, in the original setup of Maxwell's demon, that the demon creates more entropy when it observes molecules than the entropy reduction due to his feedback control.
Therefore, the observation process compensates the entropy reduction of the gas, and as a result the entire process is consistent with the second law of thermodynamics.

Brillouin assumed that, when the demon observes molecules, the demon needs to shed a probe light to molecules.
However, the demon and the system are surrounded by an environment at temperature $T$ with the blackbody radiation.
Therefore, the energy of the probe photon should be sufficiently larger than the thermal energy $\kb T$ to distinguish the probe from background noises.
Thus, the frequency of the probe photon $\nu$ satisfies
\eqn{
	h\nu \gg \kb T,
}
where $h$ is the Planck constant.
The demon observes a molecule by absorbing a photon scattered by the 
molecule.
Therefore, the entropy production of the demon by the single observation is given by
\eqn{\label{entd}
	\Delta S_{\rm demon} = \frac{h\nu}{T}\gg\kb.
}
Let $T_{\rm L}\ (T_{\rm R})$ represent the temperature of the left (right) side satisfying
\eqn{
	T_{\rm L} = T-\frac{1}{2} \Delta T,\ T_{\rm R}=T+\frac{1}{2} \Delta T,
}
where $\Delta T$ is the temperature difference satisfying $\Delta T \ll T$.
The demon transfers a fast molecule in the left box with kinetic energy $\frac{3}{2}\kb T (1+\epsilon_1)$ to the right box, and does a slow molecule in the right box with kinetic energy $\frac{3}{2}\kb T (1-\epsilon_2)$ to the left box, where $\epsilon_1,\epsilon_2>0$, and $\epsilon_1$ and $\epsilon_2$ are of the order of one.
As a result, heat
\eqn{
	Q = \frac{3}{2}\kb T (\epsilon_1+\epsilon_2)
}
is transferred from the left box to the right box, and the entropy reduction is bounded as
\eqn{
	-\Delta S_{\rm sys}\le Q\left(\frac{1}{T_{\rm L}}-\frac{1}{T_{\rm R}}\right)
	\simeq Q\frac{\Delta T}{T^2}
	= \frac{3}{2}\kb (\epsilon_1+\epsilon_2)\frac{\Delta T}{T}\ll \kb,
}
because $\Delta T/T\ll 1$ and $\epsilon_1+\epsilon_2\sim 1$.
Therefore, comparing this equation with Eq.~(\ref{entd}), we conclude that the entropy production of the demon due to the measurement is far beyond the entropy reduction achieved by the feedback control.
Thus, the second law remains valid for the entire system.
A similar discussion can be also done for the Szilard engine.

In this way, Brillouin argued that the entropy production for the measurement process exceeds the entropy reduction during the feedback process, and therefore the total entropy production of the system and the demon is positive.
However, his analysis depends on a specific model of the measurement using a photon as the probe, and the idea that the work gain by the feedback control is compensated by the work cost of the measurement is not always true.

\subsection{Landauer's principle}
\begin{figure}
\begin{center}
\includegraphics[width = 0.5\columnwidth]{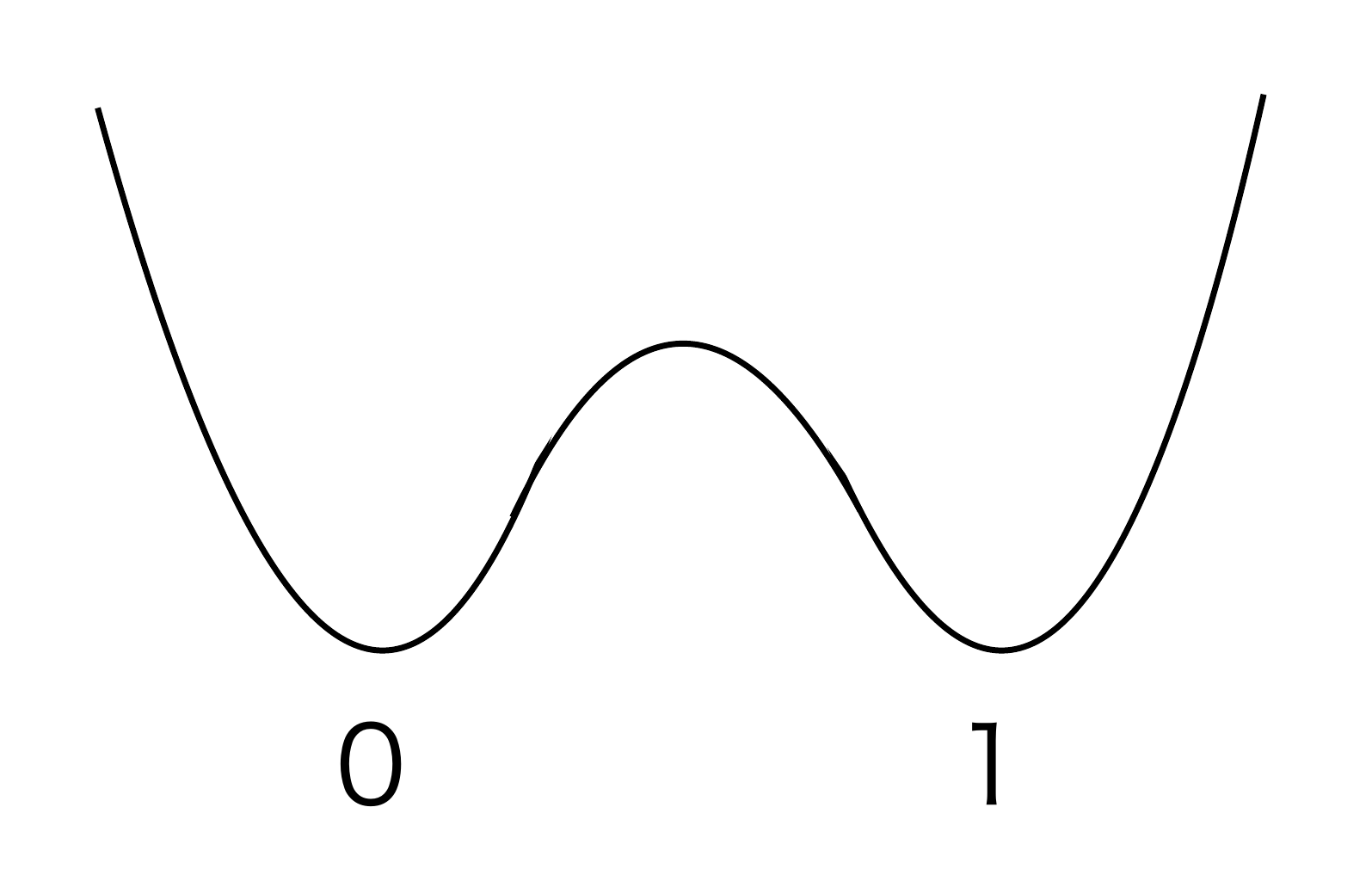}
\caption{\label{fig:dblpot}
Structure of a 1-bit memory. A particle is confined in a bistable potential.
In the zero (one) state, the particle is in the left (right) well.
}
\end{center}
\end{figure}

Landauer argued that the energy cost is needed not for measurement processes to obtain information but for erasure processes of the obtained information from the memory \cite{Lan61}.
He considered a 1-bit memory consisting of a particle in a bistable potential as shown in Fig.~\ref{fig:dblpot}.
We label the particle in the left well as the zero state, and the particle in the right well as the one state.
In the erasure process, we restore the particle to the standard state, namely the zero state.
Before the erasure, we do not know whether the particle is in the left or right well.
Therefore, the process is a two-to-one mapping, and cannot be realized by a deterministic frictionless protocol.
Thus, a protocol must involve a process with friction to erase the information.
In this way, dissipation is inevitable to erase the information stored in the memory.

The erasure process in which the system is brought to the zero state is logically irreversible because we cannot recover the state before the process from the state after the process.
Landauer argued that logical irreversibility implies physical irreversibility, which is accompanied by dissipation.
Therefore, logical irreversible operations such as erasure cause heat dissipation.

In a 1-bit symmetric memory, where the zero and one states have the same entropy, the erasure from the randomly-distributed state to the standard state means the entropy reduction by $\kb \ln 2$.
To compensate this entropy reduction, heat dissipation must occur.
Therefore, to erase the information stored in a 1-bit memory, we have inevitable heat dissipation of $\kb T\ln 2$.
This is a famous rule known as Landauer's principle.

In summary, Landauer argued that the cost for erasure of the stored information compensates the gain in feedback processes.
However, his argument is crucially dependent on the structure of the symmetric memory, and does not apply to general cases.
In fact, erasure in an asymmetric memory provides a counter-example of Landauer's principle \cite{Sag12}.

\section{Second law of information thermodynamics}
In this section, we review the second law of information thermodynamics from a modern point of view.
We restrict our attention to classical cases, because it is sufficient for the aim of this thesis.
First of all, we shortly introduce classical information quantities because they are crucial ingredients of information thermodynamics.
Then, the second law of a system under feedback control is discussed.
After that, the second law of a memory, which is a thermodynamic model of the demon, is presented.
Finally, we demonstrate that the conventional second law is recovered for the entire system.

\subsection{Classical information quantities}
In this section, we introduce three classical information quantities: the Shannon entropy, the Kullback-Leibler divergence, and the mutual information based on Refs. \cite{Sag12, NielsenChuang}.

\subsubsection{Shannon entropy}
First of all, we introduce the Shannon entropy.
Let $x$ denote a probability variable and $\mX$ denote the sample space, namely $x\in\mX$.
When $\mX$ is a discrete set, we define the Shannon entropy of a probability distribution $p(x)$ as
\footnote{
	In the context of quantum information theory, the symbol $S$ usually denotes the von Neumann entropy instead of the Shannon entropy.
	However, in this thesis, we denote the Shannon entropy by $S$ because it is the convention in the field of classical statistical mechanics.
	There would be no confusion because we do not use the von Neumann entropy in this thesis.
}
\eqn{
	S^\mX[p] = - \sum_{x\in\mX} p(x) \ln p(x).
}
On the other hand, when $\mX$ is continuous, we would naively define the Shannon entropy of a probability distribution density $p(x)$ by
\eqn{
	-\int_\mX p(x)dx \ln(p(x)dx)
	= - \int_\mX dx \ p(x)\ln p(x) -\int_\mX dx\ p(x) \ln(dx).	
}
However, the second term is divergent in the limit of $dx\to 0$.
Therefore, we define the Shannon entropy as
\eqn{\label{defcSh}
	S^\mX[p] = - \int_\mX dx \ p(x)\ln p(x).
}
Although $p(x)dx$ is invariant under a transformation of variable, $p(x)$ alone is not invariant.
Therefore, the continuous Shannon entropy as defined in Eq.~(\ref{defcSh}) is not invariant under transformation of the coordinates.

The unaveraged Shannon entropy 
\eqn{
	s(x) = -\ln p(x)
}
is an indicator of how rare an event $x$ is.
In fact, $s(x)$ increases as $p(x)$ decreases.
The Shannon entropy is the ensemble average of this rarity.
The reason why we use the logarithm is to guarantee the additivity of the Shannon entropy when we have independent events.
Let us assume $\mX=\mX_1\times\mX_2$ and $x=(x_1,x_2)$.
Then, when $p(x) = p_1(x_1)p_2(x_2)$, we have
\eqn{
	S^\mX(p) = S^{\mX_1}(p_1) + S^{\mX_2}(p_2).
}

Here, we demonstrate that the Shannon entropy is invariant under a Hamiltonian dynamics.
In this case, $\mX$ is phase space, and the dynamics is deterministic.
Let $x_{\rm i}\ (x_{\rm f})$ denote the initial (final) position in $\mX$ and $p_{\rm i}\ (p_{\rm f})$ denote the initial (final) probability distribution.
Since the probability is conserved, we have
\eqn{\label{probconvlaw}
	p_{\rm i}(x_{\rm i}) dx_{\rm i}
	= p_{\rm f}(x_{\rm f}) dx_{\rm f}.
}
In addition, Liouville's theorem states that
\eqn{
	dx_{\rm i} = dx_{\rm f},
}
which, together with Eq.~(\ref{probconvlaw}), leads to 
\eqn{
	p_{\rm i} (x_{\rm i}) = p_{\rm f} (x_{\rm f}).
}
Therefore, the initial Shannon entropy
\eqn{
	S^\mX_{\rm i} = -\int_{\mX} dx_{\rm i}\ p(x_{\rm i})\ln p(x_{\rm i})
}
has the same value as the final Shannon entropy
\eqn{
	S^\mX_{\rm f} = -\int_{\mX} dx_{\rm f}\ p(x_{\rm f})\ln p(x_{\rm f}),
}
namely,
\eqn{\label{invSh}
	S^\mX_{\rm i} = S^\mX_{\rm f}.
}
Thus, the continuous Shannon~(\ref{defcSh}) entropy is invariant in time under the Hamiltonian dynamics.

\subsubsection{Kullback-Leibler divergence}
Next, we introduce the Kullback-Leibler divergence or the relative entropy.
This quantity is defined as a relative logarithmic distance between two probability distributions $p$ and $q$ on the same sample space $\mX$.
When $\mX$ is discrete, we define the Kullback-Leibler divergence as
\eqn{
	S^\mX[p||q] = - \sum_{x\in\mX} p(x) \ln\frac{q(x)}{p(x)}.
}
On the other hand, when $\mX$ is continuous, we define the Kullback-Leibler divergence as
\eqn{
	S^\mX[p||q] = - \int_\mX dx\ p(x) \ln \frac{q(x)}{p(x)}.
}
We note that the continuous Kullback-Leibler divergence is invariant under a transformation of the coordinates.

Using an inequality
\eqn{
	\ln\frac{q(x)}{p(x)} \le \frac{q(x)}{p(x)} - 1,
}
we obtain
\eqn{
	S^\mX[p||q] &\ge& - \int_\mX dx\ p(x) \left[
		\frac{q(x)}{p(x)} - 1
	\right]
	\nonumber\\
	&=&
	- \int_\mX dx\ q(x) + \int_\mX dx\ p(x)
	\nonumber\\
	&=& 0,
}
where the equality is achieved if and only if $p(x) = q(x)$ ($p$-almost everywhere).
Therefore, the Kullback-Leibler divergence is a kind of distance to measure how different two probabilities are.

\subsubsection{Mutual information}
\begin{figure}
\begin{center}
\includegraphics[width=0.4\textwidth]{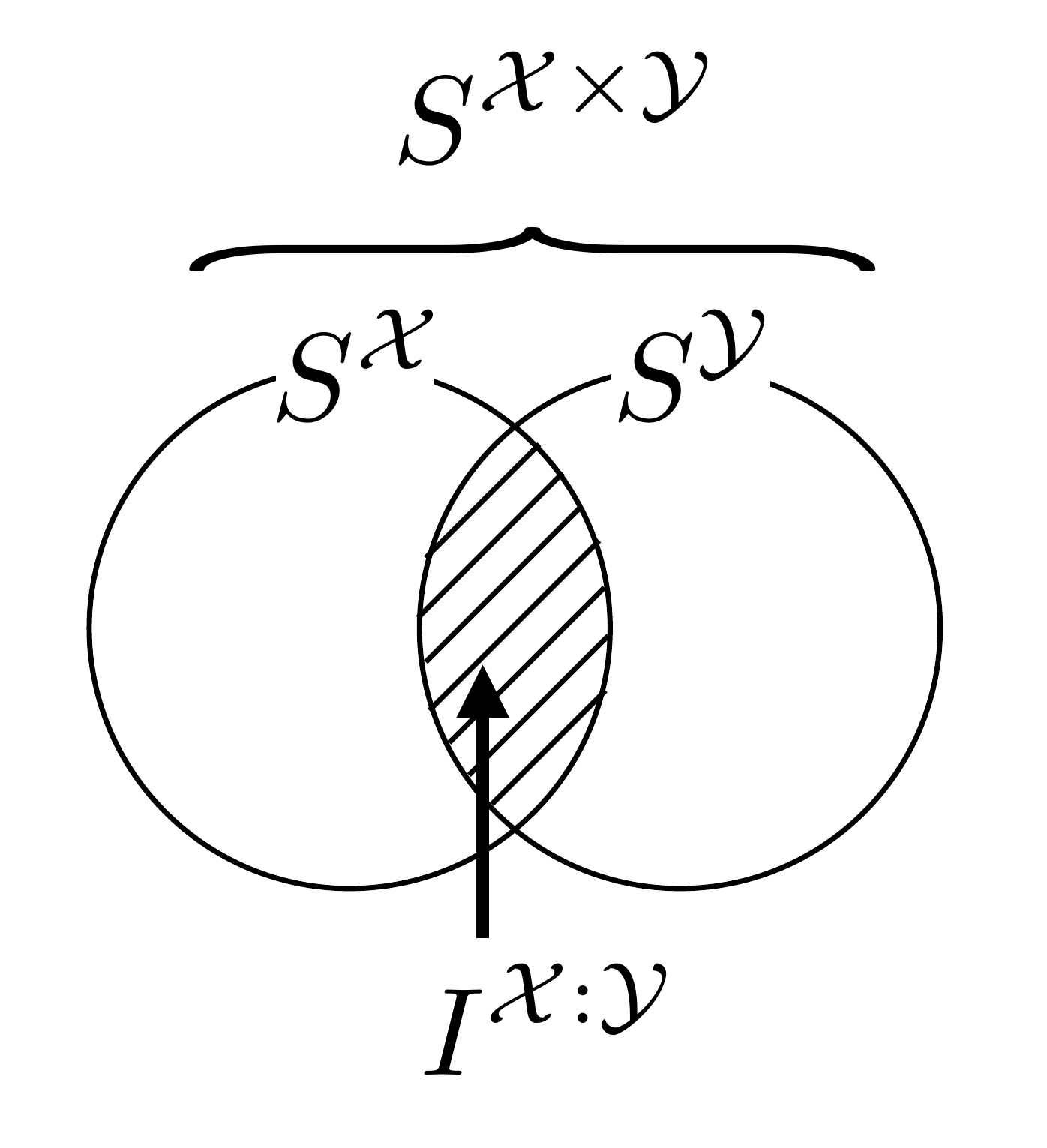}
\caption{\label{fig:VDMI}
Schematic illustration of the definition of the mutual information.
The two circles represent degrees of uncertainty of $\mX$ and $\mY$.
The union of the circles represents a degree of uncertainty of $\mX\times\mY$.
Therefore, the mutual information defined by Eq.~(\ref{defMI}) corresponds to the overlap of the degrees of uncertainty of $\mX$ and $\mY$, which describes correlations between $\mX$ and $\mY$.
}
\end{center}
\end{figure}
We consider the mutual information between two sample spaces $\mX$ and $\mY$.
Let $p^{\mX\times\mY}(x, y)$ denote a joint probability distribution of $(x,y)\in\mX\times\mY$.
The marginal probability distributions are defined as $p^\mX(x)=\int_\mY dy\ p^{\mX\times\mY}(x,y)$ and $p^\mY(y) = \int_\mX dx\ p^{\mX\times\mY}(x,y)$.
We define the mutual information as
\eqn{\label{defMI}
	I^{\mX:\mY} = S^\mX[p^\mX] + S^\mY[p^\mY] - S^{\mX\times\mY}[p^{\mX\times\mY}],
}
which represents the overlap of uncertainty of $\mX$ and $\mY$ (see Fig.~\ref{fig:VDMI}).
If the systems are independent of each other, i.e., $p(x,y)=p^\mX(x)p^\mY(y)$, we obtain $I=0$ due to the additivity of the Shannon entropy.
Moreover, we represent the mutual information in terms of the Kullback-Leibler divergence as
\eqn{
	I^{\mX:\mY} = S^{\mX\times\mY}[p^{\mX\times\mY}||p^\mX p^\mY].
}
Therefore, the mutual information quantifies how different $p^{\mX\times\mY}$ is from the non-correlated probability distribution $p^\mX p^\mY$, that is, how correlated $\mX$ and $\mY$ are.
Since the Kullback-Leibler divergence is not negative, we obtain
\eqn{
	I^{\mX:\mY}\ge 0.
}
The explicit form of the mutual information is
\eqn{
	I^{\mX:\mY} = - \int_\mX dx\int_\mY dy\ p^{\mX\times\mY}(x, y) \ln \frac{p^\mX(x)p^\mY(y)}{p^{\mX\times\mY}(x, y)}.
}
Therefore, we define the unaveraged mutual information as
\eqn{
	i(x,y) = - \ln \frac{p^\mX(x)p^\mY(y)}{p^{\mX\times\mY}(x, y)}
}
or
\eqn{\label{umi}
	i(x,y) = - \ln p^\mX(x) + \ln p^{\mX|\mY}(x|y),
}
where $p^{\mX|\mY}(x|y)$ is the probability distribution function of $x$ conditioned by $y$.
Therefore, the unaveraged mutual information quantifies the decrease of the unaveraged Shannon entropy of the system $\mX$ due to the fact that we know the system $\mY$ is in a state $y$.

\subsection{Second law under feedback control}
In this section, we formulate the second law of a system under feedback control.
The mutual information plays a crucial role in quantifying the gain of feedback control.

\subsubsection{Setup}
\begin{figure}
\begin{center}
\includegraphics[width=0.6\textwidth]{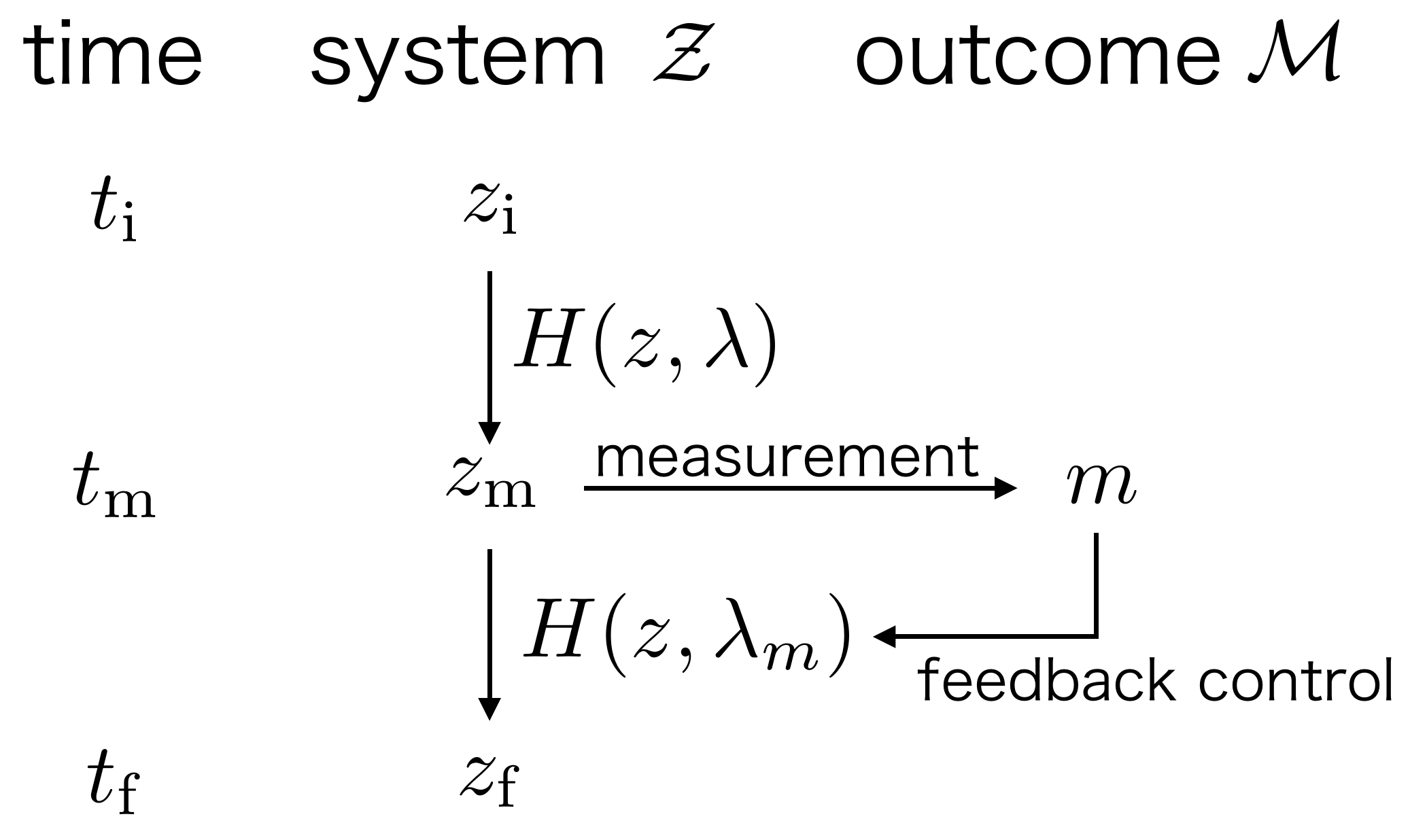}
\caption{\label{fig:FCFC}
Flow chart of feedback control.
From time $t_{\rm i}$ to $t_{\rm m}$, the system evolves in time by a Hamiltonian $H(z,\lambda)$.
At time $t_{\rm m}$, we perform a measurement on the system and obtain an outcome $m$.
We conduct feedback control from time $t_{\rm m}$ to $t_{\rm f}$.
In other words, the time evolution after the measurement is governed by a Hamiltonian $H(z,\lambda_m)$ that depends on the measurement outcome $m$.
}
\end{center}
\end{figure}

We consider feedback control on a Hamiltonian system with phase space $\mZ$ (see Fig.~\ref{fig:FCFC}).
At initial time $t_{\rm i}$, the system is at position $z_{\rm i}$ sampled from an initial probability distribution $p_{\rm i}(z_{\rm i})$.
From time $t_{\rm i}$ to $t_{\rm m}$, the system evolves under a Hamiltonian $H(z,\lambda)$, and ends up with a point $z_{\rm m}$ with a probability distribution $p_{\rm m}(z_{\rm m})$.
At time $t_{\rm m}$, we measure quantities of the system (e.g. positions, velocities and the number of particles in a given region) and obtain a measurement outcome $m$.
Let $\mathcal{M}$ denote the sample space of the outcome.
The detail of the measurement is modeled by conditional probabilities
\eqn{
	p(m|z_{\rm m}),
}
which is the probability to obtain the outcome $m$ under the condition that the system is at the position $z_{\rm m}$ at time $t_{\rm m}$.
Now that we have the outcome $m$, we know more detailed information on the system than we did before the measurement.
In fact, the probability distribution of $z_{\rm m}$ under the condition that we have obtained $m$ is calculated by the Bayes theorem as
\eqn{
	p_{\rm m}(z_{\rm m}|m) = \frac{p(m|z_{\rm m})p_{\rm m}(z_{\rm m})}{p(m)},
}
where the probability distribution of the outcome $m$ is defined by
\eqn{
	p(m) = \int dz_{\rm m}\ p(m|z_{\rm m})p_{\rm m}(z_{\rm m}).
}
From time $t_{\rm m}$ to $t_{\rm f}$, the system evolves under a Hamiltonian $H(z,\lambda_m)$.
Here, we conduct feedback control of the system, that is, adjust the Hamiltonian in accordance with $m$ via the control parameter $\lambda_m$.
Therefore, the protocol after the measurement is conditioned by $m$.
Let $z_{\rm f}$ denote the final position and $p_{\rm f}(z_{\rm f}|m)$ be the final probability distribution conditioned by $m$.
The unconditioned final probability distribution is calculated as
\eqn{
	p_{\rm f}(z_{\rm f})
	=
	\sum_{m\in\mM}\ p_{\rm f}(z_{\rm f}|m) p(m).
}

\subsubsection{Shannon entropy production}
We evaluate the Shannon entropy production in the above-described setup.
Since Hamiltonian dynamics does not change the Shannon entropy (see Eq.~(\ref{invSh})), we obtain
\eqn{\label{sism}
	S^\mZ[p_{\rm i}] = S^\mZ[p_{\rm m}].
}
By the same reason, we also obtain
\eqn{\label{smsf}
	S^\mZ[p_{\rm m}(\cdot|m)] = S^\mZ[p_{\rm f}(\cdot|m)],\ ^\forall m\in\mM.
}
To evaluate entropy production of the system, let us compare the averaged final entropy
\eqn{
	S^\mZ_{\rm f} = \sum_{m\in\mM}\ p(m) S^\mZ[p_{\rm f}(\cdot|m)]
}
with the initial entropy
\eqn{
	S^\mZ_{\rm i} = S^\mZ[p_{\rm i}].
}
By Eq.~(\ref{smsf}), the final entropy is calculated as
\eqn{
	S^\mZ_{\rm f} &=& \sum_{m\in\mM}\ p(m) S^\mZ[p_{\rm m}(\cdot|m)]
	\nonumber\\
	&=& -\int_\mZ dz_{\rm m} \sum_{m\in\mM}\ p(m)p_{\rm m}(z_{\rm m}|m)\ln p_{\rm m}(z_{\rm m}|m)
	\nonumber\\
	&=& -\int_\mZ dz_{\rm m} \sum_{m\in\mM}\ p_{\rm m}(z_{\rm m},m)\ln p_{\rm m}(z_{\rm m}|m).
}
On the other hand, by Eq.~(\ref{sism}), the initial entropy can be transformed as
\eqn{
	S^\mZ_{\rm i} = S^\mZ[p_{\rm m}] &=& - \int_\mZ dz_{\rm m}\ p_{\rm m}(z_{\rm m}) \ln p_{\rm m}(z_{\rm m})
	\nonumber\\
	&=& 
	- \int_\mZ dz_{\rm m} \sum_{m\in\mM}\ p_{\rm m}(z_{\rm m}, m) \ln p_{\rm m}(z_{\rm m}).
}
Using Eq.~(\ref{umi}), we obtain
\eqn{\label{ssi}
	S^\mZ_{\rm f} -S^\mZ_{\rm i} = - \int_\mZ dz_{\rm m} \sum_{m\in\mM}\ i(z_{\rm m}, m) = -I^{\mZ:\mM}.
}
Therefore, the Shannon entropy production of the system is the negative of the mutual information obtained by the measurement.

\subsubsection{Second law under feedback control}
We separate the degrees of freedom $z$ into that of the system $x$ and that of the heat bath $b$ as $z=(x,b)$.
Let $\mX$ and $\mB$ denote phase spaces of the system and the bath, respectively.
We assume that the total Hamiltonian reads
\eqn{
	H^\mZ(z, \lambda) = H^\mX(x, \lambda) + H^\mB(b) + H^{\rm int}(z,\lambda),
}
where the last term on the right-hand side is the interaction Hamiltonian and is assumed to vanish at the initial and final times.

First, we assume that the initial state is the product state of an initial probability distribution of the system and the canonical ensemble of the heat bath as
\eqn{
	p^\mZ_{\rm i}(z_{\rm i}) = p^\mX_{\rm i}(x_{\rm i})p^\mB_{\rm eq}(b_{\rm i}),
}
where we define
\eqn{
	p^\mB_{\rm eq}(b)
	=e^{-\beta (H^\mB(b) - F^\mB)}
}
and
\eqn{
	e^{-\beta F_\mB} = \int db\ e^{-\beta H^\mB(b)}.
}
In this case, since $x_{\rm i}$ and $b_{\rm i}$ are not correlated, the initial total Shannon entropy is calculated as
\eqn{
	S^\mZ_{\rm i} = S^\mX_{\rm i} + \beta(\ave{H^\mB}_{\rm i}-F^\mB),
}
where the initial energy of the bath is defined as
\eqn{
	\ave{H^\mB}_{\rm i} = \int_\mB db_{\rm i}\ H^\mB(b_{\rm i}) p^\mB_{\rm eq}(b_{\rm i}).
}
Using the final probability distribution of the total system $p^\mZ_{\rm f}(x_{\rm f}, b_{\rm f})$, we define the marginal final probability as
\eqn{
	p^\mX_{\rm f} (x_{\rm f}) = \int_\mB db_{\rm f}\ p^\mZ_{\rm f} (x_{\rm f}, b_{\rm f}).
}
We calculate the relative entropy between the final state of the total system and the product state of the final state of the system and the canonical state of the heat bath as
\eqn{
	S^\mZ[p^\mZ_{\rm f}||p^\mX_{\rm f}p^\mB_{\rm eq}]
	&=&
	-\int_\mX dx_{\rm f} \int_\mB db_{\rm f}\ p^\mZ_{\rm f} (x_{\rm f}, b_{\rm f}) \ln p^\mX_{\rm f}(x_{\rm f})e^{-\beta(H^\mB(b_{\rm f})-F^\mB)} - S^\mZ_{\rm f}
	\nonumber\\
	&=&
	S^\mX_{\rm f} + \beta (\ave{H^\mB}_{\rm f}-F^\mB) - S^\mZ_{\rm f}.
}
Since the relative entropy is positive, we obtain
\eqn{\label{sshf}
	S^\mZ_{\rm f} \le S^\mX_{\rm f} + \beta(\ave{H^\mB}_{\rm f}-F^\mB).
}
Therefore, comparing Eq.~(\ref{ssi}), we obtain
\eqn{
	\Delta S^\mX + \beta Q \ge -I^{\mZ:\mM},
}
where we identify the dissipated heat with the energy difference of the bath as
\eqn{
	Q=\ave{H^\mB}_{\rm f}-\ave{H^\mB}_{\rm i}.
}
Moreover, since the measurement outcome $m$ should depend only on the degrees of freedom of the system $x_{\rm m}$, we have
\eqn{
	p(m|z_{\rm m}) = p(m|x_{\rm m}),
}
and therefore
\eqn{
	i(z_{\rm m}, m) &=& -\ln p^\mZ(z_{\rm m})+\ln p(z_{\rm m}|m)
	\nonumber\\
	&=& - \ln p^\mM(m) + \ln p(m|z_{\rm m})
	\nonumber\\
	&=& - \ln p^\mM(m) + \ln p(m|x_{\rm m})
	\nonumber\\
	&=& i(x_{\rm m}, m).
}
Thus, we obtain
\eqn{\label{ssqi1}
	\Delta S^\mX + \beta Q \ge -I^{\mX:\mM}.
}
The left-hand side of Eq.~(\ref{ssqi1}) means the sum of the Shannon entropy production of the system and the entropy production of the heat bath.
Therefore, Eq.~(\ref{ssqi1}) demonstrates that the total entropy production can be reduced by the feedback control by the amount of the mutual information $I^{\mX:\mM}$.

To proceed further, we assume that the initial state of the system is in equilibrium with the inverse temperature $\beta$ given by
\eqn{
	p^\mX_{\rm i}(x_{\rm i})
	=
	p^\mX_{\rm eq}(x_{\rm i}, \lambda_{\rm i}),
}
where we define the canonical ensemble as
\eqn{
	p^\mX_{\rm eq}(x, \lambda) = e^{-\beta(H^\mX(x,\lambda) - F^\mX(\lambda))}
}
and the free energy as
\eqn{
	e^{-\beta F^\mX(\lambda)} = \int_\mX dx\ e^{-\beta H^\mX(x,\lambda)}.
}
The initial Shannon entropy is
\eqn{
	S^\mX_{\rm i} &=& - \int_\mX dx_{\rm i}\ p_{\rm i}^\mX(x_{\rm i}) \ln p_{\rm i}^\mX(x_{\rm i})
	\nonumber\\
	&=& \beta (\ave{H^\mX}_{\rm i} - F^\mX_{\rm i}),
}
where we define the initial energy of the system as
\eqn{
	\ave{H^\mX}_{\rm i} = \int_\mX dx_{\rm i}\ p^\mX_{\rm i}(x_{\rm i})H^\mX(x_{\rm i}, \lambda_{\rm i}).
}
On the other hand, since the relative entropy is positive, we obtain
\eqn{
	S^\mX[p^\mX_{\rm f}(\cdot|m)||p^\mX_{{\rm eq}}(\cdot, \lambda_{m,{\rm f}})] \ge 0.
}
This relation can be rewritten as
\eqn{
	\beta \int_\mX dx_{\rm f}\ p^\mX_{\rm f}(x_{\rm f}|m)(H^\mX(x_{\rm f}, \lambda_{m,\rm f})-F^\mX(\lambda_{m, {\rm f}})) - S^\mX[p_{\rm f}^\mX(\cdot|m)]
	\ge  0.
}
Averaging this over $m$ with the probability $p(m)$, we obtain
\eqn{
	\beta(\ave{H^\mX}_{\rm f} - F^\mX_{\rm f}) \ge S^\mX_{\rm f},
}
where we define the final internal energy as
\eqn{
	\ave{H^\mX}_{\rm f}
	= \int_\mX dx_{\rm f} \sum_{m\in\mM}\  
	p^\mX_{\rm f}(x_{\rm f}, m) H^\mX(x_{\rm f}, \lambda^\mX_{m,{\rm f}})
}
and the final free energy as
\eqn{
	F^\mX_{\rm f} = \sum_{m\in\mM}\ p(m) F^\mX(\lambda_{m, {\rm f}}).
}
Therefore, Eq.~(\ref{ssqi1}) reduces to
\eqn{
	\beta(\ave{\Delta H^\mX} - \Delta F^\mX+Q) \ge -I^{\mX:\mM}.
}
Identifying the energy change of the total system $\ave{\Delta H^\mX}+Q$ with the work $W$ performed on the system, we obtain
\eqn{
	W-\Delta F^\mX \ge -\kb TI^{\mX:\mM}.
}
We can rewrite this equation as
\eqn{
	-W \le -\Delta F^\mX + \kb T I^{\mX:\mM},
}
which means that we can extract more work from the system than the conventional second law of thermodynamics by the amount of the mutual information $I^{\mX:\mM}$ obtained by the measurement.
This form of the second law under feedback control was first formulated in Ref.~\cite{SU08} in a quantum system.
\\

In summary, we formulate the second law under feedback control, and reveal that the gain of feedback control is precisely quantified by the mutual information obtained by the measurement.
This is a rigorous formulation of Szilard's idea that the correlation made by the measurement can be utilized as a resource for the entropy reduction.

\subsection{Second laws of memories}
In this section, we formulate second laws of memories during a measurement process and during an erasure process.
The second laws of memories were discussed in quantum systems in Ref.~\cite{SU09}.
Here, we consider a classical version of this study.

\subsubsection{Measurement process}
First, we consider a measurement process.
The phase space of the total system is denoted by $\mZ$ and the system is subject to a Hamiltonian dynamics.
We assume that the total system consists of three parts: the system, the memory, and a heat bath with phase spaces $\mX$, $\mY$, and $\mB$, respectively.
Phase-space positions in $\mZ$, $\mX$, $\mY$, and $\mB$ are denoted by $z$, $x$, $y$, and $b$, respectively.
Moreover, we assume that the sample space of the memory $\mY$ is the disjoint union of $\mY_m\ (m\in\mM=\{0,\cdots,M\})$.
An outcome $m$ is stored when $y\in\mY_m$.
Therefore, $\mM$ can be regarded as a coarse-grained sample space of $\mY$.
We assume that the total Hamiltonian is decomposed into
\eqn{
	H^{\mZ}(x, y, b, \lambda)
	=
	H^{\mX}(x, \lambda)+H^{\mY}(y, \lambda)+H^\mB(b)+H^{\rm int}(x, y, b, \lambda),
}
where the last term on the right-hand side is the interaction term, which is assumed to vanish at the initial and final times.

Since the total system is a Hamiltonian system, the Shannon entropy of the total system is conserved:
\eqn{\label{secl}
	S^\mZ_{\rm i} = S^\mZ_{\rm f}.
}
We assume that the initial state is a product state as
\eqn{
	p_{\rm i}^\mZ(z_{\rm i}) = p_{\rm i}^\mX(x_{\rm i})p_{\rm i}^\mY(y_{\rm i})p^\mB_{\rm eq}(b_{\rm i}),
}
and then we obtain
\eqn{\label{ssshf}
	S^\mZ_{\rm i} = S^\mX_{\rm i} + S^\mY_{\rm i} + \beta(\ave{H^\mB}_{\rm i}-F^\mB).
}
Because of the positivity of the relative entropy, we obtain
\eqn{
	S[p^\mZ_{\rm f}||p^{\mX\times\mY}_{\rm f}p^\mB_{\rm eq}]&\ge& 0
	\nonumber\\
	\Leftrightarrow\ S^{\mX\times\mY}_{\rm f}+\beta (\ave{H^\mB}_{\rm f} - F^\mB) &\ge& S^\mZ_{\rm f}.
}
Substituting Eqs.~(\ref{secl}) and (\ref{ssshf}), we obtain
\eqn{
	S^{\mX\times\mY}_{\rm f}-S^\mX_{\rm i}-S^\mY_{\rm i} + \beta Q \ge 0,
}
where
\eqn{
	Q = \langle H^\mB\rangle_{\rm f} - \langle H^\mB\rangle_{\rm i}
}
Using the definition of the mutual information, we obtain
\eqn{\label{ssqi}
	\Delta S^\mX + \Delta S^\mY + \beta Q \ge I^{\mX:\mY}.
}
Since we perform feedback control based on $m$, we should evaluate the entropy production by $I^{\mX:\mM}$ instead of $I^{\mX:\mY}$.
The difference between these two values is calculated as
\eqn{
	I^{\mX:\mY}-I^{\mX:\mM}
	&=&
	-\int_\mX dx_{\rm f} \int_\mY dy_{\rm f}\
	p^{\mX\times\mY}_{\rm f}(x_{\rm f},y_{\rm f})
	\ln\frac{p^\mX_{\rm f}(x_{\rm f})p^\mY_{\rm f}(y_{\rm f})}
	{p^{\mX\times\mY}_{\rm f}(x_{\rm f},y_{\rm f})}
	\nonumber\\
	&&\ \ \ +\int_\mX dx_{\rm f}\sum_{m\in\mM}
	p^{\mX\times\mM}_{\rm f}(x_{\rm f},m)
	\ln\frac{p^\mX_{\rm f}(x_{\rm f})p^\mM(m)}
	{p^{\mX\times\mM}_{\rm f}(x_{\rm f},m)},
}
where the joint probability of $\mX$ and $\mM$ is defined as
\eqn{
	p^{\mX\times\mM}_{\rm f}(x_{\rm f},m) =
	\int_{\mY_m} dy_{\rm f}\ p^{\mX\times\mY}_{\rm f}(x_{\rm f}, y_{\rm f}).
}
Therefore, we obtain
\eqn{
	I^{\mX:\mY}-I^{\mX:\mM}
	=
	-\int_\mX dx_{\rm f} \sum_{m\in\mM} \int_{\mY_m} dy_{\rm f}\
	p^{\mX\times\mY}_{\rm f}(x_{\rm f},y_{\rm f})
	\ln\frac{p^\mY_{\rm f}(y_{\rm f})p^{\mX|\mM}_{\rm f}(x_{\rm f}|m)}
	{p^{\mX\times\mY}_{\rm f}(x_{\rm f},y_{\rm f})}.
}
Since the right-hand side is in the form of the relative entropy, we derive
\eqn{
	I^{\mX:\mY}-I^{\mX:\mM}\ge 0.
}
We note  that this result is natural since $\mM$ is the coarse-grained sample space of $\mY$.
Thus, Eq.~(\ref{ssqi}) reduces to
\eqn{
	\Delta S^\mX + \Delta S^\mY + \beta Q \ge I^{\mX:\mM}.
}

We assume that, before the measurement process, the system is in equilibrium and the memory is prepared to be a fixed standard state $p_{\rm st}^\mY(y_{\rm i})$, namely, a local equilibrium state in $\mY_0$ given by
\eqn{
	p^\mY_{\rm st}(y_{\rm i}) = p_{\rm eq}^{\mY_0}(y_{\rm i}, \lambda_{\rm i}),
}
where we define
\eqn{
	p_{\rm eq}^{\mY_m}(y, \lambda) = \chi_{\mY_m}(y)e^{-\beta(H^\mY(y,\lambda)-F^{\mY_m}(\lambda))},
}
and $\chi_{\mY_m}(y)$ is the characteristic function of a region $\mY_{m}$;
the conditional free energy is defined by
\eqn{
	e^{-\beta F^{\mY_m}(\lambda)}
	=
	\int_{\mY_m} dy\ e^{-\beta H^\mY(y,\lambda)}.
}
The initial entropy of the memory is
\eqn{
	S^\mY_{\rm i} = \beta(\ave{H^\mY}_{\rm i}-F^\mY_{\rm i}).
}
Using the positivity of the relative entropy, we obtain
\eqn{
	S^\mY[p^{\mY}_{\rm f}||p^{\mY}_{\rm ref}]\ge 0,
}
where we define a reference probability by
\eqn{
	p_{\rm ref}^\mY(y_{\rm f})
	=
	\sum_{m\in\mM} p(m)p_{\rm eq}^{\mY_m}(y_{\rm f}, \lambda_{m,{\rm f}}).
}
Therefore, we obtain
\eqn{
	-\int_{\mY} dy_{\rm f}\ p^\mY_{\rm f}(y_{\rm f})
	\ln\sum_{m\in\mM} p(m)p_{\rm eq}^{\mY_m}(y_{\rm f}, \lambda_{m,{\rm f}})
	&\ge&
	S^\mY_{\rm f}
	\nonumber\\
	\Leftrightarrow\ 
	-\sum_{m\in\mM} \int_{\mY_m}dy_{\rm f}\ p^\mY_{\rm f}(y_{\rm f})
	\ln p(m)p_{\rm eq}^{\mY_m}(y_{\rm f}, \lambda_{m,{\rm f}})
	&\ge&
	S^\mY_{\rm f}
	\nonumber\\
	\Leftrightarrow\ 
	S^\mM + \beta(\ave{H^\mY}_{\rm f}-F^\mY_{\rm f}) &\ge& S^\mY_{\rm f},	
}
where we use the fact that $p^{\mY_m}_{\rm eq}=0$ outside $\mY_m$, and the final free energy is defined with
\eqn{
	F^\mY_{\rm f} = \sum_{m\in\mM} p(m)F^{\mY_m}(\lambda_{m,{\rm f}}).
}
Therefore, the entropy production of the memory is bounded as
\eqn{
	\Delta S^\mY \le S^\mM + \beta (\Delta \ave{H^\mY} -\Delta F^\mY).
}
On the other hand, we can derive
\eqn{
	\Delta S^\mX &=& S^\mX_{\rm f} - S^\mX_{\rm i}
	\nonumber\\
	&\le& S^\mX_{\rm f} + S[p^\mX_{\rm f}||p^\mX_{\rm eq, f}]- S^\mX_{\rm i}
	\nonumber\\
	&=& \beta (\Delta \ave{H^\mX} -\Delta F^\mX).
}
Using Eq.~(\ref{ssqi}), we obtain
\eqn{
	\beta (\Delta\ave{H^\mX}+\Delta\ave{H^\mY} -\Delta F^\mX -\Delta F^\mY+Q) \ge I^{\mX:\mM} - S^\mM.
}
Since the energy change of the total system $\Delta\ave{H^\mX}+\Delta\ave{H^\mY}+Q$ should be identified as work $W$, we conclude
\eqn{
	W-\Delta F^\mX-\Delta F^\mY \ge \kb T(I^{\mX:\mM} - S^\mM).
}
This equality presents the minimum work needed to perform the measurement, which may be interpreted as a rigorous formulation of Brillouin's argument.

\subsubsection{Erasure process}
Next, we consider an erasure process of the stored information.
We assume that the total system of this process consists of the memory $\mY$ and the heat bath $\mB$, and the total system is subject to the Hamiltonian dynamics.
We assume that the Hamiltonian is written as
\eqn{
	H^{\mY\times\mB}(y, b, \lambda)
	=
	H^{\mY}(y,\lambda) + H^\mB(b) + H^{\rm int}(y, b, \lambda),
}
where $H^{\rm int}$ is the interaction term, which is assumed to vanish at the initial and final times.

Since the total system is under Hamiltonian dynamics, we obtain
\eqn{\label{syb}
	S^{\mY\times\mB}_{\rm i} = S^{\mY\times\mB}_{\rm f}.
}
We assume that the initial state is a product state as
\eqn{
	p_{\rm i}^{\mY\times\mB}(y_{\rm i}, b_{\rm i})
	=
	p_{\rm i}^\mY(y_{\rm i})p^\mB_{\rm eq}(b_{\rm i}),
}
and we obtain
\eqn{
	S^{\mY\times\mB}_{\rm i} = S^{\mY}_{\rm i} + \beta(\ave{H^\mB}_{\rm i} - F^\mB).
}
By the same procedure to derive Eq.~(\ref{sshf}), we obtain
\eqn{
	S^{\mY\times\mB}_{\rm f} \le S^{\mY}_{\rm f} + \beta(\ave{H^\mB}_{\rm f} - F^\mB).
}
Therefore, Eq.~(\ref{syb}) reduces to
\eqn{\label{sqf}
	\Delta S^\mY+ \beta Q \ge 0,
}
where the dissipated heat is defined as
\eqn{
	Q = \ave{H^\mB}_{\rm f} - \ave{H^\mB}_{\rm i}.
}

We assume that the memory initially stores a classical probability distribution $p(m)$ and it is in the local equilibrium state of $\mY_m$ under the condition that the stored information is $m$, i.e.,
\eqn{
	p^\mY(y_{\rm i}) = \sum_{m\in\mM} p(m) p^{\mY_m}_{\rm eq}(y_{\rm i}, \lambda_{\rm i}).
}
Therefore, the initial entropy of the memory is calculated as
\eqn{
	S^\mY_{\rm i} 
	&=& \int_{\mY} dy_{\rm i} \sum_{m\in\mM}
	p(m) p^{\mY_m}_{\rm eq}(y_{\rm i}, \lambda_{\rm i})
	\ln \sum_{m'\in\mM}
	p(m') p^{\mY_{m'}}_{\rm eq}(y_{\rm i}, \lambda_{\rm i})
	\nonumber\\
	&=&
	\sum_{m\in\mM}\int_{\mY_m} dy_{\rm i}\ p(m) p^{\mY_m}_{\rm eq}(y_{\rm i}, \lambda_{\rm i})
	\ln \sum_{m'\in\mM}
	p(m') p^{\mY_{m'}}_{\rm eq}(y_{\rm i}, \lambda_{\rm i})
	\nonumber\\
	&=&
	\sum_{m\in\mM}\int_{\mY_m} dy_{\rm i}\ p(m) p^{\mY_m}_{\rm eq}(y_{\rm i}, \lambda_{\rm i})
	\ln 	p(m) p^{\mY_m}_{\rm eq}(y_{\rm i}, \lambda_{\rm i})
	\nonumber\\
	&=&S^\mM + \beta (\ave{H^\mY}_{\rm i} - F^\mY_{\rm i}).
}
On the other hand, by the positivity of the relative entropy, we obtain
\eqn{
	S^\mY_{\rm f} \le \beta(\ave{H^\mY}_{\rm f} - F^\mY_{\rm f}),
}
which reduces Eq.~(\ref{sqf}) to
\eqn{\label{bhfqs}
	\beta(\Delta \ave{H^\mY} - \Delta F^\mY + Q) \ge S^\mM.
}
We identify $\Delta \ave{H^\mY} + Q$ as work $W$, since it is the energy increase of the total system.
Therefore, we conclude
\eqn{
	W-\Delta F^\mY \ge \kb T S^\mM,
}
which reveals the minimum work needed to erase the information stored in the memory.

When $F^{\mY_m}(\lambda_{\rm f})=F^{\mY_0}(\lambda_{\rm i})$ for arbitrary $m$, we have no free-energy difference and Eq.~(\ref{bhfqs}) reduces to
\eqn{
	W \ge \kb T S^\mM.
}
In other words, in symmetric memories, Eq.~(\ref{bhfqs}) reduces to Landauer's principle, which states that the minimum cost for erasure is the Shannon entropy of the stored information.

\subsection{Reconciliation of the demon with the conventional second law}
In this section, we summarize the results obtained in the previous sections, and demonstrate that the second law is recovered in the total process \cite{SU09}.

The second law under feedback control is
\eqn{\label{rdcsl1}
	W_{\rm fb} -\Delta F^\mX_{\rm fb} &\ge& -\kb T I^{\mX:\mM}.
}
The minus of the right-hand side quantifies the energy gain of feedback control, or the extractable work beyond the conventional second law thanks to Maxwell's demon.
In the measurement process, we have obtained
\eqn{\label{rdcsl2}
	W_{\rm meas}-\Delta F^\mX_{\rm meas}-\Delta F^\mY_{\rm meas} &\ge& \kb T(I^{\mX:\mM}-S^\mM),
}
whose right-hand side represents the additional work cost needed for the demon to perform the measurement.
In the erasure process, we have derived
\eqn{\label{rdcsl3}
	W_{\rm eras} -\Delta F^\mY_{\rm eras}&\ge& \kb T S^\mM.
}
The right-hand side is the additional work cost for the erasure of the information stored by the demon.
Therefore, the sum of the costs for both the measurement and erasure is given by
\eqn{\label{rdcsl4}
	W_{\rm m+e} - \Delta F^\mX_{\rm m+e} - \Delta F^\mY_{\rm m+e} &\ge& \kb T I^{\mX:\mM},
}
where we define $W_{\rm m+e} = W_{\rm meas} +W_{\rm eras}$ and other quantities in a similar way.
We see that the work gain by the demon is precisely compensated by the work cost that the demon pays for the measurement and erasure.
In fact, in total, the information quantities on the right-hand sides of Eqs.~(\ref{rdcsl1}) and (\ref{rdcsl4}) are cancelled out, and we obtain
\eqn{\label{rdcsl5}
	W_{\rm tot} - \Delta F^\mX_{\rm tot} - \Delta F^\mY_{\rm tot}&\ge& 0,\label{slli2}
}
where we define $W_{\rm tot} = W_{\rm meas} + W_{\rm fb} + W_{\rm eras}$ and other quantities in a similar way.
The conventional second-law-like inequality (\ref{slli2}) is recovered in the total process consisting of the measurement, feedback control, and erasure processes.
In particular, in an isothermal cycle, Eq.~(\ref{slli2}) reduces to
$
	W_{\rm tot} \ge 0,
$
which is nothing but Kelvin's principle for the isothermal composite system consisting of the system and the memory.

As reviewed in Sec. 2.1, Brillouin argued that the entropy reduction by the demon is compensated by the cost for the measurement.
On the other hand, Landauer's principle says that the work cost for the erasure of the stored information exceeds the work gain of the feedback control.
Although these views are true for some specific systems, they are not generally true.
What compensates the work gain by the demon in general is not the individual work cost for the measurement or erasure but the joint work cost for the  measurement and erasure processes.
The information-thermodynamic inequalities summarized above reveal that the reconciliation of Maxwell's demon with the second law is achieved in a rigorous manner.

In terms of the Shannon entropy, we have obtained
\eqn{\label{rdcsl6}
	\Delta S^\mX_{\rm fb} + \beta Q_{\rm fb} &\ge& - I^{\mX:\mM}, 
	\\
	\label{rdcsl7}
	\Delta S^\mX_{\rm meas} + \Delta S^\mY_{\rm meas} + \beta Q_{\rm meas}&\ge& I^{\mX:\mM},
	\\
	\label{rdcsl8}
	\Delta S^\mY_{\rm eras} + \beta Q_{\rm eras} &\ge& 0,
	\\
	\label{rdcsl9}
	\Delta S^\mX_{\rm m+e} + \Delta S^\mY_{\rm m+e} + \beta Q_{\rm m+e}&\ge& 0,
	\\
	\label{rdcsl10}
	\Delta S^\mX_{\rm tot} + \Delta S^\mY_{\rm tot} + \beta Q_{\rm tot} &\ge& 0.
}
The inequalities (\ref{rdcsl7}) and (\ref{rdcsl8}) are simpler than Eqs. (\ref{rdcsl2}) and (\ref{rdcsl3}) in that they do not have the Shannon entropy term $S^\mM$ explicitly.
We note that the inequalities on the Shannon entropy (\ref{rdcsl6}), (\ref{rdcsl7}), (\ref{rdcsl8}), (\ref{rdcsl9}), and (\ref{rdcsl10}) are stronger than the inequalities on the work (\ref{rdcsl1}), (\ref{rdcsl2}), (\ref{rdcsl3}), (\ref{rdcsl4}), and (\ref{rdcsl5}), since the latter inequalities can be derived from the former inequalities based on positivity of the relative entropy.

\section{Nonequilibrium equalities under measurements and feedback control}
The second-law-like inequality under measurements and feedback control derived in the previous section can be generalized to nonequilibrium equalities as the conventional second law is generalized to the nonequilibrium equalities reviewed in Sec 2.1.

In this section, we review information-thermodynamic nonequilibrium equalities.
Then, we derive these equalities in a unified manner similar to that of Sec. 2.2.

\subsection{Information-thermodynamic nonequilibrium equalities}
In 2010, Sagawa and Ueda generalized the Jarzynski equality in a Markov stochastic system under feedback control based on a single measurement and obtained the Sagawa-Ueda equality \cite{SU10}
\eqn{\label{SUE}
	\ave{e^{-\beta(W-\Delta F)-I}}=1,
}
where $I$ is the unaveraged mutual information obtained by the measurement.
Using Jensen's inequality, we succinctly reproduce the second law of information thermodynamics as
\eqn{
	\ave{W}-\Delta F \ge -\kb T \ave{I}.
}
Later, the Sagawa-Ueda equality is generalized to Markov systems with multiple measurements \cite{HV10}, and to non-Markov systems with multiple measurements \cite{SU12E}.

The Sagawa-Ueda equality has variants as the Jarzynski equality has the variants reviewed in Sec. 2.1.
The Hatano-Sasa relation is generalized to systems under feedback control as \cite{AS12}
\eqn{\label{GHS}
	\ave{e^{-\Delta \phi - \Delta s_{\rm ex}/\kb - I}} = 1.
}
The associate inequality is given by
\eqn{
	\ave{\Delta\phi} + \ave{\Delta s_{\rm ex}}/\kb \ge -\ave{I}.
}
Moreover, in Ref.~\cite{AS12}, the Seifert relation is generalized to
\eqn{\label{GSR}
	\ave{e^{-\Delta s_{\rm tot}/\kb-I}} = 1,
}
which leads to a second-law-like inequality
\eqn{
	\ave{\Delta s_{\rm tot}} \ge -\kb\ave{I}.
}
Reference~\cite{LRJ12} derived
\eqn{\label{GHKR}
	\ave{e^{-\Delta s_{\rm hk}-I}}=1,
}
and 
\eqn{
	\ave{\Delta s_{\rm hk}} \ge -\kb\ave{I}.
}
Equations (\ref{SUE}), (\ref{GHS}), (\ref{GSR}), and (\ref{GHKR}) are summarized in terms of the formal entropy production $\sigma$ as
\eqn{\label{GNEFB}
	\ave{e^{-\sigma -I}}=1.
}
This equality is a general nonequilibrium equalities under measurements and feedback control.
The second law of information thermodynamics is reproduced as
\eqn{
	\ave{\sigma} \ge -\ave{I}.
}

%
\subsection{Derivation of information-thermodynamic nonequilibrium equalities}
In this section, we derive the nonequilibrium equalities under measurements and feedback control.

\subsubsection{Setup}
We consider a nonequilibrium process in a classical non-Markov stochastic system with feedback control from time $t_{\rm i}$ to $t_{\rm f}$.
To formulate the dynamics of the system, we discretize the time interval into $N$ parts and define $t_n=t_{\rm i} + n(t_{\rm f}-t_{\rm i})/N\ (n=0, \cdots, N)$.
Let $\mX$ and $\mY$ denote phase spaces of the system and the memory, respectively.
Moreover, we denote the phase-space positions in $\mX$ and $\mY$ at time $t_n$ by $x_n$ and $y_n$, respectively.
An external parameter to control the system is denoted by $\lambda$, and the value of $\lambda$ at time $t_n$ is $\lambda_n$.
Initially, the system and memory are at $(x_0,y_0)$ sampled from an initial joint probability distribution $p_0(x_0, y_0)$.
Then, the system evolves from time $t_0$ to $t_1$ and a measurement is performed at time $t_1$ to obtain a measurement outcome $y_1$.
After that, the system is subject to feedback control driven by an external parameter $\lambda_1(y_1)$ from time $t_1$ to $t_2$.
In this way, we repeat measurements and feedback control.
We define $X_n=(x_0, x_1, \cdots, x_n)$ and $Y_n=(y_0,\cdots,y_n)$, and therefore $X_N$ and $Y_{N-1}$ represents the entire trajectory of the system and memory, respectively.
Since $\lambda_n$ is adjusted based on the measurement outcomes before time $t_n$, $\lambda_n$ is a function of $Y_n$.
Therefore, the non-Markov dynamics of the system is determined by transition probabilities
\eqn{
	p(x_{n+1}|X_n, \lambda_n(Y_n)).
}
On the other hand, since the measurement outcome depends on the trajectory of the system before the measurement, the dynamics of the memory is determined by
\eqn{
	p(y_n|X_n).
}
Therefore, the joint probability to realize $X_N$ and $Y_{N-1}$ is calculated as
\eqn{\label{jtrtr}
	P[X_N,Y_{N-1}]
	&=&
	\prod_{n=1}^{N-1} p(x_{n+1}|X_n, \lambda_n(Y_n)) p(y_n|X_n)
	\cdot p(x_{1}|x_0, \lambda_0(y_0))p_0(x_0, y_0)\ \ \ \ \ 
	\nonumber\\
	&=&
	P^{\rm tr}[X_N|x_0,\Lambda_{N-1}(Y_{N-1})]P^{\rm tr}[Y_{N-1}|X_{N-1}]p_0(x_0, y_0),
}
where we define the transition probabilities as
\eqn{
	P^{\rm tr}[X_N|x_0,\Lambda_{N-1}(Y_{N-1})]
	&=&
	\prod_{n=0}^{N-1} p(x_{n+1}|X_n, \lambda_n(Y_n)),
	\\
	P^{\rm tr}[Y_{N-1}|X_{N-1}]
	&=&
	\prod_{n=1}^{N-1} p(y_n|X_n).
}
Dividing Eq.~(\ref{jtrtr}) by $P[Y_{N-1}]$, we calculate the conditional probability as
\eqn{\label{cntr}
	P[X_N|Y_{N-1}] = P^{\rm tr}[X_N|\Lambda_{N-1}(Y_{N-1})]
	\frac{P^{\rm tr}[Y_{N-1}|X_{N-1}]p_0(x_0, y_0)}{P[Y_{N-1}]p_0(x_0)}.
}
Therefore, the conditional probability is different from the transition probability under measurements and feedback control.

Following Ref.~\cite{SU12E}, we define the information obtained by the measurement at time $t_n$ as the mutual information between $y_n$ and $X_n$ under the condition that we have obtained $Y_{n-1}$, that is,
\eqn{
	I_n[y_n, X_n|Y_{n-1}]
	&=&
	-\ln p(y_n|Y_{n-1}) + \ln p(y_n|X_n, Y_{n-1})
	\nonumber\\
	&=&
	-\ln p(y_n|Y_{n-1}) + \ln p(y_n|X_n).
}
The sum of the mutual information is calculated as
\eqn{
	I = \sum_{n=1}^{n-1} I_n &=& -\ln P[Y_{N-1}|y_0] + \ln P^{\rm tr}[Y_{N-1}|X_{N-1}]
	\nonumber\\
	&=& -\ln\frac{P[Y_{N-1}]}{P^{\rm tr}[Y_{N-1}|X_{N-1}]p_0(y_0)},
}
which is the total information obtained by all the measurements.
Using the total mutual information, we can transform Eq.~(\ref{cntr}) as
\eqn{\label{trcni}
	\frac{P^{\rm tr}[X_N|\Lambda_{N-1}(Y_{N-1})]}{P[X_N|Y_{N-1}]}
	=
	e^{-I-I_{\rm i}},
}
where $I_{\rm i}$ is the mutual information at the initial time defined by
\eqn{
	I_{\rm i} = -\ln\frac{p_0(x_0)p_0(y_0)}{p_0(x_0,y_0)}.
}

\subsubsection{Derivation of information-thermodynamic nonequilibrium equalities}
The formal entropy production $\sigma$ is defined as the ratio of the reference probability to the original probability under a fixed protocol as in Eq.~(\ref{DFT}).
Therefore, we obtain
\eqn{
	\frac{P^{\rm r}[X_N|\Lambda_{N-1}(Y_{N-1})]}{P^{\rm tr}[X_N|\Lambda_{N-1}(Y_{N-1})]} = e^{-\sigma[X_N|\Lambda_{N-1}(Y_{N-1})]}.
}
We define the joint probability of the reference process as
\eqn{\label{refprob}
	P^{\rm r}[X_N, Y_{N-1}]
	= P^{\rm r}[X_N|\Lambda_{N-1}(Y_{N-1})]P[Y_{N-1}],
}
which means that we sample the reference process conditioned by $Y_{N-1}$ with the same probability $P[Y_{N-1}]$ as the original process.
From Eqs.~(\ref{jtrtr}) and (\ref{refprob}), the ratio of the reference joint probability to the original joint probability is calculated as
\eqn{
	\frac{P^{\rm r}[X_N, Y_{N-1}]}{P[X_N, Y_{N-1}]}
	&=&
	\frac{P^{\rm r}[X_N|\Lambda_{N-1}(Y_{N-1})]}{P^{\rm tr}[X_N|\Lambda_{N-1}(Y_{N-1})]}
	\frac{P[Y_{N-1}]p_0(x_0)}{P^{\rm tr}[Y_{N-1}|X_{N-1}]p_0(x_0, y_0)}
	\nonumber\\
	&=&
	e^{-\sigma-I-I_{\rm i}}.
}
Multiplying both sides by ${P[X_N, Y_{N-1}]}$ and integrating over $X_N$ and $Y_{N-1}$, we obtain
\eqn{\label{esii}
	\ave{e^{-\sigma-I-I_{\rm i}}}= 1,
}
because the reference probability is normalized to unity.
In ordinary feedback protocols, the system is assumed not to be correlated with the memory at the initial time.
Therefore, Eq.~(\ref{esii}) reduces to
\eqn{\label{mneit}
	\ave{e^{-\sigma-I}}= 1.
}
The appropriate choices of the reference probability explained in Sec. 2.2 reduce Eq.~(\ref{mneit}) to Eqs.~(\ref{SUE}), (\ref{GHS}), (\ref{GSR}), and (\ref{GHKR}).

\section{Experiments}
In this section, we briefly review experimental demonstrations of Maxwell's demon.

\begin{figure}
\begin{center}
\includegraphics[width = 1.0\columnwidth]{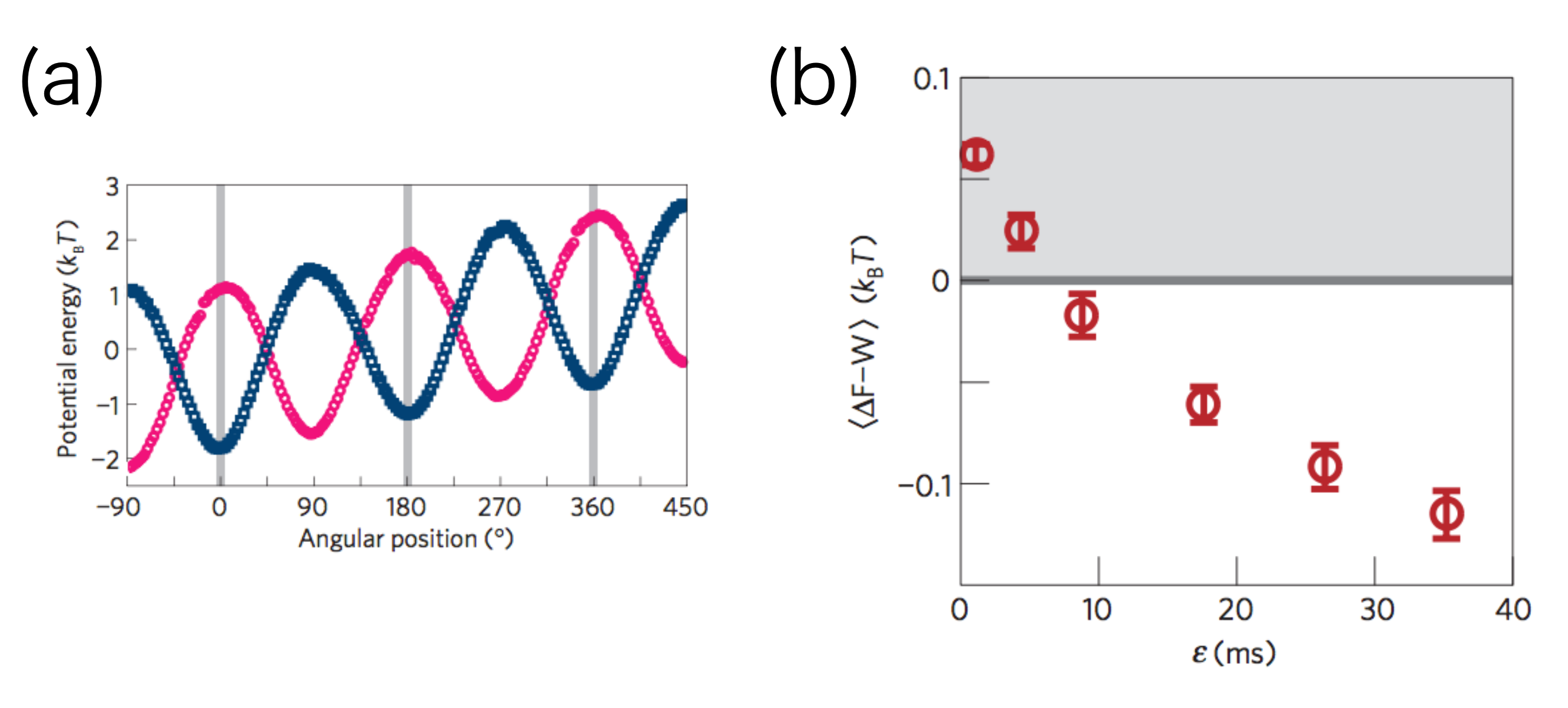}
\caption{\label{fig:Toyabe}
(a) Two washboard potentials induced by an electromagnetic field.
If the Brownian particle climbs up the potential due to thermal agitation, the potential is switched to the other potential to prevent the particle from descending.
(b) Free-energy gain of the Brownian particle subtracted by the amount of work done on it.
The abscissa represents the delay time  $\epsilon$ it takes to switch the potential after the position measurement.
The first two data points show the net free energy gain beyond the conventional second law of thermodynamics.
Reproduced from Figs.~2b and 3d in Ref.~\cite{TSUe10}.
Copyright 2010 by the Macmillan Publishers Limited.
}
\end{center}
\end{figure}

The first experimental realization of Maxwell's demon was done by Toyabe et al.~\cite{TSUe10}.
They demonstrated that the free energy of a Brownian particle can be increased by feedback control based on measurements of the position of the particle.
A couple of polystyrene beads are suspended in a water with one of them anchored to a glass plate.
The other bead can move on a ring, and is subjected to a washboard potential created by four electrodes (see Fig.~\ref{fig:Toyabe}~(a)).
At a certain instant in time, the position of the particle is measured.
After a delay time $\epsilon$, if the particle climbs up the potential due to thermal agitation, the potential is switched to the other potential to prevent the particle from descending;
otherwise the potential is left unchanged.
This protocol of feedback control is repeated.
As a result, the particle is able to gain free energy larger than the work done on it in this feedback-controlled process (see Fig.~\ref{fig:Toyabe}~(b)), namely,
it was demonstrated that information obtained by the measurements can be used as a resource for free energy.

\begin{figure}
\begin{center}
\includegraphics[width = 1.0\columnwidth]{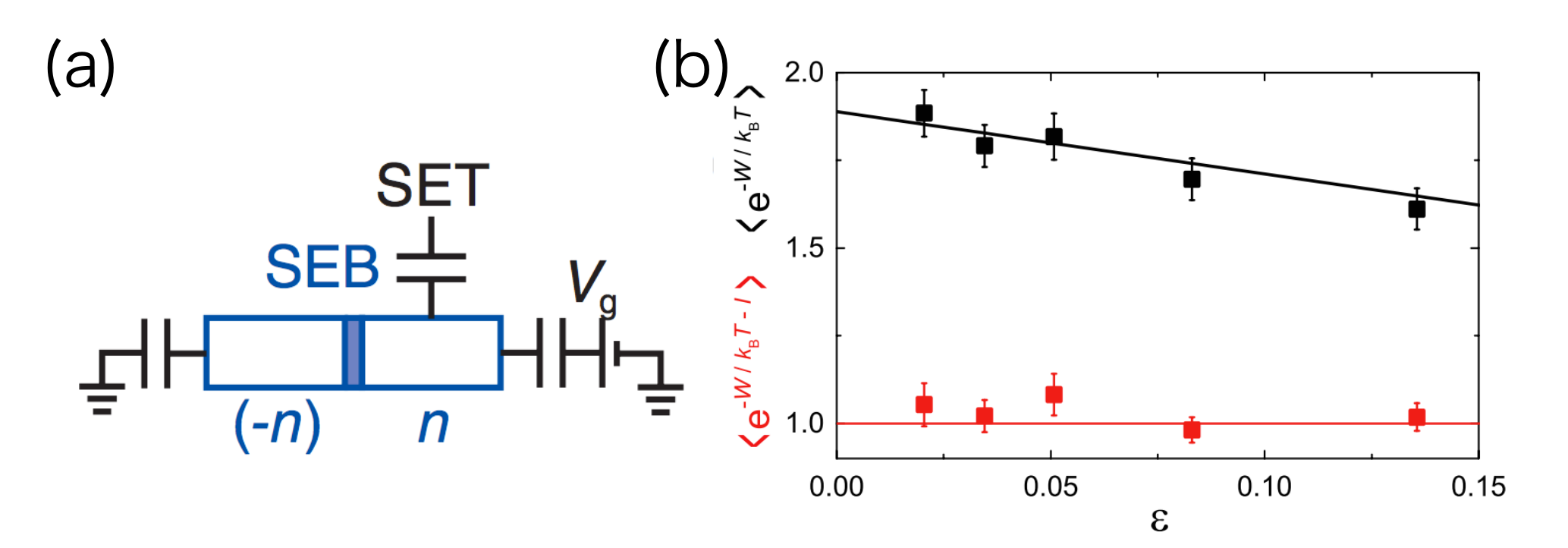}
\caption{\label{fig:Koski}
(a) Schematic illustration of the system demonstrating the information-thermodynamic nonequilibrium equality~(\ref{SUE}).
The single-electron box (SEB) consists of two metallic islands connected by a junction, through which electrons are transported by tunneling.
The number $n$ of excess electrons in the right island is monitored by a single-electron transistor (SET).
The ground-state average value $n_{\rm g}$ of $n$ is controlled by the gate voltage $V_{\rm g}$.
(b) Experimental verification of Eq.~(\ref{SUE}).
The abscissa represents the error rate of the measurement of $n$.
The data shown by black squares deviate significantly from unity, showing the breakdown of the Jarzynski equality, whereas the red squares show that Eq.~(\ref{SUE}) holds.
Reproduced from Figs.~1~(a) and 3~(b) in Ref.~\cite{KMSP14}.
Copyright 2014 by the American Physical Society.
}
\end{center}
\end{figure}

The information-thermodynamic nonequilibrium equality~(\ref{SUE}) was experimentally verified in Ref.~\cite{KMSP14}.
A feedback-controlled two-state system similar to the Szilard engine was implemented in a single-electron box (SEB) illustrated in Fig.~\ref{fig:Koski}~(a).
The gate voltage $V_{\rm g}$ in Fig.~\ref{fig:Koski}~(a) is adjusted so that the minimum charging energy is achieved when the average number $n$ of the electrons that have tunneled from the left island to the right one is $n_{\rm g}=0.5$.
The temperature of the system is low enough that the SEB is in either the $n=0$ or $n=1$ state.
Therefore, the state with $n=0$ is realized with the probability of $0.5$, so is the state with $n=1$.
The state of the SEB is monitored by a single-electron transistor (SET) (see Fig.~\ref{fig:Koski}~(a)), and the feedback control is conducted by changing $V_{\rm g}$ based on the value of $n$ to extract work from the SEB.
The average in Eq.~(\ref{SUE}) was experimentally confirmed to be unity within experimental errors as shown in Fig.~\ref{fig:Koski}~(b).

\chapter{Nonequilibrium Equalities in Absolutely Irreversible Processes}

As discussed in Chap. 2, nonequilibrium equalities apply to rather general nonequilibrium situations.
However, it is known that integral fluctuation theorems are inapplicable to some situations.
We propose a new concept of absolute irreversibility as a novel class of irreversibility that encompasses the entire range of those situations to which conventional integral nonequilibrium equalities cannot apply.
In mathematical terms, the absolute irreversibility is defined as the singular part of the reference probability measure, and is uniquely separated from the ordinary irreversible part by Lebesgue's decomposition theorem~\cite{Hal74, Bar95}.
We derive nonequilibrium equalities that are applicable to absolutely irreversible processes based on measure theory.
Inequalities derived from our nonequilibrium equalities give a positive-definite lower bound of the entropy production when a process involves absolute irreversibility.

First of all, we consider free expansion to illustrate that absolute irreversibility causes inapplicability of conventional nonequilibrium integral equalities, and define absolute irreversibility in terms of measure theory.
Then, we derive nonequilibrium equalities in absolutely irreversible processes based on Lebesgue's decomposition theorem.
Next, we verify our nonequilibrium equalities in several examples.
Finally, we compare our method with a conventional method and discuss merits of ours.

In this chapter, we restrict our attention to systems without measurements and feedback control.
This chapter is mainly based on Ref.~\cite{MFU14}.

\section{Inapplicability of conventional integral nonequilibrium equalities and absolute irreversibility}
In this section, we introduce absolute irreversibility in an example of free expansion, to which the Jarzynski equality cannot apply.
Then, we mathematically define absolute irreversibility in terms of measure theory.

\subsection{Inapplicability of the Jarzynski equality}
\begin{figure}
\begin{center}
\includegraphics[width=0.5\columnwidth]{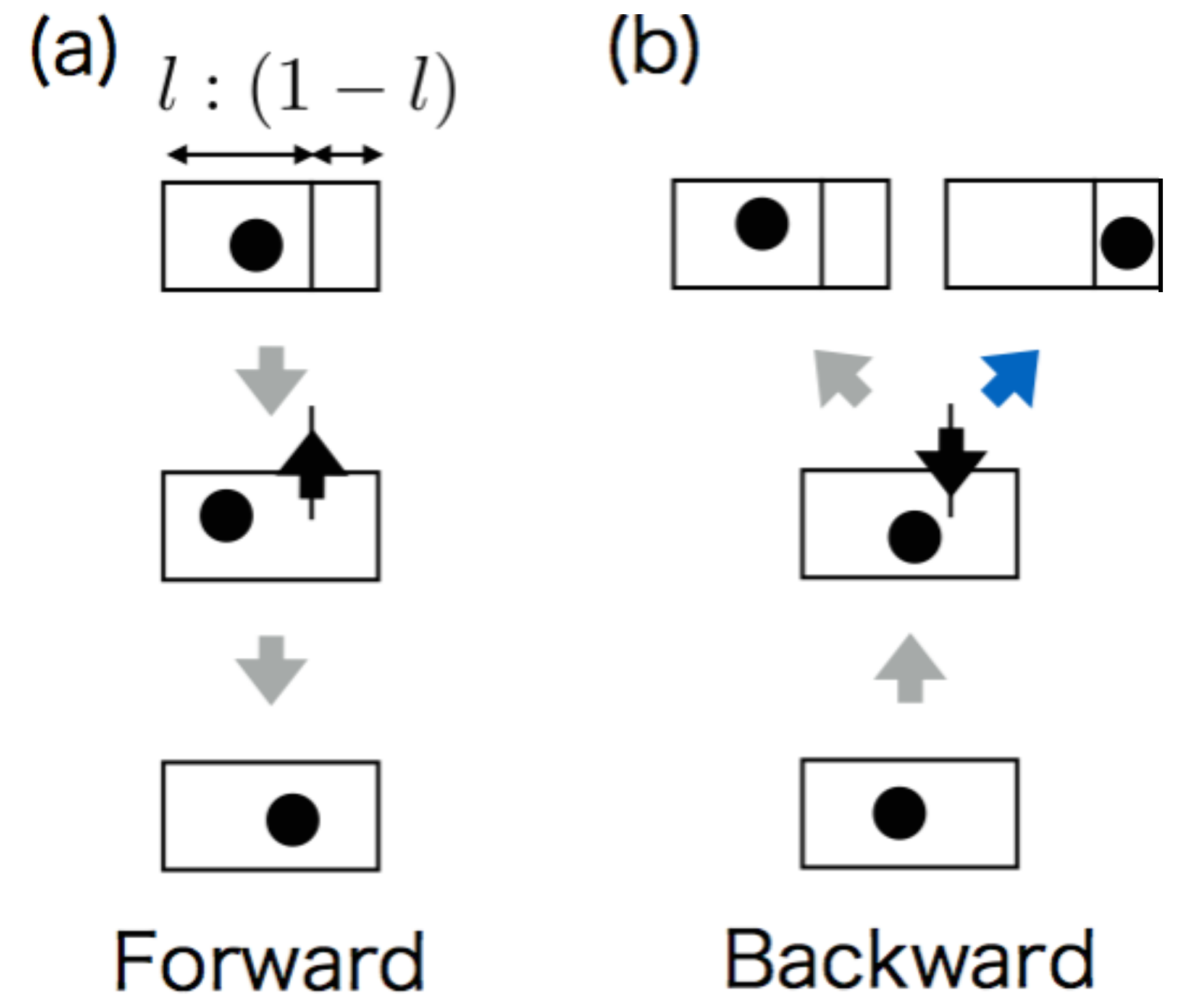}
\caption{\label{fig:FE}
(a) Forward process of free expansion.
An ideal single-particle gas is initially in a local equilibrium in the left box with temperature $T$.
Then, the partition is removed, and the gas freely expand to the entire box.
(b) Backward process of free expansion.
Initially, the single-particle gas is in a global equilibrium of the entire box with temperature $T$.
Then, the partition is inserted, and the gas particle is in either the left or right box.
The backward path ending in the right box (indicated by the blue arrow) has no corresponding forward path.
Therefore, this is a singular path with a negatively divergent entropy production.
Reproduced from Fig. 1 of Ref.~\cite{MFU14}. Copyright 2014 by the American Physical Society.
}
\end{center}
\end{figure}
The Jarzynski equality is known to be inapplicable to free expansion \cite{Gro05Aug, Jar05}.
Here, we illustrate this fact.
Suppose that an ideal single-particle gas at temperature $T$ is prepared in the left side of a box with a partition as illustrated in Fig.~\ref{fig:FE} (a).
Then, we remove the partition and let the gas expand to the entire box.
In this process, work is not extracted
($
	W=0
$),
whereas the free energy decreases
($
	\Delta F < 0
$).
Therefore, the dissipated work is always positive:
\eqn{
	W-\Delta F >0.
}
Thus, we have
\eqn{
	\langle e^{-\beta(W-\Delta F)} \rangle < 1,
}
which means that the Jarzynski equality~(\ref{Jar}) is not satisfied in this process \cite{Gro05Aug}.
In physical terms, this is because the Jarzynski equality assumes that the initial state is in a global equilibrium and this assumption is not satisfied in the present case.
Recall that, in free expansion, the initial state is not a global equilibrium state, but only a local equilibrium state \cite{Jar05}.
Therefore, the Jarzynski equality cannot apply to free expansion.
Then, it is natural to ask why a local equilibrium state cannot be assumed as an initial condition of the conventional integral nonequilibrium equality.

The mathematical reason is that we have paths with a divergent entropy production when we start from a local equilibrium state.
This statement is illustrated in free expansion as follows.
We consider a set of virtual paths $\{\Gamma_{\rm R}\}$ starting from the right box.
By assumption, the probability of these paths in the forward process vanishes: $\mP[\Gamma_{\rm R}]=0$.
On the other hand, the probability of the corresponding backward paths is nonvanishing: $\mP^\dag[\Gamma^\dag_{\rm R}]\neq 0$.
Therefore, we have
\eqn{\label{defAI}
	^\exists\Gamma,\ \mP[\Gamma]=0\ \&\ \mP^\dag[\Gamma^\dag]\neq 0.
}
For these paths, the entropy production is negatively divergent in the context of the Crooks fluctuation theorem~(\ref{CFT}) because
\eqn{
	\sigma=\beta(W-\Delta F)=-\ln\frac{\mP^\dag[\Gamma^\dag]}{\mP[\Gamma]}\to-\infty
}
and
\eqn{
	e^{-\sigma}=\frac{\mP^\dag[\Gamma^\dag]}{\mP[\Gamma]} \to \infty.
}
Due to the paths with a divergent entropy production, the conventional integral nonequilibrium equality breaks down:
\eqn{
	\langle e^{-\sigma} \rangle
	&=&\nonumber
	\int_{\mP[\Gamma]\neq0} e^{-\sigma} \mP[\Gamma] \mD\Gamma\\
	&=&\nonumber
	\int_{\mP[\Gamma]\neq0} \mP^\dag[\Gamma^\dag] \mD\Gamma^\dag\\
	&=&\nonumber
	1-\int_{\mP[\Gamma]=0} \mP^\dag[\Gamma^\dag] \mD\Gamma^\dag\\
	&<&1.
}
Therefore, we conclude that the negatively divergent entropy production of the paths starting from the region in which the initial probability vanishes is what makes the conventional integral nonequilibrium equality inapplicable to the process starting from a local equilibrium state.

This situation [Eq.~(\ref{defAI})] makes a stark contrast to ordinary irreversible processes, where every backward path has the corresponding forward path with a nonvanishing probability:
\eqn{\label{defOI}
	^\forall\Gamma,\ \mP^\dag[\Gamma^\dag]\neq0\Rightarrow
	\mP[\Gamma]\neq0,
}
or
\eqn{
	^\forall\Gamma,\ \mP[\Gamma]=0\Rightarrow\mP^\dag[\Gamma^\dag]=0.
}
Therefore, in the ordinary irreversible case, the exponentiated entropy production remains finite
\eqn{
	e^{-\sigma} = \frac{\mP^\dag[\Gamma^\dag]}{\mP[\Gamma]}<\infty,
}
and the thermodynamic irreversibility is quantitatively characterized by the entropy production.
Although these paths are thermodynamically irreversible, they are stochastically reversible in a sense of Eq.~(\ref{defOI}), namely, every backward path has the nonvanishing forward counterpart.
In contrast, if the condition~(\ref{defAI}) holds, there exist backward paths that have no counterparts in the forward process, which means that these paths are not even stochastically reversible.
Therefore, we shall call these paths absolutely irreversible paths, and call the processes with absolutely irreversible paths absolutely irreversible processes.

Mathematically, probability theory is based on measure theory, and the ratio $\mP^\dag/\mP$ is interpreted as the transformation function of the two probability measures.
Therefore, we need measure theory to formulate absolute irreversibility.
We thus give a mathematical definition of absolute irreversibility.


%
\subsection{Definition of absolute irreversibility}
\begin{figure}
\begin{center}
\includegraphics[width=0.8\columnwidth]{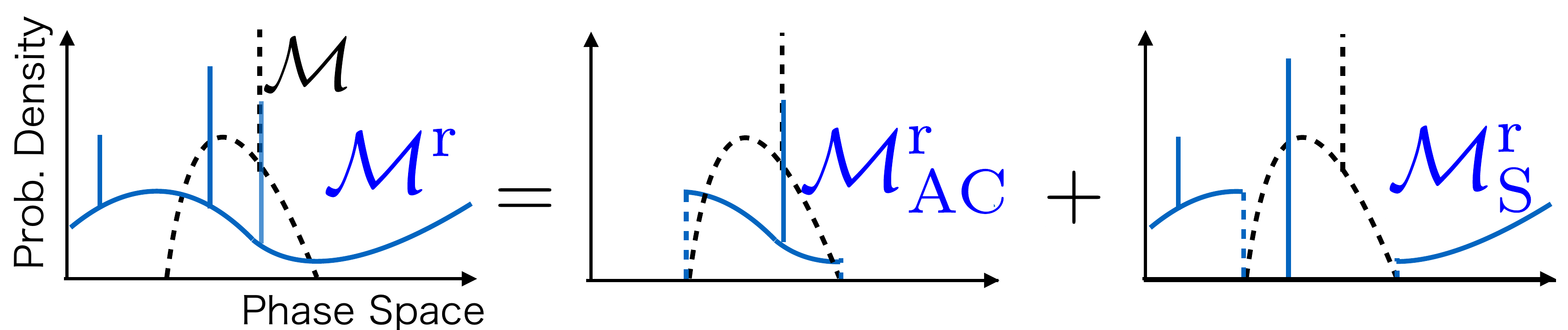}
\caption{\label{fig:LD1}
Schematic illustration of Lebesgue's decomposition theorem.
The abscissa represents coordinates of phase space and the ordinate shows the probability density.
Vertical lines represent $\delta$-function-like localization.
The reference probability measure $\mM^{\rm r}$ (blue solid curve) is decomposed into two parts with respect to the original probability measure $\mM$ (dashed curve).
The probability ratio is well-defined in the absolutely continuous part, whereas it diverges in the singular part.
Note that the $\delta$-function-like reference probability measure is absolutely continuous if its singular part coincides with that of the original probability measure as shown in the middle figure.
Reproduced from Fig. 2 of Ref.~\cite{MFU14}. Copyright 2014 by the American Physical Society.
}
\end{center}
\end{figure}

Let $\mM[\mD\Gamma]$ denote the probability measure of the original process.
This is a generalization of the description in terms of the probability density which is written as $\mP[\Gamma]\mD\Gamma$.
Let the probability measure of the reference process be denoted by $\mM^{\rm r}[\mD\Gamma]$ which is a generalization of $\mP^{\rm r}[\Gamma]\mD\Gamma$.
According to Lebesgue's decomposition theorem~\cite{Hal74, Bar95}, the reference probability measure can be uniquely decomposed into two parts as
\eqn{\label{LD1}
	\mM^{\rm r} = \mM^{\rm r}_{\rm AC} + \mM^{\rm r}_{\rm S},
}
where $\mM^{\rm r}_{\rm AC}$ and $\mM^{\rm r}_{\rm S}$ are absolutely continuous and singular with respect to $\mM$, respectively (see Fig.~\ref{fig:LD1}).
The absolute continuity of $\mM^{\rm r}_{\rm AC}$ guarantees that the probability ratio is well-defined due to the Radon-Nikod\'ym theorem~\cite{Hal74, Bar95} as
\eqn{\label{RN}
	\frac{\mD\mM^{\rm r}_{\rm AC}}{\mD\mM},
}
which is an integrable function with respect to $\mM$.
In physical terms, it is this ratio that gives the entropy production.
Therefore, in measure theory, the Crooks fluctuation theorem reads
\eqn{\label{CFT2}
	\frac{\mD\mM^{\rm r}_{\rm AC}}{\mD\mM}=e^{-\sigma}.
}
On the other hand, the probability defined by $\mM^{\rm r}_{\rm S}$ takes a nonzero value in the region where the probability defined by $\mM$ vanishes.
Therefore, the ratio of $\mM^{\rm r}_{\rm S}$ to $\mM$ is divergent, and we cannot define a finite entropy production through this ratio.
Thus, in physical terms, $\mM^{\rm r}_{\rm S}$ corresponds to the absolutely irreversible part.
We therefore identify $\mM^{\rm r}_{\rm AC}$ as the ordinary irreversible part and $\mM^{\rm r}_{\rm S}$ as the absolutely irreversible part.
If $\mM^{\rm r}_{\rm S}$ does not vanish, the conventional integral nonequilibrium equality breaks down \cite{Gro05Aug, Jar05, Sun05, LG05}.
See Appendix B for the mathematical definitions of absolute continuity and singularity, and a mathematical statement of Lebesgue's decomposition theorem.

\begin{figure}
\begin{center}
\includegraphics[width=0.85\columnwidth]{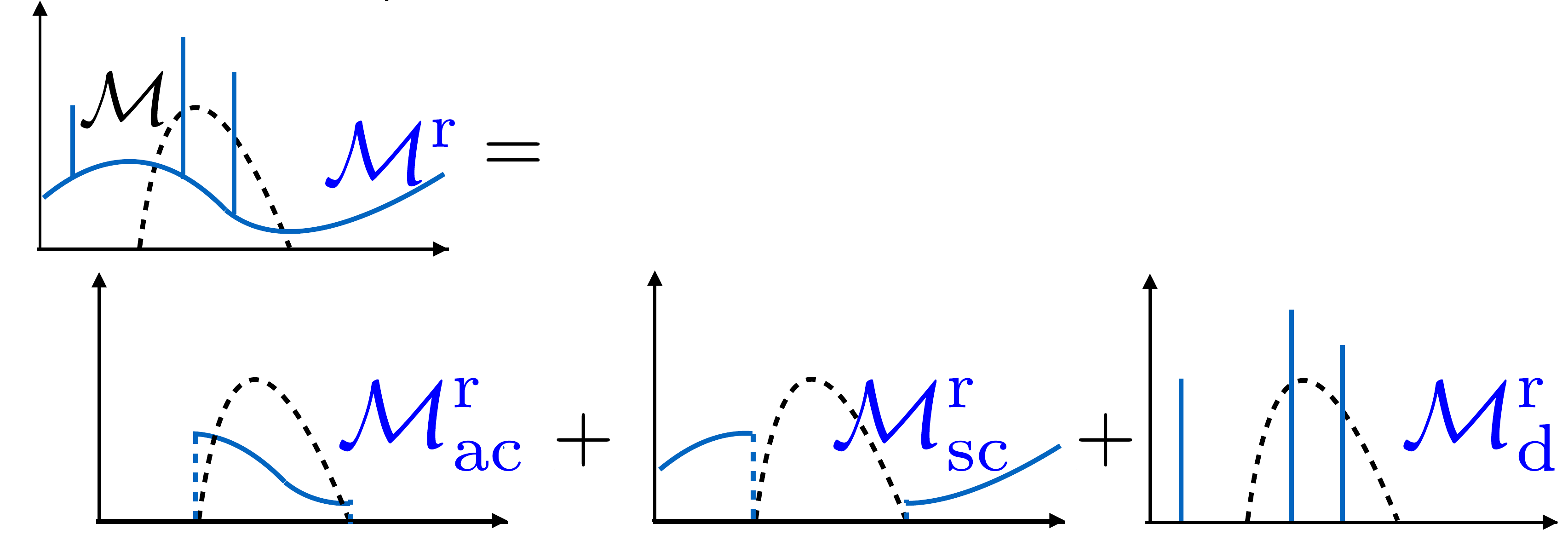}
\caption{\label{fig:LD2}
Schematic illustration of the stronger version of Lebesgue's decomposition theorem.
The reference probability measure $\mM^{\rm r}$ (blue solid curve) is decomposed into three parts with respect to the original probability measure $\mM$ (dashed curve).
In the absolutely continuous part (bottom left), the probability ratio of $\mM^{\rm r}_{\rm ac}$ to $\mM$ is well-defined.
In the singular continuous part (bottom middle), the probability ratio of $\mM^{\rm r}_{\rm sc}$ to $\mM$ is divergent because the probability density of $\mM$ vanishes and that of $\mM^{\rm r}_{\rm sc}$ remains nonvanishing.
In the discrete part (bottom right), the probability ratio of $\mM^{\rm r}_{\rm d}$ to $\mM$ is divergent because the probability density of $\mM^{\rm r}_{\rm d}$ is divergent whereas $\mM$ remains finite.
We note that $\mM$ does not involve $\delta$-function-like localization, which is the assumption of the stronger version of Lebesgue's decomposition and needed for the uniqueness of this decomposition (see Appendix B for more detail).
Reproduced from Fig. 3 of Ref.~\cite{MFU14}. Copyright 2014 by the American Physical Society.
}
\end{center}
\end{figure}

When $\mM$ can be written in terms of a probability density, a stronger version of Lebesgue's decomposition theorem holds.
In this case, $\mM^{\rm r}$ is decomposed into three parts as
\eqn{\label{LD2}
	\mM^{\rm r}
	=
	\mM^{\rm r}_{\rm ac}+\mM^{\rm r}_{\rm sc}+\mM^{\rm r}_{\rm d},
}
where $\mM^{\rm r}_{\rm ac}$ and $\mM^{\rm r}_{\rm sc}$ are absolutely continuous and singular continusous with respect to $\mM$, respectively, and $\mM^{\rm r}_{\rm d}$ is the discrete part of $\mM^{\rm r}$ (see Fig.~\ref{fig:LD2}).
The singular continuous part $\mM^{\rm r}_{\rm sc}$ corresponds to the region where the probability ratio is divergent because the denominator, which is the forward probability, vanishes.
This term represents the effect of free expansion.
The discrete part $\mM^{\rm r}_{\rm d}$ has $\delta$-function-like localization, and the probability ratio is divergent because the numerator, which is the reference probability, diverges.
This part arises when particles can localize and do not undergo thermal diffusion; such a situation occurs when there are trapping centers of particles.
In this way, the absolute irreversibility is classified into two categories.
The correspondence between the classification of irreversibility and that of probability measure is summarized in Table~\ref{tab:irrev}.

\begin{table}[b]
\caption{\label{tab:irrev}
Correspondence between the classification of irreversibility and that of probability measure.
}
{\renewcommand\arraystretch{1.1}
\begin{center}\begin{tabular}{|c|c|c|c|}
\hline
Class of irreversibility & Ordinary & Absolute I & Absolute II \\
\hline
Class of measure & absolutely continuous & singular continuous & discrete\\
\hline
\rule[-14pt]{0pt}{34pt}
Exponentiated entropy: $e^{-\sigma}$&
$\displaystyle \frac{\mP^{\rm r}[\Gamma]}{\mP[\Gamma]}=$finite &
$\displaystyle \frac{\mP^{\rm r}[\Gamma]}{0}=\infty$ &
$\displaystyle \frac{\delta(0)}{\mP[\Gamma]}=\infty$\\
\hline
\end{tabular}
\end{center}}
\end{table}

\section{Nonequilibrium equalities in absolutely irreversible processes}
In this section, we derive nonequilibrium equalities applicable to absolutely irreversible processes based on Lebesgue's decomposition theorem~(\ref{LD1}) and (\ref{LD2}).

First, we derive general nonequilibrium equalities in absolutely irreversible processes.
Then, we show that they reduce to several individual nonequilibrium equalities with their specific meanings of the entropy production by proper choices of the reference probability distribution as described in Sec. 2.2.

\subsection{General formulation}
First, we derive nonequilibrium equalities based on Lebesgue's decomposition theorem~(\ref{LD1}).
Let $\mF[\Gamma]$ denote an arbitrary functional of a path $\Gamma$, and let $\ave{\cdots}$, $\ave{\cdots}^{\rm r}$ and $\ave{\cdots}^{\rm r}_{\rm I}$ (I=AC, S) denote the average over $\mM$, $\mM^{\rm r}$ and $\mM^{\rm r}_{\rm I}$, respectively.
From Lebesgue's decomposition theorem~(\ref{LD1}), we obtain
\eqn{\label{FFF}
	\langle \mF \rangle^{\rm r}
	=
	\langle \mF \rangle^{\rm r}_{\rm AC}
	+\langle \mF \rangle^{\rm r}_{\rm S}.
}
On the other hand, using the Radon-Nikod\'ym derivative~(\ref{RN}), we evaluate the average over the absolutely continuous part in terms of the entropy production as
\eqn{\label{FFM}
	\langle \mF \rangle^{\rm r}_{\rm AC}
	&=&\nonumber
	\int \mF[\Gamma]\mM^{\rm r}_{\rm AC}[\mD\Gamma]\\
	&=&\nonumber
	\int \mF[\Gamma]
		\left.
			\frac{\mD\mM^{\rm r}_{\rm AC}}{\mD\mM}
		\right|_\Gamma
		\mM[\mD\Gamma]\\
	&=&\nonumber
	\int \mF[\Gamma] e^{-\sigma} \mM[\mD\Gamma]\\
	&=&
	\langle \mF e^{-\sigma} \rangle,
}
where we use the Crooks fluctuation theorem~(\ref{CFT2}) to obtain the third line.
Therefore, by Eqs.~(\ref{FFF}) and (\ref{FFM}), we obtain
\eqn{
	\langle \mF e^{-\sigma} \rangle
	=
	\langle \mF \rangle^{\rm r} - 
	\langle \mF \rangle^{\rm r}_{\rm S},
}
which can be regarded a generalization of the master fluctuation theorem (\ref{RMFT}).
If we set $\mF$ to unity, we obtain
\eqn{\label{NEAIP}
	\langle e^{-\sigma} \rangle
	=
	1-\lambda_{\rm S},
}
where
\eqn{
	\lambda_{\rm S}=\int\mM^{\rm r}_{\rm S}[\mD\Gamma]
}
is the probability of the singular part.
We note that $\lambda_{\rm S}$ is uniquely defined because of the uniqueness of Lebesgue's decomposition~(\ref{LD1}).
As explained in the previous section and summarized in Table~\ref{tab:irrev}, the absolute irreversibility has two classes.
Therefore, $\lambda_{\rm S}$ is calculated as the sum of two contributions of these classes of absolute irreversibility.
One is the probability of those reference paths whose corresponding original paths have vanishing probability.
The other contribution is the probability of localized reference paths.
To the best of our knowledge, the localized contribution had not been considered before our research \cite{MFU14}.
However, this part renders the conventional nonequilibrium equalities  inapplicable.
Namely, only when $\lambda_{\rm S}=0$, we reproduce
$
	\langle e^{-\sigma} \rangle = 1.
$

Using Jensen's inequality $\langle e^{-\sigma} \rangle\ge e^{-\langle \sigma \rangle}$, we obtain
\eqn{\label{2LAIP}
	\langle \sigma \rangle \ge -\ln(1-\lambda_{\rm S}).
}
Therefore, the second-law-like inequality $\langle \sigma \rangle\ge0$ is valid even with absolute irreversibility, because the right-hand side of Eq.~(\ref{2LAIP}) is nonnegative.
Moreover, if there exists absolute irreversibility $\lambda_{\rm S}>0$, then Eq.~(\ref{2LAIP}) imposes a stronger restriction on the entropy production than the conventional inequality because the right-hand side will then be strictly positive.
Thus, the average of the entropy production must be strictly positive in absolutely irreversible processes.

When the stronger version of Lebesgue's decomposition holds, by following the same procedure as before, we obtain
\eqn{
	\langle \mF e^{-\sigma} \rangle
	&=&
	\langle \mF \rangle^{\rm r}
	- \langle \mF \rangle^{\rm r}_{\rm sc}
	- \langle \mF \rangle^{\rm r}_{\rm d},\\
	\langle e^{-\sigma} \rangle
	&=&
	1-\lambda_{\rm sc}-\lambda_{\rm d},\label{NEAIP2}
}
where $\langle\cdots\rangle^{\rm r}_{\rm i}$ (i=sc, d) represents the average over $\mM^{\rm r}_{\rm i}$,
and $\lambda_{\rm sc}$ and $\lambda_{\rm d}$ are the probabilities of the singular continuous part and the discrete part, respectively.
We can calculate $\lambda_{\rm sc}$ as the sum of the probabilities of those reference paths whose corresponding counterparts in the original process vanish.
On the other hand, $\lambda_{\rm d}$ is calculated as the sum of localized reference paths.
From Eq.~(\ref{NEAIP2}), we obtain an inequality
\eqn{
	\langle \sigma \rangle \ge -\ln(1-\lambda_{\rm sc}-\lambda_{\rm d}),
}
which leads to a fundamental lower bound on the entropy production in absolutely irreversible processes.

\subsection{Physical implications}
We apply the general results in the previous section to processes starting from a restricted region.

Here, we consider time reversal as the reference dynamics.
In accordance with Table~\ref{tab}, in a Langevin or Hamiltonian system, Eq.~(\ref{CFT2}) reduces to
\eqn{\label{sqmm}
	e^{-\sigma} = e^{-\beta Q}
	\left.
	\frac{d\mu_{\rm L, AC}}{d\mu_0}
	\right|_{\Gamma_0}
	\left.
	\frac{d\mu^\dag_{0,\rm AC}}{d\mu_{\rm L}}
	\right|_{\Gamma_\tau},
}
where we rewrite the boundary term in terms of measure theory and $\Gamma$ represents the configuration coordinates in the case of an overdamped Langevin system and the phase space coordinates of the system and heat bath in the case of a Hamiltonian system. 
Here, $\mu_0$, $\mu_0^\dag$ and $\mu_{\rm L}$ represent the initial probability measure of the original process, the initial probability measure of the time-reversed process and the Lebesgue measure on the phase space $\Omega$, respectively, and $\mu_{\rm L, AC}$ is the absolutely continuous part of $\mu_{\rm L}$ with respect to $\mu_0$ and $\mu_{0,\rm AC}^\dag$ is the absolutely continuous part of $\mu_0^\dag$ with respect to $\mu_{\rm L}$.

\subsubsection{Generalization of the Jarzynski equality}
Now, let us assume that the initial state of the original process is a local equilibrium state which is restricted to a region $D_0\subset\Omega$ as
\eqn{
	\mu_0(d\Gamma_0) = e^{-\beta(H(\Gamma_0,\lambda_0)-F_0)}\chi_{D_0}(\Gamma_0)\mu_{\rm L}(d\Gamma_0),
}
where $\chi_{D_0}(\Gamma_0)$ is the characteristic function defined by
\eqn{
	\chi_{D_0}(\Gamma_0)=
	\left\{
	\begin{array}{ll}
		1	&	\Gamma_0\in D_0\\
		0	&	\Gamma_0\notin D_0,
	\end{array}
	\right.
}
and the free energy is defined by
\eqn{
	e^{-\beta F_0} = \int_{D_0} e^{-\beta H(\Gamma_0,\lambda_0)}\mu_{\rm L}(d\Gamma_0).
}
In this case, the absolutely continuous part of $\mu_{\rm L}$ with respect to $\mu_0$ is
\eqn{
	\mu_{\rm L,AC}(d\Gamma_0) = \chi_{D_0}(\Gamma_0)\mu_{\rm L}(d\Gamma_0).
}
Therefore, we obtain
\eqn{\label{RNIT}
	\left.
	\frac{d\mu_{\rm L, AC}}{d\mu_0}
	\right|_{\Gamma_0}
	=
	e^{\beta(H(\Gamma_0,\lambda_0)-F_0)}.
}
On the other hand, we set the initial probability measure of the time-reversed process to a local equilibrium distribution in a region $D_\tau$ as
\eqn{
	\mu^\dag_0(d\Gamma_\tau)=e^{-\beta(H(x_\tau,\lambda_\tau)-F_\tau)}\chi_{D_\tau}(\Gamma_\tau)\mu_{\rm L}(d\Gamma_\tau),
}
which is already absolutely continuous with respect to $\mu_{\rm L}$, namely, $\mu^\dag_{0,\rm AC}=\mu^\dag_0$.
Therefore, we obtain
\eqn{\label{RNFT}
	\left.
	\frac{d\mu^\dag_{0,\rm AC}}{d\mu_{\rm L}}
	\right|_{\Gamma_\tau}
	=
	e^{-\beta(H(x_\tau,\lambda_\tau)-F_\tau)}\chi_{D_\tau}(\Gamma_\tau).
}
Then, substituting Eqs.~(\ref{RNIT}) and (\ref{RNFT}) into Eq.~(\ref{sqmm}), we obtain
\eqn{
	e^{-\sigma} &=& e^{-\beta(Q+\Delta H-\Delta F)}\chi_{D_\tau}(\Gamma_\tau)\nonumber\\
	&=&
	e^{-\beta(W-\Delta F)}\chi_{D_\tau}(\Gamma_\tau),
}
where we use the first law of thermodynamics to obtain the last equality.
Thus, Eq.~(\ref{NEAIP}) reduces to
\eqn{
	\langle e^{-\beta(W-\Delta F)}\chi_{D_\tau}(\Gamma_\tau) \rangle
	=
	1-\lambda_{\rm S}.
}
In particular, if we set $D_\tau$ to $\Omega$, we obtain
\eqn{\label{GJar}
	\langle e^{-\beta(W-\Delta F)}\rangle
	=
	1-\lambda_{\rm S},
}
which is a generalization of the Jarzynski equality~(\ref{Jar}).

\subsubsection{Generalization of the Seifert relation}
Here, we assume that the initial probability distribution of the original process is absolutely continuous with respect to the Lebesgue measure, and therefore we have
\eqn{
	\mu_0(d\Gamma_0)=p_0(\Gamma_0)\mu_{\rm L}(d\Gamma_0).
}
Let $D_0$ denote the support of $p_0$.
Then, the absolutely continuous part of $\mu_{\rm L}$ with respect to $\mu_0$ is
\eqn{
	\mu_{\rm L, AC}(d\Gamma_0) = \chi_{D_0}(\Gamma_0)\mu_{\rm L}(d\Gamma_0).
}
Therefore, the Radon-Nikodym derivative is
\eqn{
	\left.
	\frac{d\mu_{\rm L, AC}}{d\mu_0}
	\right|_{\Gamma_0}
	=
	\frac{\chi_{D_0}(\Gamma_0)}{p_0(\Gamma_0)}.
}
We set the initial probability distribution of the reference process to be equal to the final probability distribution of the original process: $\mu^\dag_0=\mu_\tau$.
Then, the Radon-Nikodym derivative
\eqn{
	\left.
	\frac{d\mu^\dag_{0,\rm AC}}{d\mu_{\rm L}}
	\right|_{\Gamma_\tau}
	=
	\left.
	\frac{d\mu_{\tau,\rm AC}}{d\mu_{\rm L}}
	\right|_{\Gamma_\tau}
	=p_\tau(\Gamma_\tau)
}
represents the (unnormalized) final probability distribution of the original process.
Therefore, Eq.~(\ref{sqmm}) reduces to
\eqn{
	e^{-\sigma} &=& e^{-\beta Q} \frac{p_\tau(\Gamma_\tau)}{p_0(\Gamma_0)}\chi_{D_0}(\Gamma_0)\nonumber\\
	&=&
	e^{-\Delta s_{\rm bath}/\kb} e^{-\Delta s/\kb}\chi_{D_0}(\Gamma_0).
}
Thus, Eq.~(\ref{NEAIP}) reduces to
\eqn{\label{GGSR}
	\langle e^{-\Delta s_{\rm tot}} \rangle = 1-\lambda_{\rm S},
}
where we use $\chi_{D_0}(\Gamma_0)p_0(\Gamma_0)=p_0(\Gamma_0)$ because $D_0$ is the support of $p_0$.
This is a generalization of the Seifert relation~(\ref{SR}).
\\
\\
Up to here, we set the reference dynamics to the time-reversed dynamics, and have shown that our nonequilibrium equality~(\ref{NEAIP}) with absolute irreversibility derives generalized integral nonequilibrium equalities~(\ref{GJar}) and (\ref{GGSR}) involving the entropy production with the heat bath.
We note that our nonequilibrium equality~(\ref{NEAIP}) also derives generalized integral fluctuation theorems for the housekeeping and excess entropy production in a similar manner under the proper choices of the reference dynamics as summarized in Table~\ref{tab:summary}.

\begin{table}
\begin{center}
\caption{\label{tab:summary}
	Correspondence between reference probabilities, specific meanings of the entropy production and nonequilibrium equalities with absolute irreversibility.
}
{\normalsize\renewcommand\arraystretch{1.2}
\begin{tabular}{|c|c|c|c|}
\hline
Reference dynamics & Ref. ini. state & Entropy production & Nonequilibrium equality\\
\hline
time reversal & canonical & dissipated work &
$\displaystyle \langle e^{-\beta(W-\Delta F)}\rangle = 1-\lambda_{\rm S}$\\
\hline
time reversal & final state & total &
$\displaystyle \langle e^{-\Delta s_{\rm tot}/\kb}\rangle = 1-\lambda_{\rm S}$\\
\hline
time reversal & initial state & dissipation functional &
$\displaystyle \langle e^{-\Omega}\rangle = 1-\lambda_{\rm S}$\\
\hline
dual & initial state & housekeeping &
$\displaystyle \langle e^{-\Delta s_{\rm hk}/\kb}\rangle = 1-\lambda_{\rm S}$\\
\hline
time-reversed dual & steady state & excess &
$\displaystyle \langle e^{-\Delta \phi -\Delta s_{\rm ex}/\kb}\rangle = 1-\lambda_{\rm S}$\\
\hline
\end{tabular}}
\end{center}
\end{table}

\section{Examples of absolutely irreversible processes}
In this section, we verify our nonequilibrium equality in three examples: free expansion, an overdamped Langevin process starting from a local equilibrium, and an overdamped Langevin system with a trapping center.

\subsection{Free expansion}
First of all, we discuss the case of free expansion (see Fig.~\ref{fig:FE} (a)).
Initially, an ideal single-particle gas is confined in the left box with temperature $T$.
The entire box is assumed to be divided in the volume ratio $l:1-l$ by a partition.
Since the initial state can be described by the probability density, the stronger version of Lebesgue's decomposition holds.
Thus, Eq.~(\ref{GJar}) reduces to
\eqn{\label{FENE}
	\langle e^{-\beta(W-\Delta F)} \rangle = 1-\lambda_{\rm sc}-\lambda_{\rm d}.
}
We remove the partition, and the gas expands to the entire box.
In this process, work is not extracted: $W=0$,
whereas the free energy of the gas decreases due to the expansion by $\Delta F=\kb T\ln l(<0)$.
Therefore, the left-hand side of Eq.~(\ref{FENE}) is calculated to be
\eqn{
	\langle e^{-\beta(W-\Delta F)} \rangle = l.
}

We consider the time-reversed process of this free expansion to calculate the absolute irreversible probabilities (see Fig.~\ref{fig:FE} (b)).
In the time-reversed process, the system is in a global equilibrium at the initial time.
Then, the partition is inserted at the same position as the original process.
The gas particle is in either the left or the right box.
The paths ending in the right box have no corresponding forward paths in the original process.
Therefore, these paths are singular continuous paths.
The probability of these paths is proportional to the volume fraction of the right box, and therefore the singular continuous probability is
\eqn{
	\lambda_{\rm sc} = 1-l.
}
On the other hand, since there is no discrete path, which is a single path with a finite positive probability, we have
\eqn{
	\lambda_{\rm d} = 0.
}
Therefore, we obtain
\eqn{
	1-\lambda_{\rm sc}-\lambda_{\rm d} = l.
}
Thus, our nonequilibrium equality~(\ref{FENE}) is verified for the case of free expansion of an ideal single-gas particle.

\subsection{Process starting from a local equilibrium}
\begin{figure}
\begin{center}
\includegraphics[width=0.9\columnwidth]{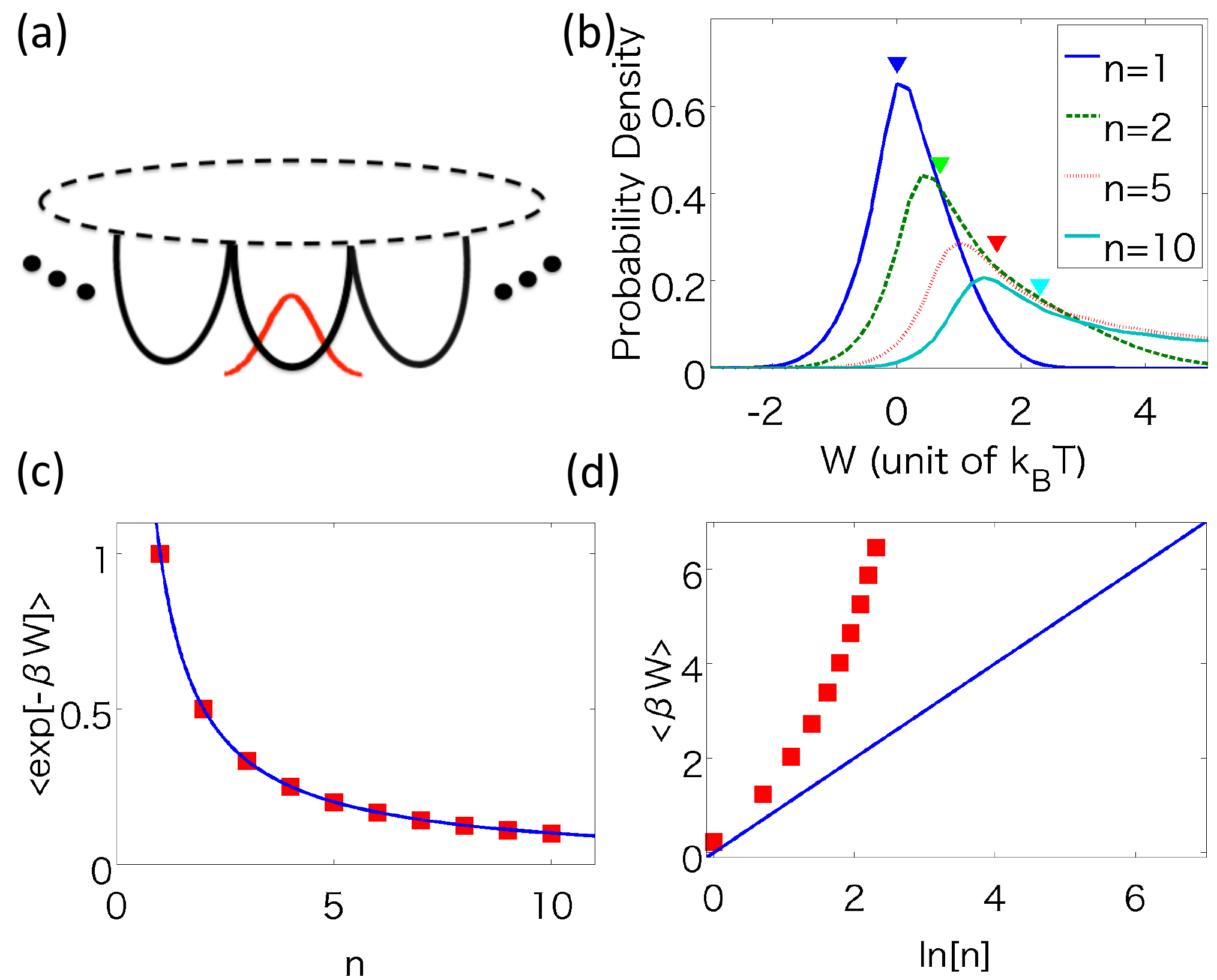}
\caption{\label{fig:SCsim}
(a) Schematic illustration of an overdamped system consisting of $n$ identical harmonic potentials confined in a one-dimensional ring.
In the original process, the system is in a local equilibrium in a given well at the initial time.
(b) Probability density of work performed on the system for several $n$ values.
The triangles indicate the points of $W=\kb T\ln n$.
(c) Values of $\langle e^{-\beta W} \rangle$ at each $n$.
Superimposed is the $1/n$ curve (with no free parameters).
(d) Values of $\langle \beta W \rangle$ at each $n$.
The line represents the minimum dissipation given by Eq.~(\ref{MDIneq}).
The parameters are chosen as follows:
diffusion constant $D=10^{-13}$m$^2$/s;
temperature $T=300$K;
duration of the process $\tau=10$sec;
half width of a single potential $a=10^{-6}$m;
the initial stiffness of the potential $K$ is chosen to satisfy $Ka^2/2=5\kb T$.
The statistical average is obtained from $10^6$ samples for each $n$.
Reproduced from Fig. 4 of Ref.~\cite{MFU14}. Copyright 2014 by the American Physical Society.
}
\end{center}
\end{figure}

Next, we consider an overdamped Langevin process starting from a local equilibrium state.
The Langevin particle is confined in a one-dimensional ring.
The potential consists of $n$ identical harmonic potential wells with the same stiffness (spring constant) $k(t)$ as illustrated in Fig.~\ref{fig:SCsim} (a).
Initially, the system is prepared in a local equilibrium state with temperature $T$ in a given well, and therefore the initial probability vanishes elsewhere.
We subject the system to a nonequilibrium process during a time interval $\tau$.
We decrease the stiffness of the potentials from $k=K$ to $0$ at a constant rate between $t=0$ and $\tau/2$, and then increase it from $k=0$ to $n^2K$ at a constant rate between $t=\tau/2$ and $\tau$.
Since the initial state can be written in terms of the probability density, the stronger version of Lebesgue's decomposition holds.
Therfore, Eq.~(\ref{GJar}) reduces to
\eqn{\label{SCNE}
	\langle e^{-\beta (W-\Delta F)} \rangle = 1-\lambda_{\rm sc}-\lambda_{\rm d}.
}

Now, we consider the time-reversed process to calculate the singular probabilities.
Initially, the system is in a global equilibrium at temperature $T$ with stiffness $k=n^2K$.
The stiffness is decreased from $k=n^2K$ to $0$ between $t=0$ and $\tau/2$, and then increased from $k=0$ to $K$ between $t=\tau/2$ and $\tau$.
Because a backward path terminates in a certain well with probability $1/n$ due to the symmetry of the potential and the initial state of the time-reversed process,
the probability that the backward path does not have the corresponding forward path is
\eqn{
	\lambda_{\rm sc}=\frac{n-1}{n}.
}
Moreover, since we have no discrete path, we obtain
\eqn{
	\lambda_{\rm d}=0.
}
Therefore, Eq.~(\ref{SCNE}) reduces to
\eqn{
	\langle e^{-\beta (W-\Delta F)} \rangle = \frac{1}{n}.
}
If we assume that $K$ is sufficiently large, $\Delta F=0$ in this process.
Thus, we obtain
\eqn{
	\langle e^{-\beta W} \rangle = \frac{1}{n}.
}
The corresponding inequality reads
\eqn{\label{MDIneq}
	\langle \beta W \rangle \ge \ln n.
}

We obtain the probability distributions of $W$ at different $n$ by numerical simulations as shown in Fig.~\ref{fig:SCsim} (b).
Based on this probability distribution, the value of $\langle e^{-\beta W} \rangle$ is obtained and confirmed to be $1/n$ (see Fig.~\ref{fig:SCsim} (c)).
We also verify that the average dissipation $\langle \beta W \rangle$ is larger than the minimum dissipation given by Eq.~(\ref{MDIneq}), namely, the fundamental lower bound due to the absolute irreversibility as demonstrated in Fig. \ref{fig:SCsim} (d).
We note that this process may be regarded as an information erasure of an $n$-digit memory.

\subsection{System with a trap}
\begin{figure}
\begin{center}
\includegraphics[width=0.9\columnwidth]{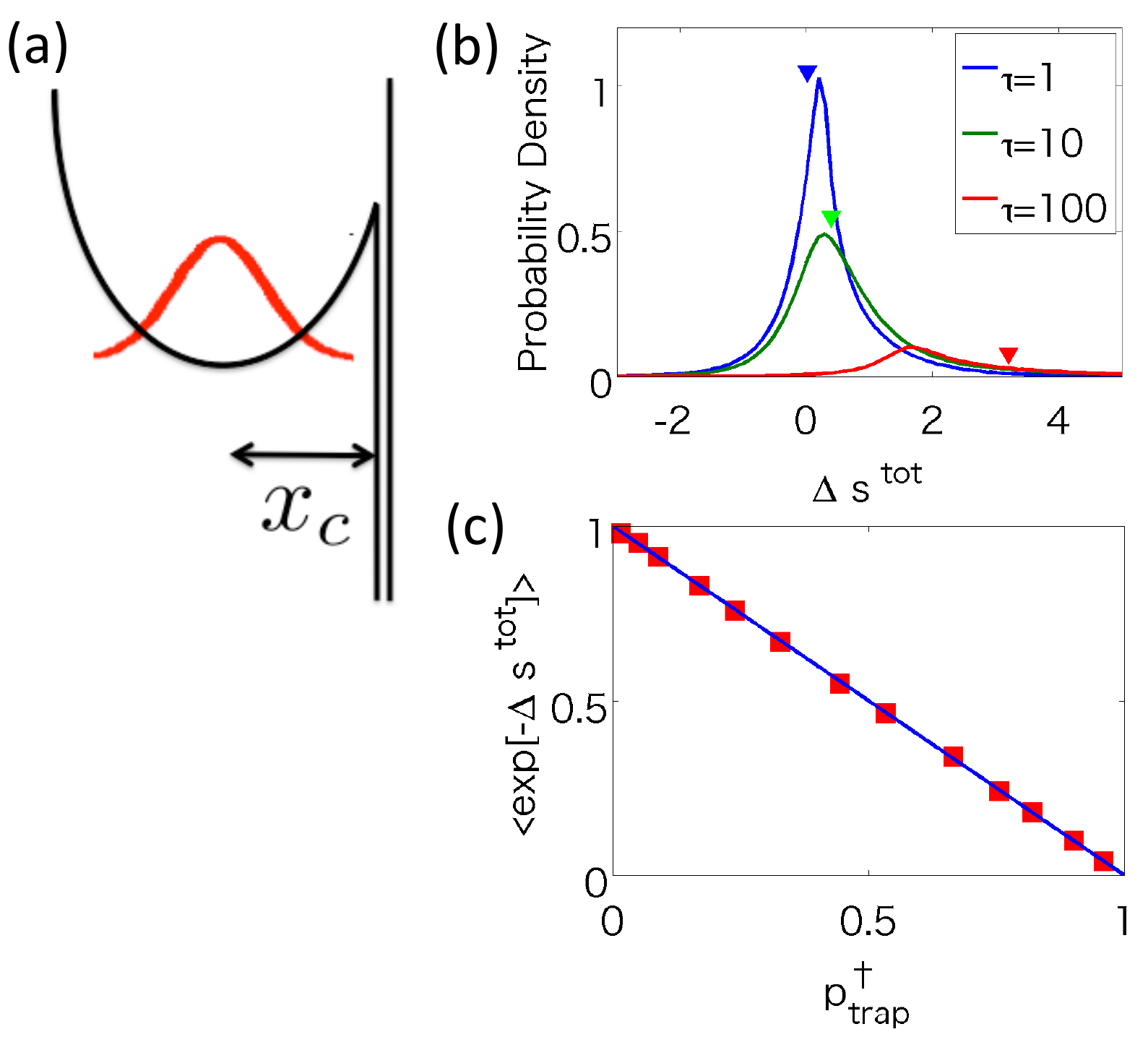}
\caption{\label{fig:Dsim}
(a) Schematic illustration of a one-dimensional overdamped system with a trapping center.
In the original process, the system is in an equilibrium of the harmonic potential.
(b) Probability density of the total entropy production.
Note that this probability density is not normalized over $\mathbb{R}$ because we have a positively divergent entropy production for the trapped paths.
The triangles indicate the points with $\Delta s_{\rm tot}=-\ln(1-p^\dag_{\rm trap})$.
(c) Values of $\langle e^{-\Delta s_{\rm tot}} \rangle$ versus the trapping probability $p^\dag_{\rm trap}$.
Superimposed is the $1-p^\dag_{\rm trap}$ line (with no free parameters).
The parameters $D$ and $T$ are the same as in Fig.~\ref{fig:SCsim}.
The distance between the center of the harmonic potential and the trapping point is $x_{\rm c}=10^{-6}$m.
The initial stiffness $K$ is set so as to satisfy $Ka^2/2=10\kb T$.
The duration of the process $\tau$ is varied between 1sec to 100sec to change the trapping probability.
The statistical average is taken over $10^6$ samples for each $\tau$.
Reproduced from Fig. 5 of Ref.~\cite{MFU14}. Copyright 2014 by the American Physical Society.
}
\end{center}
\end{figure}

Finally, we consider an overdamped Langevin system with a trap.
The system is one-dimensional, and the Langevin particle is confined in a single harmonic potential with stiffness $k(t)$.
We assume that there is a trapping point in the system and the distance between the center of the harmonic potential and the trapping point is denoted by $x_{\rm c}$ as illustrated in Fig.~\ref{fig:Dsim} (a).
If the particle reaches the trapping point, it is trapped with unit probability.
Initially, the system is prepared in an equilibrium of the harmonic potential.
We subject the system to a nonequilibrium process by changing stiffness $k$.
The stiffness is decreased from $k=K$ to $0$ at a constant rate between $t=0$ and $\tau/2$, and then increased from $k=0$ to $K$ at a constant rate between $t=\tau/2$ to $\tau$.
Since the initial probability can be written by the probability density, Eq.~(\ref{GGSR}) reduces to
\eqn{\label{DNE}
	\langle e^{-\Delta s_{\rm tot}} \rangle = 1-\lambda_{\rm sc}-\lambda_{\rm d}.
}

To evaluate the singular probabilities, we consider the time-reversed process.
The initial state of the time-reversed process is set to the final state of the original process.
Let $p_{\rm trap}$ and $p^\dag_{\rm trap}$ denote the trapping probabilities of the final state of the original and time-reversed processes, respectively.
Because the particle trapped in the original process will remain trapped in the entire time-reversed process, the probability of this single path is $p_{\rm trap}(>0)$.
Therefore, the discrete probability is 
\eqn{
	\lambda_{\rm d} = p_{\rm trap}.
}
Moreover, since the paths that fall into the trap in the time-reversed process have no corresponding counterparts in the original process, they are singular continuous.
Therefore, the singular continuous probability is
\eqn{
	\lambda_{\rm sc} = p^\dag_{\rm trap} -p_{\rm trap}.
}
Thus, Eq.~(\ref{DNE}) reduces to
\eqn{\label{DNE2}
	\langle e^{-\Delta s_{\rm tot}} \rangle = 1-p^\dag_{\rm trap},
}
which leads to an inequality
\eqn{
	\langle \Delta s_{\rm tot} \rangle\ge -\ln(1-p^\dag_{\rm trap}).
}
This inequality is automatically satisfied because the left-hand side is positively divergent due to the presence of those paths that fall into the trap in the original path with a positively divergent entropy production. 

Figure~\ref{fig:Dsim} (b) shows the probability density of the total entropy production.
Based on this probability distribution, the exponentiated average of the total entropy production is calculated and confirmed to be consistent with Eq.~(\ref{DNE2}) as demonstrated in Fig.~\ref{fig:Dsim} (c).

\section{Comparison with a conventional method}
In this section, we review a conventional method \cite{Sasa} to compare with our method based on absolute irreversibility.

\begin{figure}
\begin{center}
\includegraphics[width=0.2\columnwidth]{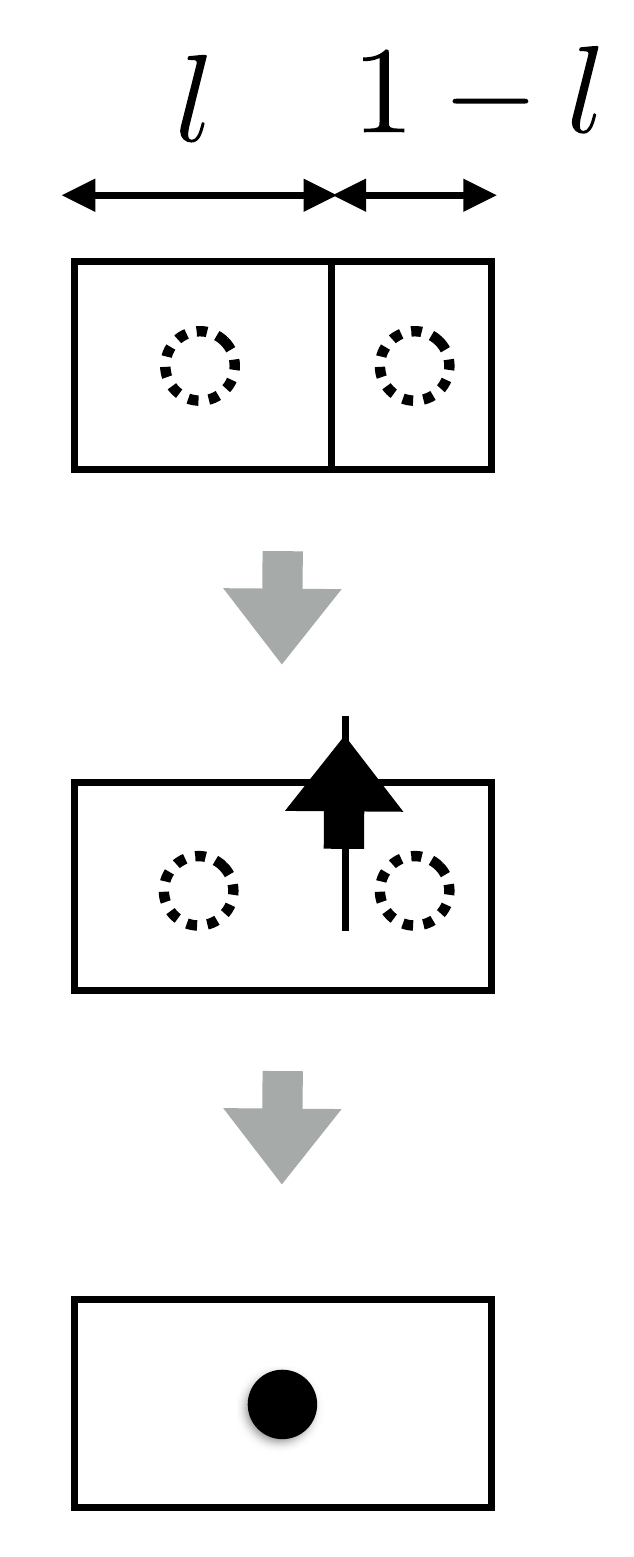}
\caption{\label{fig:VFE}
Virtual process corresponding to free expansion.
Initially, the single-particle gas is in a global equilibrium state; the particle is in the left box with probability $l$ and in the right box with probability $1-l$.
Then, the partition is removed.
Reproduced from Fig. 8 of Ref.~\cite{MFU14}. Copyright 2014 by the American Physical Society.
}
\end{center}
\end{figure}

As an illustration, we rederive our nonequilibrium equality in free expansion, namely, Eq.~(\ref{FENE}) by using the conventional method.
We consider a virtual process starting from a global equilibrium state with temperature $T$ as shown in Fig.~\ref{fig:VFE}.
Namely, the particle is in the left box with probability $l$ and in the right box with probability $1-l$.
Therefore, the probability of paths starting from the right box does not vanish, and absolutely irreversible paths do not exist.
Thus, the conventional nonequilibrium (\ref{MFT}) applies as
\eqn{\label{convne}
	\langle\langle \mF[\Gamma] e^{-\beta(\hat W-\Delta\hat F)} \rangle\rangle
	= \langle\langle \mF[\Gamma] \rangle\rangle^\dag,
}
where $\mF[\Gamma]$ is an arbitrary functional, and $\langle\langle\cdots\rangle\rangle$ denotes the statistical average of the virtual process, and the caret \^{} means that the accompanying quantity is the one in the virtual process.
In this simple case, we can easily obtain relations between the physical quantities in the virtual process and the corresponding quantities in the original process as
\eqn{
	\hat W &=& W,\label{VRWW}\\
	\hat F_0 &=& F_0 + \kb T\ln l,\\
	\hat F_\tau &=& F_\tau,\\
	\Delta \hat F &=& \Delta F -\kb T \ln l\label{VRFF}.
}
To derive a nonequilibrium equality in the original process, we set $\mF[\Gamma]$ to the characteristic functional whose value is equal to one only if the path starts from the left box and zero otherwise.
Then, considering the normalization of probabilities properly, we can express the average in the original process in terms of that in the virtual process as
\eqn{\label{VRAR}
	\langle\langle \mF[\Gamma]\cdots\rangle\rangle = l\langle\cdots\rangle.
}
In the time-reversed process, the probability of paths ending in the right box is $l$, and therefore we obtain
\eqn{
	\langle\langle\mF[\Gamma]\rangle\rangle^\dag = l.
}
Therefore, Eq.~(\ref{convne}) reduces to
\eqn{
	\langle e^{-\beta(W-\Delta F)} \rangle = l,
}
which agrees with Eq.~(\ref{FENE}) derived from the nonequilibrium equalities with absolute irreversibility.
Following a similar procedure, we can also rederive Eqs.~(\ref{SCNE}) and (\ref{DNE}) in principle.

Let us examine the meaning of this conventional procedure.
First, we extend the initial probability distribution to the global canonical distribution of the system to circumvent the problem of the vanishing probability.
Then, we derive the nonequilibrium equality (\ref{convne}) in this artificial process.
Next, we express physical quantities of the artificial process by those of the original process as in Eqs.~(\ref{VRWW}) and (\ref{VRFF}).
Finally, we erase some paths irrelevant to the original dynamics by setting $\mF$ to the characteristic function, and relate the average of the virtual process to that of the original process as in Eq.~(\ref{VRAR}).

In this way, the conventional method must introduce the artificial process to derive the nonequilibrium equality to avoid the problem arising from absolute irreversibility.
In contrast, our method directly deals with the original process.
Moreover, the reason why the right-hand sides of Eqs.~(\ref{FENE}), (\ref{SCNE}), and (\ref{DNE}) deviate from one is clearer in our method, that is, the probability of absolutely irreversible paths should be subtracted from one.
Incidentally, we note that although Eq.~(\ref{DNE}) can in hindsight be derived by the conventional method, we naturally find this absolutely irreversible example by virtue of the stronger version of Lebesgue's decomposition~(\ref{LD2}).

\chapter{Information-Thermodynamic Nonequilibrium Equalities in Absolutely Irreversible Processes}
In this chapter, we generalize the nonequilibrium equalities in the presence of absolute irreversibility obtained in Chap. 4 to situations under measurements and feedback control.
This generalization is of vital importance when error-free measurements are preformed, because they project the probability distribution of the measured state onto a localized region in the phase space.
The subsequent time evolution of this confined post-measurement state, in general, involves expansion into an initially unoccupied region, which provides yet another example of absolute irreversibility.
The generalized nonequilibrium equalities under measurements and feedback control enable us to identify the unavailable information, which results from the inevitable loss of information that we cannot utilize under a fixed feedback protocol.
The unavailable information provides a fundamental limit of performance of the feedback protocol.

First, we derive information-thermodynamic nonequilibrium equalities in the presence of absolute irreversibility.
Next, we introduce a notion of unavailable information and derive a different type of nonequilibrium equalities that involve the unavailable information.
Finally, we demonstrate our nonequilibrium equalities in several examples.

This chapter is partly based on Refs.~\cite{MFU14, AFMU14}.

\section{Inforamtion-thermodynamic equalities}
In this section, we derive information-thermodynamic nonequilibrium equalities in the presence of absolute irreversibility.
This section is mainly based on Ref.~\cite{MFU14}.

Under feedback control, it is important to consider the effect of the singular part of the reference probability measure discussed in Sec. 4.1.2 because high-precision measurements such as error-free measurements localize the probability distribution.
Since the feedback control starts from this post-measurement state confined in a narrow region, the subsequent time evolution involves expansion into an initially unoccupied region unless the feedback protocol is fine-tuned, and therefore the process exhibits absolute irreversibility.
In experiments, since we have access to only a few parameters, the fine-tuning is a difficult task in general.
Hence, absolute irreversibility has experimental relevance in a system under measurements and feedback control.

We consider a nonequilibrium process in a classical system with measurements and  feedback control from time $t=t_{\rm i}$ to $t_{\rm f}$ as in Sec. 3.3.2.
Let $x(t)$ and $y(t)$ denote phase-space points at time $t$ of the system and the measurement outcomes, respectively.
The external control parameter $\lambda(t)$ is determined based on the measurement outcome before $t$, namely, $\{y(s)\}_{s=t_{\rm i}}^t$.
Let $X$, $Y$ and $\Lambda[Y]$ denote the entire paths of the system, the measurement outcomes and the control parameter, respectively.
Moreover, let $\mM_{|Y}$ denote the conditional probability measure of $X$ under given measurement outcomes $Y$, and let $\mM^{\rm r}_{|\Lambda[Y]}$ denote the reference probability measure of $X$ under a given feedback protocol $\Lambda[Y]$.
We apply Lebesgue's decomposition theorem to $\mM^{\rm r}_{|\Lambda[Y]}$ with respect to $\mM_{|Y}$, and obtain
\eqn{\label{LDCY}
	\mM^{\rm r}_{|\Lambda[Y]} = \mM^{\rm r}_{{\rm AC}|\Lambda[Y]} +  \mM^{\rm r}_{{\rm S}|\Lambda[Y]},
}
where the first and second terms on the right-hand side give the absolutely continuous and singular parts of the reference probability measure, respectively.
Lebesgue's decomposition theorem ensures that this decomposition is unique.
Let $\mF[X,Y]$ denote an arbitrary path functional.
It follows from Eq.~(\ref{LDCY}) that
\eqn{
	\langle\mF\rangle^{\rm r}_{|\Lambda[Y]}
	=
	\langle\mF\rangle^{\rm r}_{{\rm AC}|\Lambda[Y]}
	+
	\langle\mF\rangle^{\rm r}_{{\rm S}|\Lambda[Y]},
}
where $\langle\cdots\rangle^{\rm r}_{{\rm I}|\Lambda[Y]}\ ({\rm I}=\emptyset,{\rm AC},{\rm S)}$ denotes the average over $\mM^{\rm r}_{{\rm I}|\Lambda[Y]}$.
Moreover, the average over the absolutely continuous part can be transformed by the Radon-Nikod\'ym derivative because
\eqn{
	\langle\mF\rangle^{\rm r}_{{\rm AC}|\Lambda[Y]}
	&=&\nonumber
	\int \mF[X,Y] \mM^{\rm r}_{{\rm AC}|\Lambda[Y]}[\mD X]\\
	&=&\nonumber
	\int \mF[X,Y]
	\left. \frac{\mD\mM^{\rm r}_{{\rm AC}|\Lambda[Y]}}{\mD\mM_{|Y}} \right|_X
	{\mM_{|Y}}[\mD X]\\
	&=&
	\langle \mF e^{-R} \rangle_{|Y},
}
where $\langle\cdots\rangle_{|Y}$ denotes the average over $\mM_{|Y}$, and we define
\eqn{\label{RRN}
	e^{-R} = \left. \frac{\mD\mM^{\rm r}_{{\rm AC}|\Lambda[Y]}}{\mD\mM_{|Y}} \right|_X .
}
Therefore, we obtain
\eqn{\label{UAMRE}
	\langle \mF e^{-R} \rangle_{|Y}
	=
	\langle\mF\rangle^{\rm r}_{|\Lambda[Y]}
	-
	\langle\mF\rangle^{\rm r}_{{\rm S}|\Lambda[Y]}.
}
If we set $\mF$ to unity, we obtain
\eqn{\label{UARE}
	\langle e^{-R} \rangle_{|Y}
	=
	1-\lambda_{{\rm S}|\Lambda[Y]},
}
where
\eqn{
	\lambda_{{\rm S}|\Lambda[Y]} = \int \mM^{\rm r}_{{\rm S}|\Lambda[Y]}[\mD X]
}
is the probability of the singular part of the reference probability measure conditioned by $\Lambda[Y]$.

Here, we consider the physical meaning of $R$ defined in Eq.~(\ref{RRN})\footnote{
Here, we assume that $\mM^{\rm r}_{{\rm AC}|\Lambda[Y]}$ is absolutely continuous with respect to $\mM^{\rm tr}_{|\Lambda[Y]}$ and that $\mM^{\rm tr}_{|\Lambda[Y]}$ is absolutely continuous with respect to $\mM_{|Y}$.
Otherwise, we cannot define the entropy production and mutual information individually, because their values are divergent and mathematically ill-defined.
However, the sum of these quantities remains finite due to the absolute continuity of $\mM^{\rm r}_{{\rm AC}|\Lambda[Y]}$ with respect to  $\mM_{|Y}$.
}.
The Crooks fluctuation theorem~(\ref{CFT2}) applies to the ratio of the transition probabilities under a fixed protocol:
\eqn{\label{FNref}
	e^{-\sigma} = \left.\frac{\mD\mM^{\rm r}_{{\rm AC}|\Lambda[Y]}}{\mD\mM^{\rm tr}_{|\Lambda[Y]}}\right|_X,
}
where $\mM^{\rm tr}_{|\Lambda[Y]}$ is the transition probability under the fixed protocol $\Lambda[Y]$.
Therefore, $R$ is different from $\sigma$ because the conditional probability is different form the transition probability as discussed in Sec. 3.2.2.
The difference is represented by the Radon-Nikodym derivative as
\eqn{
	e^{-R} = e^{-\sigma}
	\left. \frac{\mD\mM^{\rm tr}_{|\Lambda[Y]}}{\mD\mM_{|Y}}\right|_X.
}
Comparing this with Eq.~(\ref{trcni}), we find that the Radon-Nikodym derivative in this equation may be written as $e^{-I-I_{\rm i}}$, where $I$ is the total mutual information obtained by the measurements and $I_{\rm i}$ is the initial correlation between the system and the measurement outcomes.
Therefore, we have
\eqn{
	R=\sigma +I+I_{\rm i},
}
and Eqs.~(\ref{UAMRE}) and (\ref{UARE}) reduce to
\eqn{
	\langle \mF e^{-\sigma-I-I_{\rm i}} \rangle_{|Y}
	&=&
	\langle\mF\rangle^{\rm r}_{|\Lambda[Y]}
	-
	\langle\mF\rangle^{\rm r}_{{\rm S}|\Lambda[Y]},\\
	\langle e^{-\sigma-I-I_{\rm i}} \rangle_{|Y}
	&=&
	1-\lambda_{{\rm S}|\Lambda[Y]},\label{UAITNE}
}
respectively.
Averaging these over measurement outcomes $Y$, we obtain
\eqn{
	\langle \mF e^{-\sigma-I-I_{\rm i}} \rangle
	&=&
	\langle\mF\rangle^{\rm r}
	-
	\langle\mF\rangle^{\rm r}_{\rm S},\\
	\langle e^{-\sigma-I-I_{\rm i}} \rangle
	&=&
	1-\overline \lambda_{\rm S},\label{ITNE}
}
where $\overline\lambda_{\rm S}$ is the average over $Y$ of the $\Lambda[Y]$-conditioned singular probability $\lambda_{{\rm S}|\Lambda[Y]}$.
Equation~(\ref{ITNE}) derives a second-law-like inequality as
\eqn{\label{2LITIC}
	\langle \sigma \rangle \ge -\langle I+I_{\rm i} \rangle-\ln(1-\overline\lambda_{\rm S}).
}
Therefore, the lower bound of the entropy production is determined not only by the mutual information but also by the term arising from the absolute irreversibility.
If the feedback protocol is so poorly designed that $-\ln(1-\overline\lambda_{\rm S})>\langle I+I_{\rm i} \rangle$, Eq.~(\ref{2LITIC}) implies that the entropy production is positive and the feedback protocol does not work.
In the case of vanishing initial correlations, we obtain
\eqn{
	\langle \mF e^{-\sigma-I} \rangle
	&=&
	\langle\mF\rangle^{\rm r}
	-
	\langle\mF\rangle^{\rm r}_{\rm S},\\
	\langle e^{-\sigma-I} \rangle
	&=&
	1-\overline \lambda_{\rm S},\\
	\langle \sigma \rangle &\ge& -\langle I \rangle-\ln(1-\overline\lambda_{\rm S}).\label{2LITAI}
}
Moreover, if the conditioned initial states $\mM_{|Y}$ satisfy the assumption of the stronger version of Lebesgue's decomposition~(\ref{LD2}), we obtain
\eqn{
	\langle \mF e^{-\sigma-I} \rangle
	&=&
	\langle\mF\rangle^{\rm r}
	-
	\langle\mF\rangle^{\rm r}_{\rm sc}
	-
	\langle\mF\rangle^{\rm r}_{\rm d},\\
	\langle e^{-\sigma-I} \rangle
	&=&
	1-\overline \lambda_{\rm sc}-\overline \lambda_{\rm d},\\
	\langle \sigma \rangle &\ge& -\langle I \rangle-\ln(1-\overline\lambda_{\rm sc}-\overline\lambda_{\rm d}),
}
where the subscripts ``sc'' and ``d'' represent the singular continuous and discrete parts, respectively.

In the same manner as the case without feedback control, proper choices of the reference probability lead to nonequilibrium equalities with specific meanings of the entropy production as listed in Table~\ref{tab:summary}, and we obtain the corresponding equalities
\eqn{
	\langle e^{-\beta(W-\Delta F) -I} \rangle  &=& 1-\overline\lambda_{\rm S},\\
	\langle e^{-\Delta s_{\rm tot}/\kb -I} \rangle  &=& 1-\overline\lambda_{\rm S},\\
	\langle e^{-\Omega -I} \rangle  &=& 1-\overline\lambda_{\rm S},\\
	\langle e^{-\Delta s_{\rm hk}/\kb -I} \rangle  &=& 1-\overline\lambda_{\rm S},\\
	\langle e^{-\Delta \phi -\Delta s_{\rm ex}/\kb -I} \rangle  &=& 1-\overline\lambda_{\rm S},
}
under the proper assumptions described in Sec. 2.2.

\section{Unavailable information and associated equalities}
In this section, we define a concept of unavailable information based on the information-thermodynamic nonequilibrium equalities obtained in the previous section, and derive new nonequilibrium equalities that involve the unavailable information.
Inequalities derived from the new equalities give an achievable lower bound of the entropy production in the case of error-free measurements.

First of all, we point out that the equality of Eq.~(\ref{2LITAI}) cannot be achieved in general even in the quasi-static limit.
The equality condition of Jensen's inequality $\langle e^{-x} \rangle\ge e^{-\langle x\rangle}$ is that the quantity $x$ does not fluctuate.
On the other hand, in the quasi-static limit, $\sigma$ is expected to have a single definite value.
Therefore, in situations without feedback control, the equality in the inequality
\eqn{
	\langle \sigma \rangle \ge -\ln(1-\lambda_{\rm S})
}
is achieved.
In contrast, the equality in the inequality under feedback control
\eqn{\label{2LITAI2}
	\langle \sigma \rangle \ge -\langle I \rangle -\ln(1-\overline\lambda_{\rm S})
}
cannot be achieved in general even in the quasi-static limit due to fluctuations of $I$, that is, $I$  depends on the measurement outcomes unless all outcomes are obtained with the same probability.
Hence, Eq.~(\ref{2LITAI2}) gives only a loose lower bound of the entropy production, although it is tighter than the conventional inequality ($\langle\sigma\rangle\ge-\langle I\rangle$) \cite{SU08}.

To find an achievable lower bound of the entropy production, we start from Eq.~(\ref{UAITNE}).
If we assume that we have no initial correlations, Eq.~(\ref{UAITNE}) reduces to
\eqn{\label{UAITNE2}
	\langle e^{-\sigma - I } \rangle_{|Y} = 1-\lambda_{{\rm S}|\Lambda[Y]}.
}
From Jensen's inequality, we obtain
\eqn{\label{AIneq}
	\langle \sigma \rangle_{|Y} \ge -\langle I \rangle_{|Y} -\ln(1-\lambda_{{\rm S}|\Lambda[Y]}).
}
What is noteworthy on this inequality is that the equality is achievable under error-free measurements in the quasi-static limit, because $I$ reduces to the unaveraged Shannon entropy of $Y$, which has a single definite value under fixed measurement outcomes $Y$.
Therefore, the right-hand side of Eq.~(\ref{AIneq}) gives an achievable lower bound of the entropy production for a fixed $Y$.
Hence, the last term in Eq.~(\ref{AIneq}) represents the inevitable dissipation due to the incompleteness of the feedback protocol $\Lambda[Y]$.
In other words, the feedback  protocol $\Lambda[Y]$ cannot fully utilize the mutual information obtained by the measurements.
Thus, we define unavailable information in the protocol $\Lambda[Y]$ as
\eqn{\label{Iudef}
	I_{{\rm u}|\Lambda[Y]} = - \ln (1-\lambda_{{\rm S}|\Lambda[Y]})\ (\ge0).
}
We note that, under error-free measurements and in the quasi-static limit. we have
\eqn{\label{efqseq}
	\sigma = -I+I_{\rm u}\ {\rm (error\mathchar"712D free, quasi\mathchar"712D static)}.
}
Using Eq.~(\ref{Iudef}), we rewrite Eq.~(\ref{UAITNE2}) as
\eqn{
	\langle e^{-\sigma-I+I_{\rm u}} \rangle_{|Y} = 1.
}
Averaging this equality over $Y$, we obtain a new nonequilibrium equality:
\eqn{\label{IuNE}
	\langle e^{-\sigma-I+I_{\rm u}}\rangle = 1.
}
This equality leads to
\eqn{\label{IuIneq}
	\langle \sigma \rangle \ge -\langle I -I_{\rm u}\rangle.
}
We note that this inequality is stronger than Eq.~(\ref{2LITAI2}) due to the convexity of the logarithmic function $-\ln x$, that is,
\eqn{
	\langle I_{{\rm u}} \rangle = \langle -\ln(1- \lambda_{{\rm S}|\Lambda[Y]}) \rangle \ge -\ln(1-\overline\lambda_{\rm S}).
}
Furthermore, Eq.~(\ref{IuIneq}) gives an achievable lower bound of the  entropy production under error-free measurements and in the quasi-static limit because of Eq.~(\ref{efqseq}).
Therefore, we can quantitatively characterize the incompleteness of the feedback protocol by calculating the unavailable information.

Here, we assume that the initial state is an equilibrium state, and set the reference probability to the time-reversed one starting from an equilibrium state.
Then, Eqs.~(\ref{IuNE}) and (\ref{IuIneq}) reduce to Jarzynski-type equations as
\eqn{
	\langle e^{-\beta(W-\Delta F)-I+I_{\rm u}} \rangle&=& 1,\\
	-\langle W \rangle& \le& -\Delta F + \kb T \langle I -I_{\rm u} \rangle,\label{IuWIneq}
}
respectively.
Equation~(\ref{IuWIneq}) gives an achievable upper bound of extractable work in the feedback process.
The bound is reduced due to the unavailable information compared with the conventional result \cite{SU08}.

The other choices of the reference probability summarized in Table~\ref{tab} also give nonequilibrium equalities with their individual meanings.

\section{Examples of absolutely irreversible processes}
In this section, we verify the information-thermodynamic nonequilibrium equalities derived in the previous sections.
Here, we only consider relations with dissipated work, namely,
\eqn{
	\label{ourJar}
	\langle e^{-\beta(W-\Delta F)-I}\rangle &=& 1-\overline\lambda_{\rm S},\\
	\label{ourJarIneq}
	\langle W\rangle-\Delta F& \ge& -\kb T \langle I\rangle-\kb T\ln(1-\overline\lambda_{\rm S}),\\
	\label{AsJar}
	\langle e^{-\beta(W-\Delta F)-I+I_{\rm u}} \rangle &=& 1,\\
	\label{AsJarIneq}
	\langle W\rangle-\Delta F &\ge& -\kb T\langle I-I_{\rm u} \rangle.
}

First, we consider a measurement and subsequent trivial feedback control.
Second, we discuss the two-particle Szilard engine.
Finally, the multi-particle Szilard engine is considered.

\subsection{Measurement and trivial feedback control}
\begin{figure}
\begin{center}
\includegraphics[width=0.8\columnwidth]{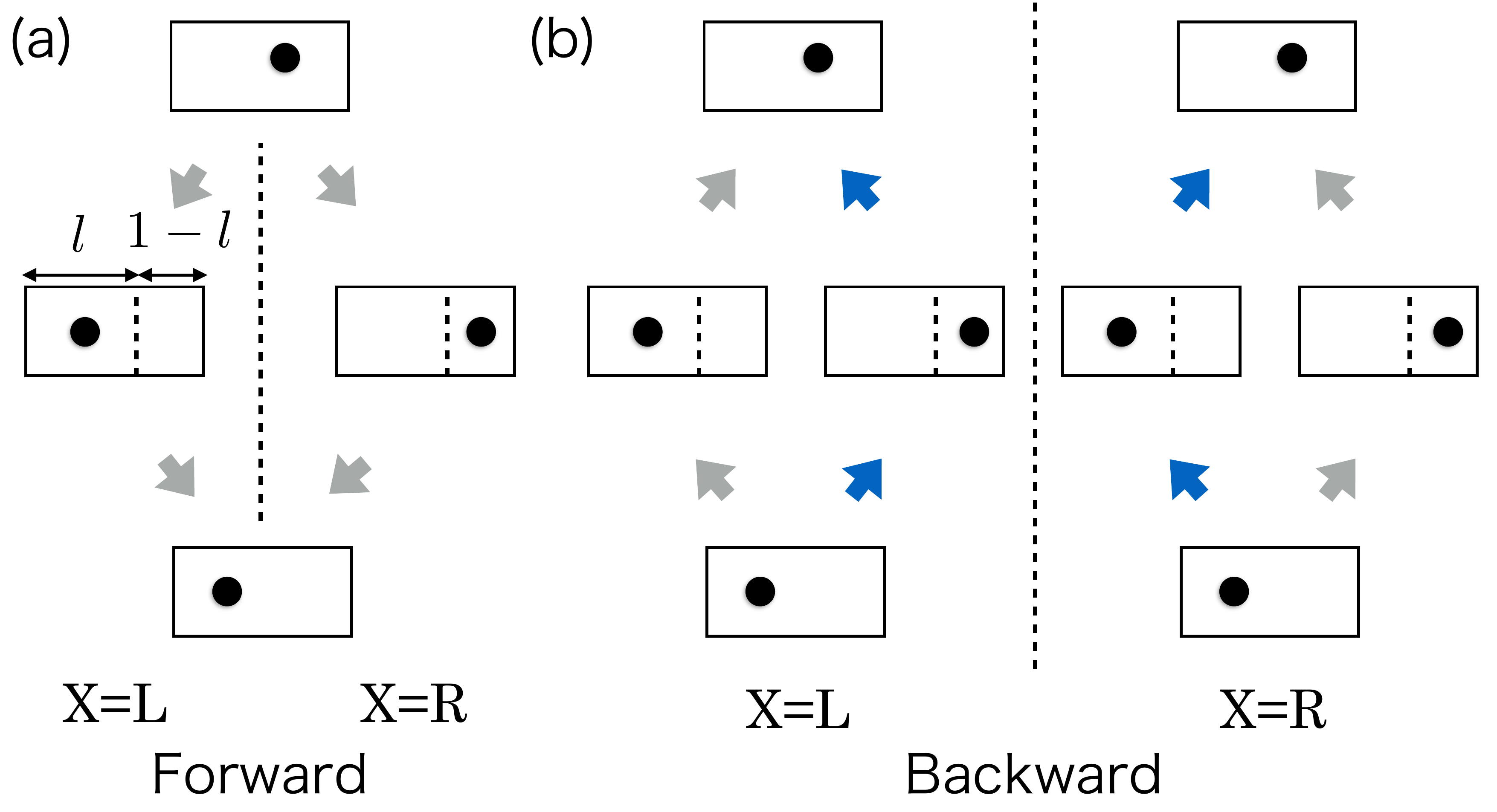}
\caption{\label{fig:MwoFC}
(a) A measurement and the subsequent trivial feedback control.
Initially, an ideal single-particle gas is in a global equilibrium state of the entire box.
At a certain time, we perform an error-free position measurement to determine whether the particle is in the left or right side.
After the measurement, we do nothing as the feedback control.
(b) The time-reversed protocol.
Initially, the particle is in the global equilibrium of the entire box and we do nothing regardless of the outcome of the forward process.
At the time of the measurement of the forward process, the particle is stochastically in the left or right side.
For the outcome $X={\rm L}\ ({\rm R})$, the case in which the particle is in the right (left) side is singular (blue arrows). 
Reproduced from Fig. 6 of Ref.~\cite{MFU14}. Copyright 2014 by the American Physical Society.
}
\end{center}
\end{figure}

We consider a measurement and the subsequent feedback control.
Initially, an ideal single-particle gas is in a global equilibrium state of the box as illustrated in Fig.~\ref{fig:MwoFC} (a).
At a certain time, we perform an instantaneous error-free measurement to determine the position $X$ of the particle.
An outcome $X={\rm L}$ is obtained when the particle is in the left side, that is, the length from the left-end wall is shorter than the length of the whole box multiplied by $l$ ($0<l<1$).
On the other hand, an outcome $X={\rm R}$ is obtained when the particle is in the right side.
In both cases, we conduct trivial feedback control, i.e., we leave the system as it is.
Therefore, the gas expands to the entire box.

In this process, work is not extracted: $W=0$, and the free energy does not change: $\Delta F=0$.
Let $p({\rm L})$ and $p({\rm R})$ denote the probabilities to obtain outcomes $\rm L$ and $\rm R$, respectively. 
If the measurement outcome is $X={\rm L}$, we obtain mutual information $I=-\ln p({\rm L})=-\ln l$.
If $X={\rm R}$, we obtain mutual information $I=-\ln p({\rm R})=-\ln(1-l)$.

Next, we consider the time-reversed process to find the singular probabilities.
In the time-reversed process, we do nothing for both $X={\rm L}$ and $\rm R$, because we do nothing in the forward process except for the measurement.
The time-reversed process corresponding to the outcome $X={\rm L}$ is illustrated in the left half of Fig.~\ref{fig:MwoFC} (b).
The particle is in the left side with probability $l$ and in the right side with probability $1-l$.
The case of the particle being found in the right side has no corresponding forward event under $X={\rm L}$, and therefore it is a singular event.
Therefore, we obtain
\eqn{
	\lambda_{{\rm S}|\Lambda[{\rm L}]}=1-l.
}
On the other hand, in the time-reversed process corresponding to the outcome $X={\rm R}$ illustrated in the right half of Fig.~\ref{fig:MwoFC} (b), we obtain
\eqn{
	\lambda_{{\rm S}|\Lambda[{\rm R}]}=l.
}
Therefore, the unavailable information is obtained as
\eqn{
	I_{{\rm u}|\Lambda[{\rm L}]} &=& -\ln (1-\lambda_{{\rm S}|\Lambda[{\rm L}]}) = -\ln l,\\
	I_{{\rm u}|\Lambda[{\rm R}]} &=& -\ln (1-\lambda_{{\rm S}|\Lambda[{\rm R}]}) = -\ln (1-l).
}
We note that the unavailable information coincides with the mutual information in this case, because all the information is lost since we do nothing as feedback control.
The values of the physical quantities are summarized in Table~\ref{tab:MwoFC}.

\begin{table}[b]
\begin{center}
\caption{\label{tab:MwoFC}
	Values of physical quantities corresponding to the outcomes L and R.
}
{\renewcommand\arraystretch{1.2}
\begin{tabular}{|c|c|c|c|c|c|c|}
\hline
$X$ & $p(X)$ & $W$ & $\Delta F$ & $I=-\ln p({\rm X})$ &$\lambda_{{\rm S}|\Lambda[X]}$ & $I_{{\rm u}|\Lambda[X]}=-\ln(1-\lambda_{{\rm S}|\Lambda[X]})$\\
\hline
L & $l$ & 0 & 0 & $-\ln l$ & $1-l$ & $-\ln l$\\
\hline
R & $1-l$ & 0 & 0 & $-\ln(1-l)$ & $l$ & $-\ln (1-l)$\\
\hline
\end{tabular}
}
\end{center}
\end{table}

The left-hand side of Eq.~(\ref{ourJar}) is calculated as
\eqn{
	\langle e^{-\beta(W-\Delta F)-I} \rangle
	&=&\nonumber
	p({\rm L})\cdot e^{\ln l} + p({\rm R})\cdot e^{\ln (1-l)}\\
	&=&
	l^2+(1-l)^2.
}
On the other hand, the averaged singular probability is
\eqn{
	\overline \lambda_{\rm S}
	&=& p({\rm L})\lambda_{{\rm S}|\Lambda[{\rm L}]}
	+ p({\rm R})\lambda_{{\rm S}|\Lambda[{\rm R}]}
	\nonumber\\
	&=& 2l(1-l).
}
Therefore, by a simple calculation, we can verify Eq.~(\ref{ourJar}).
Next, let us verify Eq.~(\ref{ourJarIneq}).
The average of the dissipated work is
\eqn{
	\langle W \rangle -\Delta F = 0.
}
The right-hand side of Eq.~(\ref{ourJarIneq}) is
\eqn{\label{analy}
	-\kb T \langle I \rangle -\kb T \ln (1-\overline\lambda_{\rm S})
	&=&
	\kb T [ l\ln l + (1-l)\ln(1-l) - \ln (l^2+(1-l)^2)].\ \ \ 
}
By analytic calculation, we can show that the right-hand side of Eq.~(\ref{analy}) is nonpositive, that is,
\eqn{
	\kb T [ l\ln l + (1-l)\ln(1-l) - \ln (l^2+(1-l)^2)] \le 0,
}
and it is zero if and only if $l=1/2$.
Thus, Eq.~(\ref{ourJarIneq}) is verified.
Moreover, the necessary and sufficient condition for the equality is $l=1/2$, which is consistent with the observation made in Sec. 5.2, namely, the equality in the inequality~(\ref{2LITAI2}) is achieved only when all outcomes are obtained with the equal probability.

Now, we verify Eqs.~(\ref{AsJar}) and (\ref{AsJarIneq}) in the presence of the unavailable information.
The left-hand side of Eq.~(\ref{AsJar}) is calculated as
\eqn{
	\langle e^{-\beta(W-\Delta F)-I+I_{\rm u}} \rangle
	=
	p({\rm L})\cdot e^{\ln l - \ln l}
	+
	p({\rm R})\cdot e^{\ln (1-l) - \ln (1-l)}
	= 1,
}
which verifies Eq.~(\ref{AsJar}).
Moreover, because $I_{\rm u}=I$ in this case, the right-hand side of Eq.~(\ref{AsJarIneq}) is
\eqn{
	-\kb T \langle I-I_{\rm u} \rangle = 0.
}
Therefore, we obtain
\eqn{
	W-\Delta F = -\kb T \langle I-I_{\rm u} \rangle.
}
This equality means that the equality in the inequality~(\ref{AsJarIneq}) is achieved regardless of the value of $l$, which is what we expect because Eq.~(\ref{AsJarIneq}) gives an achievable bound.

\subsection{Two-particle Szilard engine}
Next, we consider the two-particle Szilard engine.
The reason why we do not consider the single-particle Szilard engine as in Refs.~\cite{SU10, HV10} is that the singular part does not arise in the single-particle Szilard engine since we can fully utilize the information obtained by the measurements.
Therefore, we should consider a Szilard engine with two or more particles to observe effects of the absolute irreversibility.
We begin by the two-particle Szilard engine.
\begin{figure}
\begin{center}
\includegraphics[width=0.7\columnwidth]{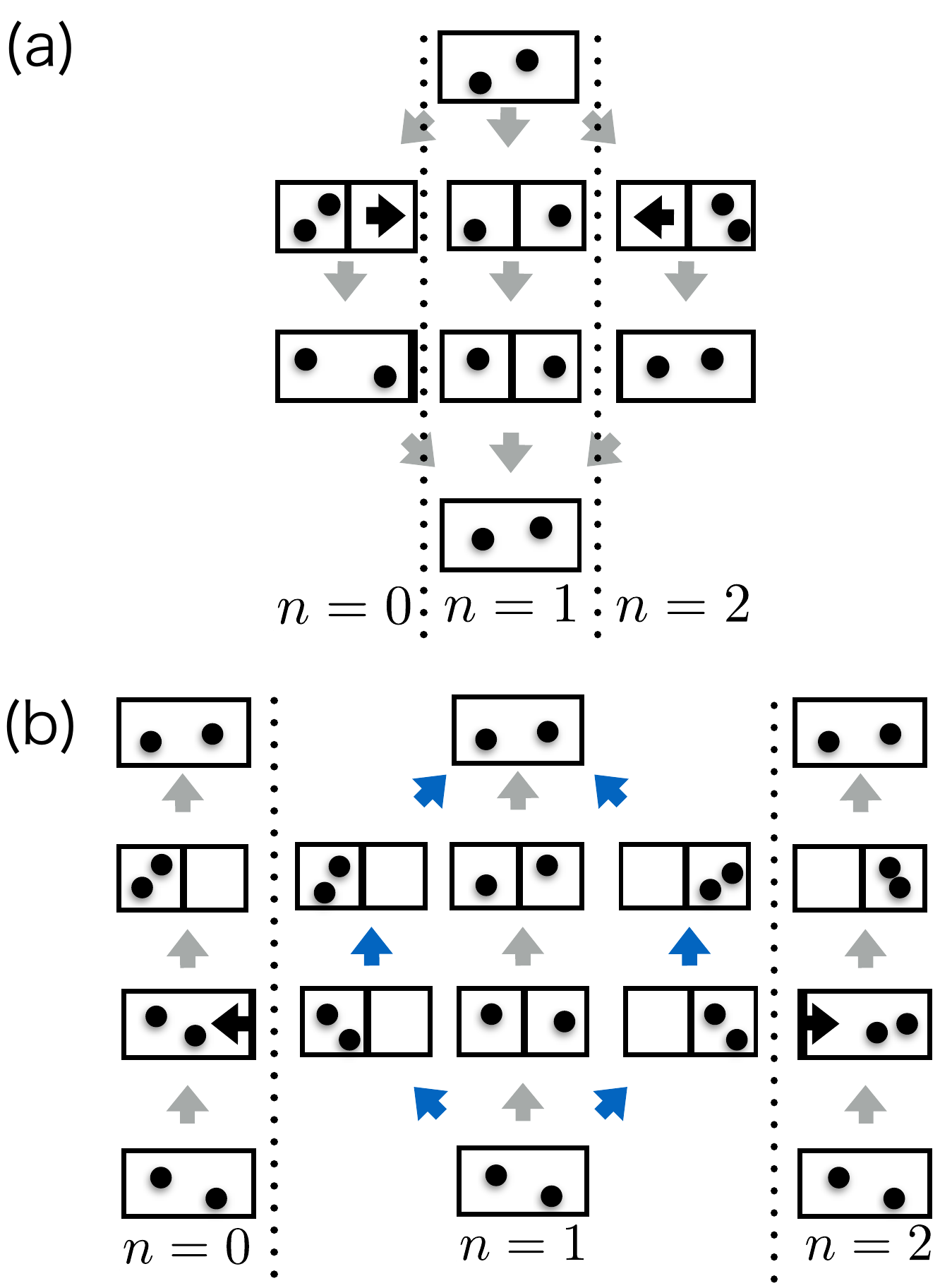}
\caption{\label{fig:2SZ}
(a) Two-particle Szilard engine.
Initially, an ideal two-particle gas is in a global equilibrium of the entire box.
Then, we insert a partition in the middle of the box and perform an error-free measurement to find the number $n$ of the particles in the right box.
Depending on the outcome, we isothermally and quasi-statically shift the partition to such a position that the extracted work is maximal.
Finally, we remove the partition.
(b) Time-reversed protocol for each $n$.
The partition is inserted and then shifted in accordance with the value of $n$.
We have singular paths without the corresponding forward paths in the case of $n=1$ indicated by blue arrows.
Reproduced from Fig. 7 of Ref.~\cite{MFU14}. Copyright 2014 by the American Physical Society.
}
\end{center}
\end{figure}

Initially, two identical ideal-gas particles are confined in a box as illustrated in Fig.~\ref{fig:2SZ}.
We insert a partition in the middle of the box.
Then, we perform an error-free measurement to find the number $n$ of the particles in the right side.
If $n=0$, we isothermally and quasi-statically shift the wall to the right end to extract work, and then remove the partition.
If $n=2$, we shift the wall to the opposite direction to obtain the same amount of work, and then remove the partition.
If $n=1$, we just remove the wall because we cannot extract work by shifting the partition.
For all $n$, the system returns to the initial state after these protocols.
The values of extracted work are summarized in Table~\ref{tab:2SZ}.

\begin{table}[b]
\begin{center}
\caption{\label{tab:2SZ}
	Values of physical quantities corresponding to the outcome $n$.
}
{\renewcommand\arraystretch{1.2}
\begin{tabular}{|c|c|c|c|c|c|c|}
\hline
$n$ & $p(n)$ & $W$ & $\Delta F$ & $I=-\ln p(n)$ & $\lambda_{{\rm S}|\Lambda[n]}$ & $I_{{\rm u}|\Lambda[n]}=-\ln(1-\lambda_{{\rm S}|\Lambda[n]})$\\
\hline
\rule[-10pt]{0pt}{28pt}
0 & $\displaystyle \frac{1}{4}$ & $-2\kb T\ln 2$ & 0 & $\ln 4$ & 0 & 0\\
\hline
\rule[-10pt]{0pt}{28pt}
1 & $\displaystyle \frac{1}{2}$ & 0 & 0 & $\ln 2$ & $\displaystyle\frac{1}{2}$ & $\ln 2$\\
\hline
\rule[-10pt]{0pt}{28pt}
2 & $\displaystyle \frac{1}{4}$ & $-2\kb T\ln 2$ & 0 & $\ln 4$ & 0 & 0\\
\hline
\end{tabular}
}
\end{center}
\end{table}

We consider the time-reversed process to obtain the singular probabilities (see Fig.~\ref{fig:2SZ}).
Initially, the two particles are in a global equilibrium regardless of $n$.
When $n=0$, we insert the partition in the right end of the box.
Then, we isothermally and quasi-statically shift the partition to the middle of the box and remove the partition.
Because this path has the counterpart in the forward process, we have no singularity when $n=0$:
\eqn{
	\lambda_{{\rm S}|\Lambda[0]}=0.
}
Similarly, we have no singular part when $n=2$:
\eqn{
	\lambda_{{\rm S}|\Lambda[2]}=0.
}
Now, let us consider the time-reversed process when $n=1$.
In this case, we insert the partition in the middle of the box and then remove it.
The probability of the two particles being found in the left box has nonvanishing probability, so does the probability of both particles being found in the right box.
However, we have no corresponding forward paths in the protocol of $n=1$.
Therefore, these time-reversed paths are singular, and the singular probability is
\eqn{
	\lambda_{{\rm S}|\Lambda[1]}=\frac{1}{2}.
}
Thus, the unavailable information is
\eqn{
	I_{{\rm u}|\Lambda[0]}&=&0,\\
	I_{{\rm u}|\Lambda[1]}&=&\ln 2,\\
	I_{{\rm u}|\Lambda[2]}&=&0.
}

Now, we are ready to verify our nonequilibrium equalities.
The left-hand side of Eq.~(\ref{ourJar}) is calculated as
\eqn{
	\langle e^{-\beta(W-\Delta F)-I} \rangle
	=
	p(0)\cdot e^{2\beta\kb T\ln 2 -\ln 4}
	+ p(1) \cdot e^{-\ln 2}
	+ p(2) \cdot e^{2\beta\kb T\ln 2 -\ln 4}
	= \frac{3}{4}.
}
On the other hand, the average of the singular part is
\eqn{
	\overline \lambda_{\rm S} =
	p(0)\lambda_{{\rm S}|\Lambda[0]}
	+p(1)\lambda_{{\rm S}|\Lambda[1]}
	+p(2)\lambda_{{\rm S}|\Lambda[2]}= \frac{1}{4}.
}
Therefore, Eq.~(\ref{ourJar}) is verified.
The left-hand side of Eq.~(\ref{ourJarIneq}) is
\eqn{
	\langle W\rangle -\Delta F = -\kb T\ln 2.
}
On the other hand, the right-hand side of Eq.~(\ref{ourJarIneq}) is
\eqn{
	-\langle I \rangle -\ln (1-\overline\lambda_{\rm S})
	=-\ln\frac{3}{\sqrt 2}.
}
Hence, we obtain
\eqn{
	\langle W\rangle -\Delta F > -\kb T\langle I \rangle -\kb T\ln (1-\overline\lambda_{\rm S}),
}
and Eq.~(\ref{ourJarIneq}) is verified.
We note that the equality in the inequality~(\ref{ourJarIneq}) is not achieved because the probability of finding a particular value of $n$ varies for each $n$.

The left-hand side of Eq.~(\ref{AsJar}) is calculated as
\eqn{
	\langle e^{-\beta(W-\Delta F)-I+I_{\rm u}} \rangle
	=
	p(0)\cdot e^{2\beta\kb T\ln 2 -\ln 4}
	+p(1)\cdot e^{-\ln 2 + \ln 2}
	+p(2)\cdot e^{2\beta\kb T\ln 2 -\ln 4}
	= 1,\ \ \ 
}
which verifies Eq.~(\ref{AsJar}).
The right-hand side of Eq.~(\ref{AsJarIneq}) is
\eqn{
	-\kb T\langle I-I_{\rm u} \rangle = -\kb T \ln 2.
}
Thus, we obtain
\eqn{
	\langle W \rangle -\Delta F = -\kb T\langle I-I_{\rm u} \rangle,
}
verifying Eq.~(\ref{AsJarIneq}).
Because Eq.~(\ref{AsJarIneq}) gives an achievable bound of work in the quasi-static limit, the equality in Eq.~(\ref{AsJarIneq}) is achieved.

\subsection{Multi-particle Szilard engine}
We generalize the two-particle Szilard engine to the multi-particle case.
Initially, an ideal $N$-particle gas is in a global equilibrium of the entire box.
We insert a partition in the middle of the box.
Then, we measure the number $n$ of the particles in the right box.
Then, we isothermally and quasi-statically shift the partition to the position that divides the entire box according to the ratio $(N-n):n$.
Finally, we remove the partition.
The probability of the outcome $n$ being found is
\eqn{
	p(n) = \frac{{}_N C_n}{2^N},
}
where $_N C_n$ represents the number of combination that we choose $n$ objects from $N$ objects.
Therefore, the information obtained by the measurement is
\eqn{
	I(n) = N\ln 2 -\ln {}_N C_n.
}
The work during the process can be calculated as
\eqn{
	W &=& -n\kb T\ln\frac{2n}{N} -(N-n)\kb T\ln\frac{2(N-n)}{N}\nonumber\\
	&=& -n\kb T\ln\frac{n}{N}-(N-n)\kb T\ln\frac{(N-n)}{N}-N\kb T\ln 2.
}
Since the system returns to the initial state, we have
\eqn{
	\Delta F = 0.
}
We calculate the singular probabilities by considering the time-reversed process, and obtain
\eqn{
	\lambda_{{\rm S}|\Lambda[n]} =
	1- {}_NC_n \left(\frac{n}{N}\right)^n
	\left(\frac{N-n}{N}\right)^{N-n},
}
and
\eqn{
	I_{{\rm u}|\Lambda[n]} = -\ln {}_NC_n -n \ln\frac{n}{N}-(N-n) \ln\frac{N-n}{N}.
}

The left-hand side of Eq.~(\ref{ourJar}) is 
\eqn{
	\langle e^{-\beta(W-\Delta F)-I} \rangle
	&=&\nonumber
	\sum_{n=0}^N p(n) \exp\left[
		n\ln\frac{n}{N}+(N-n)\ln\frac{N-n}{N}+\ln {}_NC_n
	\right]\\
	&=&\nonumber
	\sum_{n=0}^N p(n)
	\left(\frac{n}{N}\right)^n
	\left(\frac{N-n}{N}\right)^{N-n}
	{}_NC_n\\
	&=&
	\sum_{n=0}^N \frac{({}_NC_n)^2}{2^N}
	\left(\frac{n}{N}\right)^n
	\left(\frac{N-n}{N}\right)^{N-n}.
}
On the other hand, the averaged singular probability is
\eqn{
	\overline \lambda_{\rm S}
	&=&
	\sum_{n=0}^N p(n) \lambda_{{\rm S}|\Lambda[n]}\\
	&=&
	\sum_{n=0}^N p(n)\left[
	1- {}_NC_n \left(\frac{n}{N}\right)^n
	\left(\frac{N-n}{N}\right)^{N-n}
	\right]\\
	&=&
	1
	-\sum_{n=0}^N \frac{({}_NC_n)^2}{2^N}
	\left(\frac{n}{N}\right)^n
	\left(\frac{N-n}{N}\right)^{N-n}.
}
Therefore, Eq.~(\ref{ourJar}) is verified.

By a simple calculation, we obtain
\eqn{
	W-\Delta F = -\kb T(I-I_{\rm u}).
}
Hence, Eqs.~(\ref{AsJar}) and (\ref{AsJarIneq}) are automatically satisfied.
Moreover, the equality in Eq.~(\ref{AsJarIneq}) is achieved, because the measurement is error-free and the process is quasi-static.

\subsubsection{Large-$N$ limit}
We consider the large-$N$ limit of the $N$-particle Szilard engine.
The fluctuation of the particle number $n$ in the right box after the partition is inserted is of the order of $\sqrt{N}$.
Therefore, the dominant contributions to the average of physical quantities come from events with
\eqn{
	n = \frac{N}{2} (1+x),\ x=\mathcal{O}\left(\frac{1}{\sqrt N}\right).
}
We consider approximate formulae in this range of $n$.

The logarithm of the probability $p(n)$ is calculated as
\eqn{
	\ln p(n)
	&=&
	\ln N! -\ln n! -\ln (N-n)! -N\ln 2.
}
By using the Stirling formula
\eqn{
	\ln N! = \frac{1}{2}\ln (2\pi N) + N\ln\frac{N}{e} + \mathcal{O}\left(\frac{1}{N}\right),
}
we obtain
\eqn{
	\ln p(n)
	&=&
	\frac{1}{2}\ln \frac{N}{2\pi n(N-n)} + N\ln N -n\ln n -(N-n)\ln (N-n)-N\ln 2
	+ \mathcal{O}\left(\frac{1}{N}\right)
	\nonumber\\
	&=&
	\frac{1}{2} \ln \frac{2}{\pi N(1-x^2)}-\frac{N}{2}(1+x)\ln (1+x) -\frac{N}{2}(1-x)\ln (1-x)
	+ \mathcal{O}\left(\frac{1}{N}\right)
	\nonumber\\
	&=&
	\frac{1}{2} \ln \frac{2}{\pi N} -\frac{N}{2} x^2
	+ \mathcal{O}\left(\frac{1}{N}\right).
}
Therefore, we obtain
\eqn{
	p(n) = \sqrt{\frac{2}{\pi N}} \exp\left[-\frac{N}{2}x^2 + \mathcal{O}\left(\frac{1}{N}\right)\right]
}
and
\eqn{
	I(n) = \frac{1}{2}\ln \frac{\pi N}{2} + \frac{N}{2}x^2 + \mathcal{O}\left(\frac{1}{N}\right).
}
In a similar manner, the unavailable information is calculated as
\eqn{
	I_{{\rm u}|\Lambda[n]} &=& - \frac{1}{2}\ln \frac{N}{2\pi n(N-n)}
	+ \mathcal{O}\left(\frac{1}{N}\right)
	\nonumber\\
	&=&
	\frac{1}{2} \ln \frac{\pi N}{2} + \mathcal{O}\left(\frac{1}{N}\right).
}

The average of the mutual information is evaluated as
\eqn{
	\langle I \rangle
	&=&
	\sum _{n=0}^N p(n)I(n)
	\nonumber\\
	&=&
	\frac{N}{2} \int_{-\infty}^\infty p(n) I(n) dx 
	\nonumber\\
	&=&
	\frac{1}{2}\ln\frac{\pi N}{2} + \frac{1}{2} + \mathcal{O}\left(\frac{1}{N}\right).
}
Therefore, the conventional second law of information thermodynamics
\eqn{
	\langle W \rangle - \Delta F \ge -\kb T \langle I \rangle
}
reveals that the effect of feedback control is sub-extensive and of the order of $\ln N$.

On the other hand, the average of the unavailable information is obtained as
\eqn{
	\langle I_{\rm u} \rangle
	&=&
	\frac{1}{2}\ln\frac{\pi N}{2}  + \mathcal{O}\left(\frac{1}{N}\right).
}
Therefore, the amount of the available information is
\eqn{
	\langle I - I_{\rm u} \rangle = \frac{1}{2} + \mathcal{O}\left(\frac{1}{N}\right).
}
Thus, our inequality
\eqn{
	\langle W \rangle - \Delta F \ge -\kb T \langle I - I_{\rm u} \rangle
}
limits the effect of feedback control to the order of unity,
which is a qualitatively new restriction compared with that of the conventional second law of information thermodynamics.
The same behavior was derived by another method \cite{KK11} based on the Kawai-Parrondo-Van den Broeck equality \cite{KPvB07, PvBK09}.

\chapter{Gibbs' Paradox Viewed from Absolute Irreversibility}
In this chapter, we apply our nonequilibrium equalities in the presence of absolute irreversibility to the problem of gas mixing.
This problem is what Gibbs' paradox concerns \cite{Gibbs1875}.
Gibbs' paradox is qualitatively resolved once we realize that there are many entropies \cite{Grad61, Kampen84, Jaynes92}.
The most prevalent quantitative resolution based on quantum mechanics is in fact irrelevant to Gibbs' paradox \cite{Kampen84, Jaynes92}.
Another well recognized resolution is based on the extensivity of the thermodynamic entropy \cite{Pauli73,Jaynes92}.
However, this resolution is valid only in the thermodynamic limit, and cannot deal with any sub-leading effects.
We propose a new quantitative resolution of Gibbs' paradox based on our nonequilibrium equalities.

First of all, we review the original Gibbs' paradox and later qualitative discussions.
Then, we present two widespread quantitative resolutions.
Finally, we resolve Gibbs' paradox from the viewpoint of absolute irreversibility.

\section{History and Resolutions}
In this section, we review a brief history of Gibbs' paradox and discuss its resolutions.
We first review the original discussion and interpretation given by Gibbs.
Next, we review works by later researchers that clarify Gibbs' interpretation and argue that Gibbs' paradox is qualitatively resolved even in classical theory.
Then, we discuss the quantum resolution of Gibbs' paradox, which is the standard resolution of Gibbs' paradox.
Finally, we examine yet another resolution given by Pauli.

\subsection{Original Gibbs' paradox}
\begin{figure}
\begin{center}
\includegraphics[width=0.7\columnwidth]{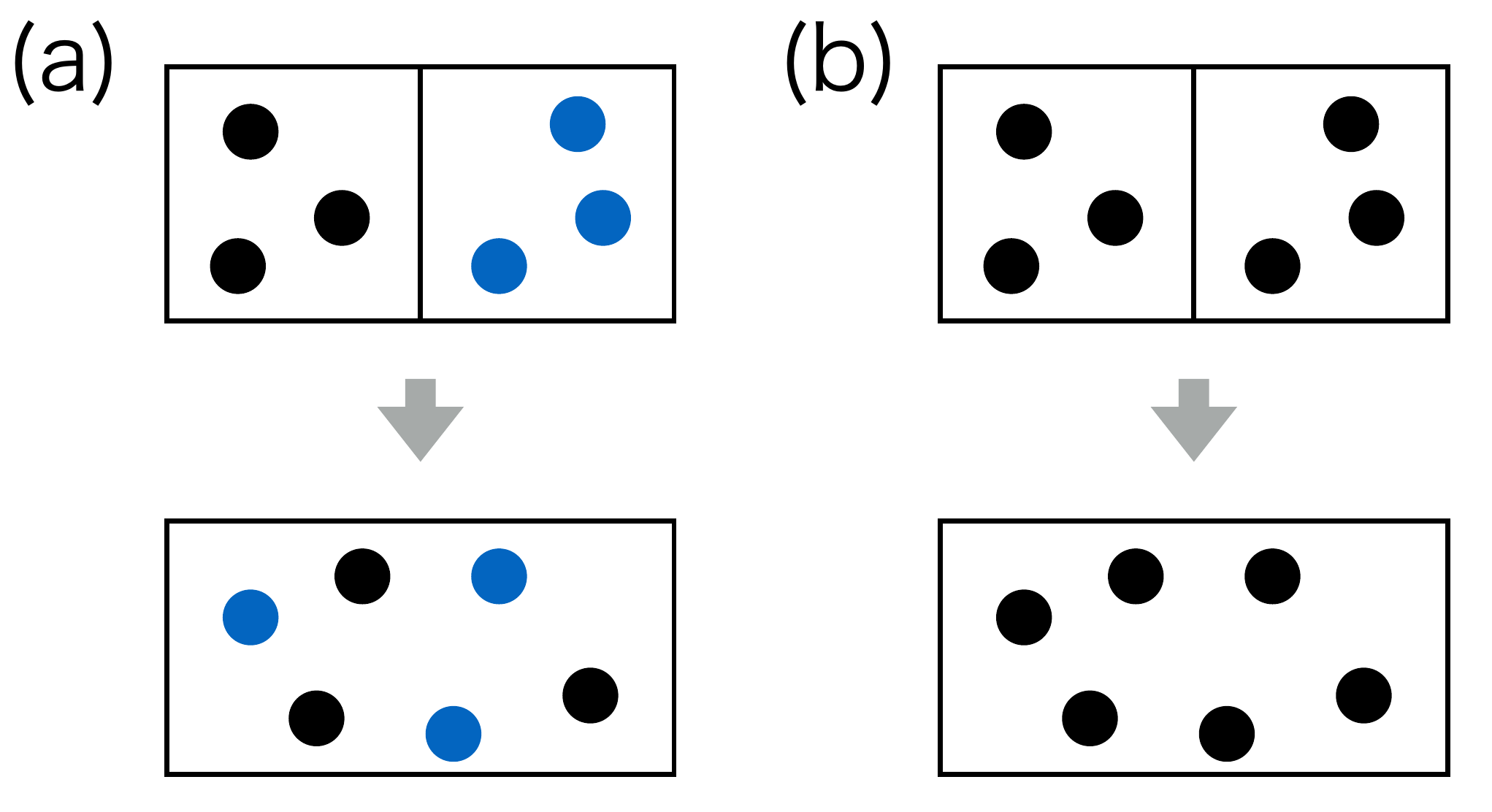}
\caption{\label{fig:OGP}
(a) Mixing of two different gases.
Initially, an ideal $N$-particle gas is confined in the left side and another ideal $N$-particle gas of the other kind is confined in the right side.
Then, we remove the partition, and the two gases expand to the entire box.
(b) Mixing of two identical gases.
Initially, two ideal $N$-particle gases of the same kind are confined both in the left and right sides.
Then, we remove the partition, and the gases expand to the entire box.
}
\end{center}
\end{figure}

First of all, we review Gibbs' original discussion which appeared in his writing titled ``On the Equilibrium of Heterogeneous Substances'' \cite{Gibbs1875}.
We consider mixing of two different gases (see Fig.~\ref{fig:OGP}~(a)).
The gases are confined in a box partitioned into equal halves.
Initially, an ideal $N$-particle gas of one kind is confined in the left side with temperature $T$, and an ideal $N$-particle gas of the other kind is in the right side with the same temperature.
Then, we remove the partition and the two gases expand to the entire box.
Let $S(T, V, N)$ denote the thermodynamic entropy of an ideal gas with temperature $T$, volume $V$ and the number of particles $N$.
The entropy production of this mixing process can be calculated by the Clausius definition of the thermodynamic entropy.
The entropy production of one gas is given by
\eqn{\label{CLdef}
	S(T, 2V, N) - S(T, V,N)
	=
	-\int_{(T, V, N)}^{(T, 2V, N)} \frac{\delta Q}{T},
}
where $Q$ denotes the heat transferred from the system to a heat bath, and the integral is conducted along an arbitrary virtual quasi-static process that connects the state $(T, V, N)$ to $(T, 2V, N)$.
We set the virtual process to the quasi-static isothermal process.
Due to the first law of thermodynamics and the fact that the internal energy of an ideal gas does not change in an isothermal process, we have
\eqn{
	\delta Q = -p dV,
}
where $p$ is the pressure of the gas.
Therefore, we obtain
\eqn{
	S(T, 2V, N) - S(T, V,N)
	=
	\int_V^{2V} \frac{pdV}{T}.
}
Using the equation of state $pV=N\kb T$, we obtain
\eqn{
	S(T, 2V, N) - S(T, V,N)
	=
	N\kb\int_V^{2V} \frac{dV}{V}
	=
	N\kb\ln 2.
}
Due to the additivity of the thermodynamic entropy, the entropy production of two different gases is the sum of their individual entropy productions.
Therefore, the entropy production of the mixing of two different gases is given by
\eqn{\label{EPdif}
	\Delta S^{\rm dif} = 2N\kb\ln 2.
}

After Gibbs derived Eq.~(\ref{EPdif}), he argued \cite{Gibbs1875}
\begin{quote}
``It is noticeable that the value of this expression does not depend upon the kinds of gas which are concerned, if the quantities are such as has been supposed, except that the gases which are mixed must be of different kinds.
If we should bring into contact two masses of the same kind of gas, they would also mix, but there would be no increase of entropy.''
\end{quote}
In fact, when we consider mixing of two identical gases (see Fig.~\ref{fig:OGP}~(b)), the initial entropy of the system is given by
\eqn{\label{iniitalentropy}
	2S(T,V,N)
}
due to the additivity of the thermodynamic entropy.
On the other hand, the final entropy is
\eqn{
	S(T,2V,2N),
}
which is equal to the initial entropy~(\ref{iniitalentropy}) due to the extensivity of the thermodynamic entropy:
\eqn{
	S(T, qV, qN)=qS(T,V,N),
}
where $q$ is an arbitrary positive real number.
Therefore, the entropy production of  the mixing of two identical gases is
\eqn{
	\Delta S^{\rm id} = 0,
}
which is different from the entropy production of the mixing of two different gases~(\ref{EPdif}), although Eq.~(\ref{EPdif}) is
``independent of the degrees of similarity or dissimilarity between them.''
In particular, if we consider
\begin{quote}
``the case of two gases which should be absolutely identical in all the properties (sensible and molecular) which come into play while they exist as gases whether pure or mixed with each other, but which should differ in respect to the attractions between their atoms and the atoms of some other substances,''
\end{quote}
the thermodynamic entropy increases in the mixing process of two different gases, although
\begin{quote}
``the process of mixture, dynamically considered, might be absolutely identical in its minutest details (even with respect to the precise path of each atom) with processes which might take place without any increase of entropy,''
\end{quote}
i.e., with processes of the mixing of identical gases.
This paradoxical consequence of the gas mixing is referred to as Gibbs' paradox.

To explain this fact, Gibbs stressed,
\begin{quote}
	``if we ask what changes in external bodies are necessary to bring the system to its original state, we do not mean a state which shall be undistinguishable from the previous one in its sensible properties.
	It is to states of systems thus incompletely defined that the problems of thermodynamic relate.''
\end{quote}
In other words, the thermodynamic entropy is not a property of a microstate, but a property of a set of microstates indistinguishable from each other, i.e., a property of a macrostate.
In this way, ``entropy stands strongly contrasted with energy.''
Therefore,
\begin{quote}
``the mixture of gas-masses of the same kind stands on a different footing from the mixture of gas-masses of different kinds,''
\end{quote}
This explanation given by Gibbs was clarified by later researchers as we see in the next section.

\subsection{Gibbs' paradox is not a paradox}
In this section, we review discussions by three researchers, and argue that Gibbs' paradox is not a paradox even within classical statistical mechanics.

\subsubsection{Grad \cite{Grad61}}
Grad discussed Gibbs' paradox in the introduction of his article titled ``The Many Faces of Entropy'' \cite{Grad61}.
In the introduction, he emphasized,
\begin{quote}
``A given object of study cannot always be assigned a unique value, its `entropy.' It may have many different entropies, each one worthwhile.''
\end{quote}
and argued,
\begin{quote}
``much of the confusion in the subject is traceable to the ostensibly unifying belief (possibly theological in origin!) that there is only one entropy.''
\end{quote}
In fact, Gibbs' paradox is resolved once we renounce this belief.
Grad continued,
\begin{quote}
``Whether or not diffusion occurs when a barrier is removed depends not on a difference in physical properties of the two substances but on a decision that we are or are not interested in such a difference (which is what governs the choice of an entropy function).
There is no paradox to any observer.
When he is aware of a difference in properties, he observes diffusion together with an increase in entropy.
When he is unaware of any difference, he observes no diffusion and no increase in the entropy which he is using.
If two observers disagree, they must be interested in different phenomena, and there is no conflict.''
\end{quote}

To clarify his idea, Grad raised different situations that describe states of an $n$-particle gas.
In the first situation, we label particles in the gas by $1,\cdots, n$.
In the second situation, we count the particle number of the gas in a certain region of space.
These two situations are different and give rise to two different non-comparable entropies.
In the first situation, when we remove the partition and mix the gas, the entropy $S_1$ increases by $n\ln 2$.
Then, when we reinsert the partition, the $S_1$ decreases by $n\ln 2$ and returns to its original value.
This is because we have complete information whether each particle is in the left or right side since we label all the particles.
In the second situation, when we remove the partition, the entropy $S_2$ increases by $n\ln 2$, if we distinguish the particles originally in the two different sides when we count the number of particles.
On the other hand, $S_2$ does not increase, if we are not interested in any difference of the particles.
In both cases, when we reinsert the partition, the entropy does not change.
In this way, the entropies in these descriptions differ from each other.
Although $S_1$ is completely impractical from the viewpoint of thermodynamics, it does exist and give one description of the system.

In summary, the notion of entropy depends on our interest or description of a system.
If entropies differ from each other, it just implies a difference of our interest, and there is no conflict.
As to Gibbs' paradox, the difference of the entropy production reflects our situation of whether or not we are interested in a difference of the particles.

\subsubsection{van Kampen \cite{Kampen84}}
In his essay, van Kampen discussed Gibbs' paradox.
The aim of his essay is to refute such statements as
\begin{quote}
``It is not possible to understand classically why we must divide $N!$ to obtain the correct counting of states,'' and ``Classical statistics thus leads to a contradiction with experience even in the range in which quantum effects in the proper sense can be completely neglected.''
\end{quote}

First, he calculated the entropy $S(P,T)$ of an ideal $N$-particle gas with pressure $P$ and temperature $T$, and obtained
\eqn{
	S(P,T) = \frac{5}{2}N\kb \ln T - N\kb\ln P + C.
}
Then, he emphasized,
\begin{quote}
``the second law defines only entropy differences between states that can be connected by reversible change.''
\end{quote}
Therefore, the constant $C$ is only required to be independent of $P$ and $T$.
Thus, at this point, there is no way to compare entropy with different $N$, unless we introduce a new reversible process that varies $N$.
We consider a process in which we attach two boxes containing identical gases with the same $(P,T,N)$, and open a channel between them.
Then, the constant $C$ should be proportional to $N$ and we obtain
\eqn{\label{vKsame}
	S(P,T,N) = \frac{5}{2}N\kb \ln T - N\kb\ln P + cN,
}
where $c$ does not depend on $(P,T,N)$.
If two boxes contain two different gases A, B, the process in which the channel is opened is no longer reversible.
Instead, we need to consider a process with semi-permeable walls.
We combine two boxes using the semi-permeable walls in a reversible way, and then isothermally and quasi-statistically expand the box so that the pressure returns to its original value.
Together with the convention of the additivity of the thermodynamic entropy, we obtain
\eqn{
	S(P, T, N_{\rm A}=N/2, N_{\rm B}=N/2)
	=
	\frac{5}{2}N\kb\ln T - N\kb\ln\frac{P}{2}+\frac{N}{2}(c_{\rm A}+c_{\rm B}).\label{vKdif}
}
Even when we set A$=$B, Eq.~(\ref{vKdif}) does not reduce to Eq.~(\ref{vKsame}).
This difference is what Gibbs' paradox concerns.
Then, van Kampen argued,
\begin{quote}
``The origin of the difference is that two different process had to be chosen for extending the definition of entropy.
They are mutually exclusive:
the first one cannot be used for two different gases and the second one does not apply to a single one.''
\end{quote}
Next, he considered the case in which A and B are so similar that an experimenter cannot distinguish them operationally, namely, he does not have the semi-permeable walls needed in the second process.
Then, he will conclude
\eqn{
	S(P, T, N_{\rm A}=N/2, N_{\rm B}=N/2)
	=
	\frac{5}{2}N\kb\ln T - N\kb\ln P+cN.
}
This seems to be contradictory at a first glance, but
\begin{quote}
``The point is, that this is perfectly justified and that he will not be led to any wrong results.
If you tell him that `actually' the entropy increased when he opened the channel he will answer that this is a useless statement since he cannot utilize the entropy increase for running a machine.
The entropy increase is no more physical to him than the one that could be manufactured by taking a single gas and mentally tagging the molecules A or B. [...] The expression for the entropy (which he constructs by one or the other processes mentioned above) depends on whether or not he is able and willing to distinguish between the molecules A and B.
This is a paradox only for those who attach more physical reality to the entropy than is implied by its definition.''
\end{quote}
Van Kampen concluded
\begin{quote}
``The question is not whether they are identical in the eye of God, but merely in the eye of the beholder.''
\end{quote}

\subsubsection{Jaynes \cite{Jaynes92}}
Jaynes argued that the writing of Gibbs \cite{Gibbs1875} contains a correct analysis of Gibbs' paradox.
However, this analysis has been lost due to ambiguity in the writing and the fact that Gibbs did not include it in his later renowned textbook \cite{Gibbs1902}.
Jaynes presented a ``half direct quotation'' of Gibbs' explanation \cite{Jaynes92}.

First of all, Jaynes stressed that we have to be circumspect about what we mean by the words ``state'' and ``reversible.''
He wrote
\begin{quote}
``But by the word `original state' we do not mean that every molecule has been returned to its original position, but only a state which is indistinguishable from the original one in the macroscopic properties that we are observing.''
\end{quote}
Therefore, in the mixing of two different gases, the particles originally in the left (right) side must return to the left (right) when we say that the state returns to the original state.
In contrast, in the mixing of two identical gases, we do not mean that the particles originally in the left (right) side should return to the left (right) when we say that the mixing is reversible and accompanied by no entropy production.
We say that the mixing is reversible because all macroscopic properties (e.g. the chemical composition, the number of particles) return to their original value after we reinsert the partition.
Thus,
\begin{quote}
``Trying to interpret the phenomenon as a discontinuous change in the physical nature of the gases (i.e. in the behavior of their microstates) when they become exactly the same, misses the point. [...]
We might put it thus: when the gases become exactly the same, the discontinuity is in what you and I mean by the words `restore' and `reversible'.''
\end{quote}

To clarify this point, Jaynes continued his discussion.
He noted that a thermodynamic state is defined by specifying a small number of macroscopic quantities $\{X_1, X_2,\cdots,X_n\}$.
The entropy is defined as a property of a macrostate specified by these quantities: $S=S(X_1,X_2,\cdots,X_n)$.
As indicated by the discussion of the gas mixing, the entropy is not a property of the microstate, whereas other thermodynamic variables as the total mass and the total energy are physically real properties of a microstate.
In fact, the thermodynamic entropy is a property of a macrostate $C(X)$, namely, a set of microstates compatible with $X=\{X_1,X_2,\cdots,X_n\}$.
Thus, it is possible to assign different entropies $S, S'$ to the same microstate, if we choose different sets of macroscopic variables and embed the microstate in two different macrostates $C,C'$. 
This implies that we have to specify macroscopic variables that we can measure and control in advance to define the thermodynamic entropy.
Then, the thermodynamic entropy obeys the second law of thermodynamics as long as all experimental manipulations are within the set of macroscopic variables that we have chosen beforehand.
Because this choice connotes whether we regard the gases as different or identical, the behavior of the entropy under the gas mixing hinges upon this choice.

\subsubsection{Summary}
As we have seen, the thermodynamic entropy is not an intrinsic property of a microstate, and the definition of the thermodynamic entropy involves arbitrariness.
Whether the gases are identical or different is predetermined within our thermodynamic framework that we have chosen beforehand.
Thus, it is meaningless to discuss the discontinuity in the thermodynamic entropy when we consider the infinitely similar gases.
In this sense, Gibbs' paradox is not a paradox.
\\
\\
Now that we qualitatively understand that Gibbs' paradox is not a paradox even within classical thermodynamics, the remaining question is how to derive the factorial in the statistical definition of the thermodynamic entropy, namely, the factor $N!$ in the definition of the entropy
\eqn{
	S = \ln \frac{W}{N!},
}
where $W$ denotes the number of microstates, or in the definition of the partition function
\eqn{
	Z= \frac{1}{N!}\int e^{-\beta H(\Gamma)} d\Gamma,
}
where $H(\Gamma)$ is the Hamiltonian of the system.
In the following two sections, we review two well-known quantitative resolutions of Gibbs' paradox.

\subsection{Quantum resolution}
The standard resolution of Gibbs' paradox is based on quantum mechanics.
In quantum theory, the interchange of identical particles does not lead to another state, and identical particles are indistinguishable in principle.
This indistinguishability naturally leads to the factor $N!$.
Many textbooks (for example, see Refs.~\cite{Feynman72,Zubarev74,LanLif75,Kubo78,Callen85}) resolve Gibbs' paradox in this way.

Following Ref.~\cite{Feynman72}, we consider an $N$-particle quantum system with a Hamiltonian
\eqn{
	\hat H = \sum_i \frac{{\bf p}_i^2}{2m} + V({\bf x}_1,\cdots,{\bf x}_N),
}
where $m$ is the mass of the particles.
Let $|i\rangle$ and $E_i$ denote the $i$-th eigenstate and eigenenergy, respectively.
Then, the (unnormalized) density matrix of the canonical ensemble is
\eqn{
	\rho_\beta = \sum_i e^{-\beta E_i} |i\rangle\langle i|.
}
The configurational representation reads
\eqn{
	\rho_\beta({\bf x}_1,\cdots,{\bf x}_N;{\bf x}'_1,\cdots,{\bf x}'_N)
	= \sum_i e^{-\beta E_i} \psi_i({\bf x}_1,\cdots,{\bf x}_N)
	\psi^*_i({\bf x}'_1,\cdots,{\bf x}'_N),
}
where $\psi_i({\bf x}_1,\cdots,{\bf x}_N) = \langle{\bf x}_1,\cdots,{\bf x}_N|i\rangle$.
When the particles are bosons, this density matrix must be symmetrized as
\eqn{
	\rho_\beta^{\rm sym}({\bf x}_1,\cdots,{\bf x}_N;{\bf x}'_1,\cdots,{\bf x}'_N)
	= \frac{1}{N!} \sum_{\sigma\in S_N} \rho_\beta
	({\bf x}_1,\cdots,{\bf x}_N;{\bf x}'_{\sigma(1)},\cdots,{\bf x}'_{\sigma(N)}),
}
where $S_N$ is the symmetry group of degree $n$.
The partition function is given by
\eqn{
	Z^{\rm sym}_\beta
	&=& \int d{\bf x_1}\cdots d{\bf x}_N\ 
	\rho_\beta^{\rm sym}({\bf x}_1,\cdots,{\bf x}_N;{\bf x}_1,\cdots,{\bf x}_N)\nonumber\\
	&=&
	\frac{1}{N!} \sum_{\sigma\in S_N}
	\int d{\bf x_1}\cdots d{\bf x}_N\ 
	\rho_\beta({\bf x}_1,\cdots,{\bf x}_N;{\bf x}_{\sigma(1)},\cdots,{\bf x}_{\sigma(N)})\label{symZ}
}
The integrand in this equation involves factors
\eqn{
	\exp\left[
		-\frac{m({\bf x}_i-{\bf x}_{\sigma(i)})^2}{2\beta\hbar^2}
	\right].
}
In the classical (i.e., high-temperature) limit, due to this exponential decay, the dominant contribution in Eq.~(\ref{symZ}) is the term in which $\sigma$ is the identity permutation.
Therefore, Eq.~(\ref{symZ}) is approximated as
\eqn{
	Z_\beta^{\rm sym} \simeq \frac{1}{N!} 
	\int d{\bf x_1}\cdots d{\bf x}_N\ 
	\rho_\beta({\bf x}_1,\cdots,{\bf x}_N;{\bf x}_{1},\cdots,{\bf x}_{N}).
}
In the case of fermions, a similar discussion leads to the same conclusion.
In this way, the factor $N!$ can naturally be derived from quantum mechanics and this factor leads to the extensive thermodynamic entropy.

Although this resolution is the standard resolution of Gibbs' paradox, it involves two crucial problems.
First, this resolution cannot apply to mesoscopic particles.
Let us consider colloidal particles in liquid.
Because colloidal particles have vast internal degrees of freedom, we cannot expect that these particles have the same internal states.
Therefore, the wave function of this system should not be symmetrized.
Hence, we cannot derive the factor $N!$, and thermodynamic quantities (e.g. entropy) fail to be extensive.
This implies that the quantum resolution is inapplicable to a mesoscopic regime.
The second point is more fundamental and to be explained at the end of the next section, because this point is closely linked to the topic described in the next section.

\subsection{Pauli's resolution based on the extensivity}
As we see in the review of van Kampen's work \cite{Kampen84}, the Clausius definition of thermodynamic entropy reveals nothing about the dependence of the entropy on the particle number.
Pauli recognized this fact and gave a resolution within classical theory \cite{Pauli73, Jaynes92}.
Pauli's analysis is based on the extensivity of the thermodynamic entropy.

Let $S(T, V, N)$ be a phenomenological thermodynamic entropy of an ideal $N$-particle gas with temperature $T$ and volume $V$ defined by the Clausius definition~(\ref{CLdef}).
Then, from the definition, we obtain
\eqn{\label{SbyCD}
	S(T,V,N) = \frac{3}{2}N\kb\ln T + N\kb \ln V + \kb f(N),
}
where $f(N)$ is not an arbitrary constant, but an arbitrary function of $N$.
This is because the Clausius definition does not involve the $N$-dependence of the thermodynamic entropy.
To determine $f(N)$, we require the extensivity of the entropy as an additional condition.
The extensivity means
\eqn{\label{extensivity}
	S(T,qV,qN) = qS(T,V,N),
}
where $q$ is an arbitrary positive real number.
Substituting Eq.~(\ref{SbyCD}) into Eq.~(\ref{extensivity}), we obtain
\eqn{
	qN\ln q + f(qN) = qf(N).
}
Differentiating with respect to $q$ and setting $q=1$, we obtain
\eqn{
	N+Nf'(N)=f(N).
}
This equation can be rewritten as
\eqn{
	\frac{d}{dN}\left(\frac{f(N)}{N}\right)
	=
	-\frac{1}{N}.
}
Therefore, we obtain
\eqn{
	f(N)=Nf(1)-N\ln N,
}
where the second term is the factor $N!$ due to the Stirling formula in the large-$N$ limit, i.e., in the thermodynamic limit.
Thus, the thermodynamic entropy is
\eqn{
	S(T,V,N) = \frac{3}{2}N\kb\ln T+ N\kb\ln\frac{V}{N} + N\kb f(1).
}
By this form, we see that the entropy is extensive and $f(1)$ is essentially the chemical potential.
In this way, the requirement of the extensivity leads to the factor $N!$ in the phenomenological entropy defined by the Clausius in the thermodynamic limit.

The same argument applies to the entropy defined by classical statistical mechanics.
The reason why we identify the entropy defined by classical statistical mechanics as the thermodynamic entropy is that these two entropies have the same response to any variation of such macroscopic variables as the temperature and volume.
In mathematical terms, these two entropies have the identical differential form.
Therefore, what we can conclude about relations between the thermodynamic entropy $S(T,V,N)$ and the entropy in classical statistical mechanics $S^{\rm C}(T,V,N)$ is
\eqn{\label{SSCf}
	S(T,V,N) = S^{\rm C}(T,V,N) + \kb f^{\rm C}(N).
}
For an $N$-particle ideal gas, we have
\eqn{\label{SbyCSM}
	S^{\rm C}(T,V,N) = \frac{3}{2}N\kb[\ln (2\pi m\kb T)+1] + N\kb\ln V-3N\kb\ln\xi,
}
where $\xi$ is a constant of the dimension of action.
Then, the requirement of the extensivity~(\ref{extensivity}) leads to
\eqn{
	f^{\rm C}(N) = Nf^{\rm C}(1)-N\ln N.
}
The extensivity reproduces the second term on the right-hand side, i.e., the factor $N!$ again.

Quantum statistical mechanics is in the the same position as classical statistical mechanics.
We have
\eqn{
	S(T,V,N)=S^{\rm Q}(T,V,N) + \kb f^{\rm Q}(N).
}
In the classical limit, since we have
\eqn{
	S^{\rm Q}(T,V,N)\simeq S^{\rm C}(T,V,N)-\ln N!,
}
the requirement of the extensivity leads to
\eqn{
	f^{\rm Q}(N) = Nf^{\rm Q}(1)+\ln\frac{N!}{N^N},
}
where the second term on the right-hand side vanishes in the thermodynamic limit.
Hence, $f^{\rm Q}(N)$ has only a trivial dependence on $N$ as
\eqn{
	f^{\rm Q}(N) \simeq Nf^{\rm Q}(1),
}
which corresponds to the contribution from the chemical potential.
In this manner, the procedure to determine an arbitrary function of $N$ is 
needed for quantum statistical mechanics as well, although the result is simpler than that of classical statistical mechanics.
Therefore, \cite{Kampen84}
\begin{quote}
``the Gibbs paradox is no different in quantum mechanics, it is only less manifest.''
\end{quote}

As we have seen, the quantum resolution is irrelevant to Gibbs' paradox.
The resolution based on the extensivity is a better and logical resolution, and applicable to the phenomenological entropy, the entropy in classical statistical mechanics and the entropy in quantum statistical mechanics.
However, this resolution still suffers a problem.
The resolution is only applicable to systems in the thermodynamic limit in which we are entitled to require the extensivity of the entropy.
Namely, it ignores deviations from the extensivity, which are essential to deal with mesoscopic physics and surface effects.

\section{Resolution from absolute irreversibility}
In this section, we resolve Gibbs' paradox based on the nonequilibrium equalities with absolute irreversibility.
We explain why the entropy productions of the two mixing processes are different from each other, and then derive the factor $N!$.
Our resolution is valid even for non-extensive entropy, for which the resolution given by Pauli breaks down.

\subsection{Requirement and Results}
In the resolution by Pauli, we require the extensivity of the thermodynamic entropy.
Instead of this requirement, we require the additivity of the thermodynamic entropy.
In mesoscopic systems, the extensivity breaks down, whereas the additivity remains valid as long as the interaction is short-range.
The additivity plays a crucial role when we compare the thermodynamic entropy with different $N$.

Under the requirement of the additivity, in the context of our nonequilibrium equality, we show that the entropy production of the mixing of two identical $N$-particle gases is
\eqn{
	\Delta S^{\rm id} = 0
}
and that the entropy production of the mixing of two different $N$-particle gases is
\eqn{
	\Delta S^{\rm dif} = 2N\kb \ln 2
}
in the thermodynamic limit.
This difference of the two processes will be explained in terms of absolute irreversibility.
Moreover, we will derive the factor $N!$, that is, we show that the arbitrary function $f^{\rm C}(N)$ in classical statistical mechanics should take the following form:
\eqn{
	f^{\rm C}(N) = Nf^{\rm C}(1)-\ln N!.
}
This result is valid for a finite $N$ without the thermodynamic limit.

\subsection{Difference of the two processes}
We consider a difference between the mixing of two identical gases and that of two different gases in terms of absolute irreversibility.
\begin{figure}
\begin{center}
\includegraphics[width=1.0\textwidth]{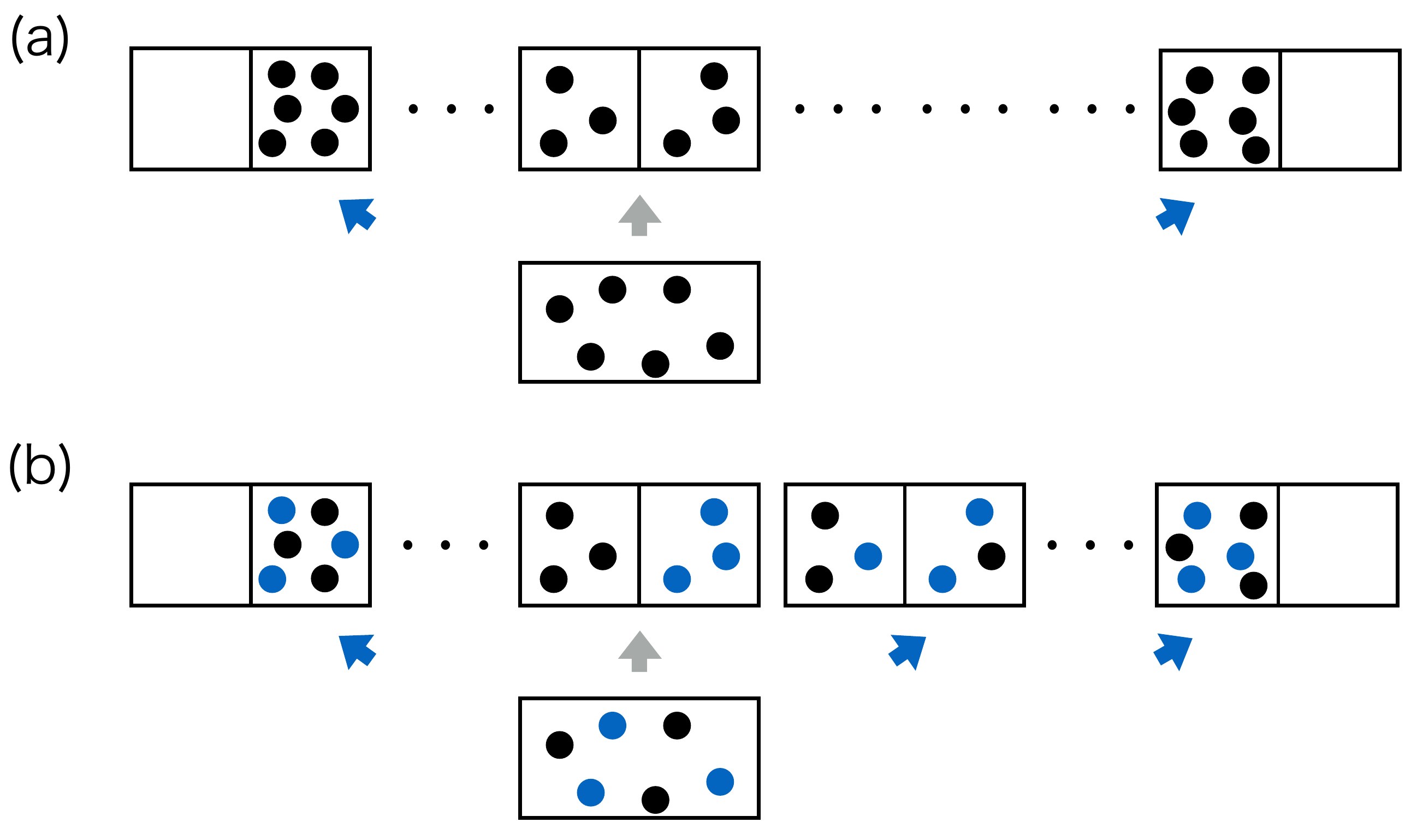}
\caption{\label{fig:GPrev}
(a) Reverse process of the mixing of two identical gases.
Initially, an ideal $2N$-particle gas is at thermal equilibrium in the entire box.
Then, we insert the partition in the middle.
The particle number in the left side $n$ varies from $0$ to $2N$ according to the binomial distribution.
The events of $n\neq N$ indicated by the blue arrows are singular because they have no counterparts in the original (i.e., forward) process.
(b) Reverse process of the mixing of two different gases.
Initially, two ideal $N$-particle gases of different kinds are at thermal equilibrium of the entire box.
Then, we insert the partition in the middle.
To return to the original state, not only must the particle number in the left side be $N$, but also the composition of gas must be the same as the original state.
Therefore, the mixing of two different gases has much more absolute irreversibility than the mixing of two identical gases.
}
\end{center}
\end{figure}
First, we consider the reverse process of the mixing of two identical gases illustrated in Fig.~\ref{fig:GPrev} (a) to evaluate the absolute irreversibility.
Initially, an ideal $2N$-particle gas is in the equilibrium of the entire box.
Then, we insert the partition in the middle.
Let $n$ denote the number of the particles found in the left side after the insertion.
The event of $n=N$ is the only event that has the corresponding event in the original process (see also Fig.~\ref{fig:OGP}~(b)).
Therefore, the events of $n\neq N$ are singular because they have no counterparts.
Hence, the singular probability is calculated as
\eqn{\label{LamId}
	\lambda^{\rm id}_{\rm S} = 1-\frac{{}_{2N}C_N}{2^{2N}}.
}
Secondly, we consider the reverse process of the mixing of two different gases illustrated in Fig.~\ref{fig:GPrev}~(b).
Initially, two ideal $N$-particle gases of different kinds are at thermal equilibrium in the entire box.
Then, we insert the partition in the middle.
To recover the original state, the particle number in the left side after the insertion must be $N$.
Moreover, the chemical composition must return to the original state.
Namely, the particles from the left (right) side in the original process must return to the left (right) side.
This fact makes sharp contrast to the case of two identical gases, in which the particles from the left (right) may go to the right (left) sides as long as the particle number returns to $N$.
The only non-singular event is the one in which all the particles of one kind return to the left side and the rest particles return to the right side.
All the other events indicated by blue arrows are singular.
Thus, the singular probability is
\eqn{\label{LamDif}
	\lambda^{\rm dif}_{\rm S}=1-\frac{1}{2^{2N}}.
}
In this way, the difference of the intuitive physical descriptions in the reversed processes is quantitatively characterized by the difference of the probabilities of the absolutely irreversible paths.

Next, we connect these singular probabilities to the thermodynamic entropy production.
The Jarzynski-type nonequilibrium equality~(\ref{GJar}) in the presence of absolute irreversibility reads
\eqn{
	\langle e^{-\beta(W-\Delta F)} \rangle = 1-\lambda_{\rm S}.
}
Since work is zero ($W=0$) in the mixing process, we obtain
\eqn{
	\Delta F = \kb T\ln(1-\lambda_{\rm S}).
}
Thus, in terms of the thermodynamic entropy, we obtain
\eqn{\label{SLam}
	\Delta S = -\kb\ln (1-\lambda_{\rm S}).
}
Substituting Eq.~(\ref{LamId}) into Eq.~(\ref{SLam}), we obtain
\eqn{\label{EPid}
	\Delta S^{\rm id} = 2N\kb\ln 2 - \kb\ln {}_{2N}C_N.
}
When $N$ is sufficiently large, the combination in Eq.~(\ref{EPid}) is approximated as
\eqn{
	_{2N}C_N \simeq \frac{2^{2N}}{\sqrt{\pi N}}.
}
Therefore, Eq.~(\ref{EPid}) reduces to
\eqn{
	\Delta S^{\rm id}\simeq\frac{1}{2}\kb\ln\pi N.
}
Since this value is sub-extensive, we have
\eqn{
	\Delta S^{\rm id} = 0
}
in the thermodynamic limit.
This is consistent with the fact that the removal of the partition becomes reversible in the large-$N$ limit due to the law of large numbers, namely, the reinsertion of the partition results in the state of $n=N$ with almost unit probability.
In the case of two different gases, from Eqs.~(\ref{LamDif}) and (\ref{SLam}), we obtain
\eqn{
	\Delta S^{\rm dif} = 2N\kb\ln 2.
}
The difference of these entropy productions of the two processes originates from the difference of the degree of absolute irreversibility, namely, the difference in the behaviors under the reinsertion of the partition.

\subsection{Derivation of the factor $N!$}
\begin{figure}
\begin{center}
\includegraphics[width=1.0\textwidth]{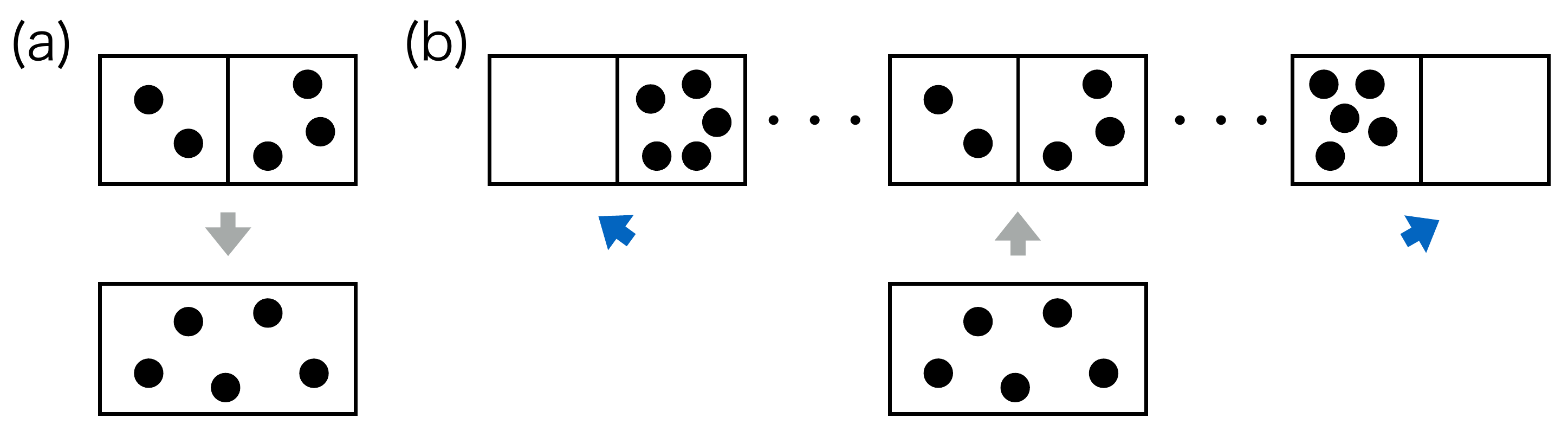}
\caption{\label{fig:NMmix}
(a) Mixing of an $N$-particle gas and an $M$-particle gas of the same kind.
A box is divided into equal halves by a partition.
Initially, the $N$-particle gas is in the left side, and the $M$-particle gas is in the right side.
We remove the partition and the gases expand to the entire box.
(b) The reverse process.
Initially, an $(N+M)$-particle gas is at thermal equilibrium in the entire box.
Then, we insert the partition.
The number of the particles in the left side $n$ varies from $0$ to $N+M$ according to the binomial distribution.
}
\end{center}
\end{figure}
To determine the arbitrary function $f^{\rm C}(N)$ in Eq.~(\ref{SSCf}), we consider a mixing process of an $N$-particle gas and an $M$-particle gas of the same kind illustrated in Fig.~\ref{fig:NMmix} (a).
To evaluate the probability of absolute irreversibility, we consider the reverse process (see Fig.~\ref{fig:NMmix} (b)).
We insert the partition in the middle in an $(N+M)$-particle gas.
The number of particles in the left side, $n$, after the reinsertion must be $N$ to recover the original state.
This is the only non-singular event.
Therefore, the singular probability is
\eqn{
	\lambda_{\rm S} = 1- \frac{{}_{N+M}C_N}{2^{N+M}}.
}
Thus, from Eq.~(\ref{SLam}), the thermodynamic entropy production is
\eqn{
	\Delta S = (N+M)\kb\ln 2 -\kb \ln {}_{N+M}C_N.
}
On the other hand, from Eq.~(\ref{SbyCSM}) and the additivity of the thermodynamic entropy, we obtain
\eqn{
	\Delta S = (N+M)\kb\ln 2+\kb f^{\rm C}(N+M)-\kb f^{\rm C}(N)-\kb f^{\rm C}(M)
}
Therefore, we obtain
\eqn{
	f^{\rm C}(N+M)- f^{\rm C}(N)- f^{\rm C}(M) = -\ln {}_{N+M}C_N.
}
When set $M=1$, we have
\eqn{\label{funcEq}
	f^{\rm C}(N+1)- f^{\rm C}(N)- f^{\rm C}(1) = -\ln (N+1).
}
Let us define
\eqn{
	 g^{\rm C}(N) = \exp[ f^{\rm C}(N)],
}
and then Eq.~(\ref{funcEq}) reduces to
\eqn{
	g^{\rm C}(N+1)=g^{\rm C}(N)\frac{g^{\rm C}(1)}{N+1}.
}
This equation can be rewritten as
\eqn{
	{(N+1)!}\cdot{g^{\rm C}(N+1)}=g^{\rm C}(1)\cdot{N!}\cdot g^{\rm C}(N).
}
Therefore, we obtain
\eqn{
	{N!} \cdot{g^{\rm C}(N)}= \{g^{\rm C}(1)\}^N
}
and 
\eqn{
	g^{\rm C}(N) = \frac{\{g^{\rm C}(1)\}^N}{N!}.
}
Hence, we conclude
\eqn{
	f^{\rm C}(N) = Nf^{\rm C}(1)-\ln N!.
}
Thus, the desired factor $N!$ is reproduced.
\\
\\
In summary, based on our nonequilibrium equality that is applicable in the presence of absolute irreversibility, we have shown that the difference of the entropy productions in the two mixing processes originates from the difference of the degree of absolute irreversibility.
Furthermore, we have reproduced the factor $N!$ in the relation between the thermodynamic entropy and the classical statistical mechanical entropy.
Our new resolution automatically takes account of the sub-leading term and mesoscopic effects in the thermodynamic entropy, which was ignored in the resolution based on the extensivity of the thermodynamic entropy.

\chapter{Conclusions and Future Prospects}
\section{Conclusions}
In this thesis, we have investigated the situations to which the conventional integral nonequilibrium equalities cannot apply, and proposed a new concept of absolute irreversibility to describe these situations in a unified manner.
In absolutely irreversible processes, some of time-reversed paths have no counterpart in the original forward process, and the entropy production diverges in the context of the detailed fluctuation theorems.
In mathematical terms, absolute irreversibility is defined as the singular part of the time-reversed probability measure with respect to the forward probability measure.
Lebesgue's decomposition enables us to separate the absolutely irreversible part from the ordinarily irreversible part.
As a consequence, we have obtained the integral nonequilibrium equalities in the presence of absolute irreversibility.
The obtained equalities involve two physical quantities related to irreversibility: the entropy production representing ordinary irreversibility and the singular probability describing absolute irreversibility.
The corresponding inequalities give tighter fundamental restrictions on the entropy production in nonequilibrium processes than the conventional second-law like inequalities.
Our nonequilibrium equalities have been verified in free expansion and in numerical simulations of the two Langevin systems.

Moreover, we have generalized our nonequilibrium equalities in absolutely irreversible processes to the situations in which the system is subject to measurement-based feedback control.
We have transformed the obtained nonequilibrium equalities and introduced a concept of unavailable information, which characterizes the inevitable inefficiency of feedback protocols.
As a result, we have derived inequalities that give an achievable lower bound of the entropy production.
We have verified our information-thermodynamic absolutely irreversible nonequilibrium equalities in the process with a measurement and trivial feedback control and in the two- and multi-particle Szilard engines.

We have applied the notion of absolute irreversibility to the gas-mixing problem of Gibbs' paradox.
The difference between the entropy production of the mixing process of two different gases and that of two identical gases originates from the difference of absolute irreversibility, i.e., different behaviors under the reinsertion of the partition.
Moreover, we have reproduced the factorial in the particle-number dependence of the thermodynamic entropy.
Our quantitative resolution of Gibbs' paradox applies to a classical mesoscopic regime, where Pauli's resolution based on the extensivity of the thermodynamic entropy breaks down.

\section{Future prospects}
As future prospects, I enumerate several outstanding issues.

First of all, I intend to generalize our absolutely irreversible integral nonequilibrium equalities to the quantum regime.
A part of this extension has already been done, and we have shown that absolute irreversibility is essential under inefficient feedback control and projective measurements \cite{FMU14}.
However, nonequilibrium equalities under the condition that the initial state has such quantum correlations as entanglement are elusive.
Absolute irreversibility may play an important role in these quantum nonequilibrium situations.

Next, I intend to apply our formulation based on measure theory to the chaotic systems.
Although the nonequilibrium equalities were first proven in chaotic systems, most of recent researches of nonequilibrium equalities are restricted to such simple systems as the Langevin systems.
I expect our formulation is compatible with the chaotic system that has a singular continuous probability measure with respect to the Lebesgue measure.
This issue may be related to a topic of thermalization because the relaxed state after a nonequilibrium process starting from the canonical distribution sometimes exhibits singular behaviors.

Finally, I contemplate applying our nonequilibrium equality to finite-time thermodynamics.
Finite-time thermodynamics is a field of thermodynamics that puts emphasis on power of thermodynamic engines and seeks to their optimal efficiency.
Therefore, thermodynamic engines under consideration are subject to finite-time nonequilibrium processes.
Recently, in the context of nonequilibrium equalities, the efficiency of finite-time engines has been studied \cite{VEWB14}.
I expect that our nonequilibrium equalities with absolute irreversibility give new restrictions on the efficiency.

\begin{appendix}
\chapter{From the Langevin Dynamics to Other Formulations}
In this Appendix, we derive the path integral formula (\ref{PIform}) and the Fokker-Planck equation (\ref{FPeq})  from the overdamped Langevin equation
\eqn{
	\dot x (t) = \mu F(x(t),\lambda(t)) + \zeta (t),
}
where $\zeta(t)$ is a white Gaussian noise satisfying
\eqn{
	\langle \zeta(t) \rangle &=& 0,\\
	\langle \zeta(t)\zeta(t') \rangle &=& 2D\delta(t-t').
}

\section{Path-integral formula}
To calculate the path probability, we first discretize the time interval $[0,\tau]$ into $N$ sections with the same length $\Delta t=\tau/N$.
We define $t_i=i\Delta t\ (i=0,1,\cdots,N)$, $x_i =x (t_i)$, and $\lambda_i = \lambda(t_i)$.
Then, the discretized Langevin equation reads\footnote{We use the Stratonovich convention.}
\eqn{
	x_{i+1}-x_i = \mu \frac{F(x_i,\lambda_i)+F(x_{i+1},\lambda_{i+1})}{2} \Delta t+ \Delta W_{i+1},
}
where
\eqn{
	\Delta W_{i+1}= \int_{t_{i}}^{t_{i+1}}\zeta(t)dt
}
for $i=0,\cdots,N-1$.
The discretized noise $\Delta W_i$ obeys the Gaussian distribution with the average
\eqn{
	\langle \Delta W_i \rangle = \int_{t_{i-1}}^{t_{i}}\langle \zeta(t)\rangle dt=0
}
and the variance
\eqn{
	\langle \Delta W_i^2 \rangle
	&=&
	\int_{t_{i-1}}^{t_{i}}\int_{t_{i-1}}^{t_{i}}\langle \zeta(t)\zeta(t') \rangle dtdt'
	\nonumber\\
	&=&
	\int_{t_{i-1}}^{t_{i}}\int_{t_{i-1}}^{t_{i}} 2D\delta(t-t') dtdt'
	\nonumber\\
	&=&
	\int_{t_{i-1}}^{t_{i}}2Ddt
	\nonumber\\
	&=&
	2D\Delta t.
}
Therefore, the probability distribution of $\Delta W_i$ is given by
\eqn{
	P(\Delta W_i) = \frac{1}{\sqrt{4\pi D\Delta t}}\exp\left[
		- \frac{\Delta W_i^2}{4D\Delta t}
	\right].
}
Thus, the joint probability distribution of all $\Delta W_i$ is calculated as
\eqn{
	\mP[\{\Delta W\}] &=& \frac{1}{(\sqrt{4\pi D\Delta t})^n}
	\exp\left[
		- \frac{1}{4D\Delta t}\sum_{i=1}^{N}\Delta W_i^2
	\right]
	\nonumber\\
	&=&
	\frac{1}{(\sqrt{4\pi D\Delta t})^n}
	\exp\left[
		- \frac{1}{4D}\sum_{i=0}^{N-1}\left(
			\frac{x_{i+1}-x_i}{\Delta t}-\mu\frac{F(x_i,\lambda_i)+F(x_{i+1},\lambda_{i+1})}{2}
		\right)^2\Delta t
	\right].
	\nonumber\\
}
In the large-$N$ limit, we obtain
\eqn{
	\mP[\{\Delta W\}] = \mathcal{N}
	\exp\left[
		- \frac{1}{4D}\int_0^\tau\left(
			\dot x(t)-\mu F(x(t),\lambda(t))
		\right)^2dt
	\right],
}
where $\mathcal N$ is the normalization constant.

The path probability $P[\{x\}|x_0]$ is defined in terms of the probability distribution $P[\{\Delta W\}]$ as
\eqn{
	\mP[\{x\}|x_0]\prod_{i=1}^N dx_i = \mP[\{\Delta W\}] \prod_{i=1}^{N} d\Delta W_i.
}
To obtain the path probability, we calculate the Jacobian as
\eqn{
	\mathcal{J} &=&
	\det \left(
		\frac{\partial\Delta W_i}{\partial x_j}
	\right)
	\nonumber\\
	&=&
	\det \left(
		\delta_{i,j}\left(1-\frac{\mu}{2}\dl_x F(x_i,\lambda_i)\Delta t\right)
		-\delta_{i-1,j}\left(1+\frac{\mu}{2}\dl_x F(x_{i-1},\lambda_{i-1})\Delta t\right)
	\right)
	\nonumber\\
	&=&
	\prod_{i=1}^N \left(1-\frac{\mu}{2}\dl_x F(x_i,\lambda_i)\Delta t\right)
	\nonumber\\
	&=&
	\prod_{i=1}^N \exp\left[
		-\frac{\mu}{2}\dl_x F(x_i,\lambda_i)\Delta t
	\right]
	\nonumber\\
	&=&
	\exp\left[
		-\sum_{i=1}^N\frac{\mu}{2}\dl_x F(x_i,\lambda_i)\Delta t
	\right].
}
Thus, in the continuous limit ($\Delta t\to 0$), we obtain
\eqn{
	\mathcal{J} = \exp\left[
		-\int_0^\tau \frac{\mu}{2}\dl_x F(x(t),\lambda(t)) dt
	\right]
}
and therefore
\eqn{
	\mP[\{x\}|x_0] &=&
	\mP[\{\Delta W\}]\mathcal{J} 
	\nonumber\\
	&=&
	\mathcal{N} \exp\left[
		-\int_0^\tau\left(
			\frac{(\dot x(t) -\mu F(x(t),\lambda(t)))^2}{4D} + \frac{\mu}{2}\dl_xF(x(t),\lambda(t))
		\right)dt
	\right],
	\nonumber\\
}
which is nothing but the path integral formula for the overdamped Langevin equation (\ref{PIform}).

\section{Fokker-Planck equation}
First of all, we evaluate the time evolution of an arbitrary function $f(x(t))$ as
\eqn{\label{dfTaylor}
	df(x(t))&=& f(x(t+dt))-f(x(t))\nonumber\\
	&=&
	f'(x(t))dx + \frac{1}{2}f''(x(t))dx^2+\mathcal{O}((dx)^3),
}
where
\eqn{\label{dxdef}
	dx = x(t+dt)-x(t) = \mu F(x(t),\lambda(t))dt+\int_t^{t+dt}\zeta(t')dt' + \mathcal{O}((dt)^2).
}
We take the statistical average of Eq.~(\ref{dfTaylor}).
Substituting Eq.~(\ref{dxdef}) into the average of the first term on the right-hand side of Eq.~(\ref{dfTaylor}), we obtain
\eqn{\label{firsttermave}
	\langle f'(x(t)) dx \rangle
	=
	\langle\mu F(x(t),\lambda(t)) f'(x(t)) \rangle dt
	+ \int_t^{t+dt} \langle f'(x(t))\zeta(t')\rangle dt'
	+\mathcal{O}((dt)^2).
}
Since the stochastic quantity $f'(x(t))$ at time $t$ is independent of the noise $\zeta(t')$ for $t'>t$, we obtain
\eqn{
	\langle f'(x(t))\zeta(t')\rangle = \langle f'(x(t))\rangle\langle\zeta(t')\rangle = 0\ \ \ (t'>t).
}
Therefore, Eq.~(\ref{firsttermave}) reduces to
\eqn{
	\langle f'(x(t)) dx \rangle
	=
	\langle\mu F(x(t),\lambda(t)) f'(x(t)) \rangle dt + \mathcal{O}((dt)^2).
}
In a similar way, the average of the second term on the right-hand side of Eq.~(\ref{dfTaylor}) reduces to
\eqn{
	\frac{1}{2}\langle f''(x(t))dx^2 \rangle &=&
	\frac{1}{2}\langle\mu^2 F(x(t),\lambda(t))^2 f''(x(t)) \rangle dt^2
	+ \int_t^{t+dt} \langle \mu F(x(t),\lambda(t))f''(x(t))\zeta(t') \rangle dt\nonumber\\
	&&+ \frac{1}{2}\int_t^{t+dt} \int_t^{t+dt} \langle \zeta(t')\zeta(t'')f''(x(t)) \rangle dt' dt''
	+\mathcal{O}((dt)^3)\nonumber\\
	&=& D\int_t^{t+dt} \int_t^{t+dt} \delta(t'-t'') \langle f''(x(t)) \rangle dt' dt''+\mathcal{O}((dt)^2)\nonumber\\
	&=& D\langle f''(x(t)) \rangle dt+\mathcal{O}((dt)^2).
}
Therefore, the average of Eq.~(\ref{dfTaylor}) is
\eqn{\label{Ito}
	d\langle f(x(t)) \rangle
	=
	\langle\mu F(x(t),\lambda(t)) f'(x(t)) \rangle dt
	+D\langle f''(x(t)) \rangle dt+\mathcal{O}((dt)^2).
}

To derive the Fokker-Planck equation, we note that the probability $p(x,t)$ to find the Langevin particle at position $x$ at time $t$ is given by
\eqn{
	p(x,t) = \langle \delta(x-x(t)) \rangle.
}
If we set $f(x(t))=\delta(x-x(t))$, Eq.~(\ref{Ito}) reduces to
\eqn{
	dp(x,t) &=& -\langle \mu F(x(t),\lambda(t))\delta'(x-x(t)) \rangle dt + D\langle \delta''(x-x(t)) \rangle dt+\mathcal{O}((dt)^2)
	\nonumber\\
	&=& - \dl_x\langle \mu F(x(t),\lambda(t))\delta(x-x(t)) \rangle dt
	 + D\dl_x^2\langle \delta(x-x(t)) \rangle dt+\mathcal{O}((dt)^2)
	\nonumber\\
	&=& - \dl_x [\mu F(x,\lambda(t))p(x,t)]dt + D\dl_x^2 p(x,t) dt + \mathcal{O}((dt)^2).
}
Thus, we obtain the following Fokker-Planck equation
\eqn{\label{FPap}
	\dl_t p(x,t) = -\dl_x[(\mu F(x, \lambda(t))-D\dl_x)p(x,t)].
}

\chapter{Measure Theory and Lebesgue's Decomposition}
In this Appendix, we briefly review measure theory and Lebesgue's decomposition theorem.
This Appendix is mainly based on \cite{Bar95}.

\section{Preliminary subjects}
\begin{defi}[$\sigma$-algebra]
A family $\mX$ of subsets of $X$ is said to be a {\bf $\sigma$-algebra} if the following three conditions are met:
\begin{description}
\item{(i)} $\emptyset$ and $X$ belong to $\mX$;
\item{(ii)} If $A$ belongs to $X$, then $X\verb+\+A$ belongs to $\mX$;
\item{(iii)} If $(A_n)$ is a sequence of sets in $\mX$, then $\bigcup_{n=1}^\infty A_n$ belongs to $\mX$.
\end{description}
\end{defi}

\begin{defi}[measurable space]
An ordered pair $(X,\mX)$ consisting of a set $X$ and a $\sigma$-algebra $\mX$ of subsets of $X$ is called a {\bf measurable space}.
\end{defi}

\begin{defi}[measure]
A {\bf measure} is an extended real-valued function $\mu$ defined on a $\sigma$-algebra $\mX$ of subsets of $X$ satisfying the following conditions:
\begin{description}
\item{(i)} $\mu(\emptyset) = 0$;
\item{(ii)} $\mu(E)\ge0$ for all $E\in \mX$;
\item{(iii)} $\mu$ is {\bf countably additive}, i.e., if $(E_n)$ is any disjoint sequence of sets in $\mX$, then
\eqn{
		\mu \left( \bigcup_{n=1}^\infty E_n \right)
		= \sum_{n=1}^\infty \mu (E_n).
}
\end{description}
\end{defi}

\begin{defi}[measure space]
A {\bf measure space} is an ordered triad $(X, \mX, \mu)$ consisting of a set $X$, a $\sigma$-algebra $\mX$ of subsets of $X$ and a measure $\mu$ defined on $\mX$.
\end{defi}

\begin{defi}[($\sigma$-)finite measure]
Let $(X,\mX,\mu)$ be a measure space.
If $\mu$ does not take on an infinite value, we say that $\mu$ is {\bf finite}. If there exists a sequence $(E_n)$ of sets in $\mX$ with $X=\bigcup_{n=1}^\infty E_n$ and $\mu(E_n)<\infty$ for all $n$, then we say $\mu$ is {\bf$\sigma$-finite}.
\end{defi}

\begin{defi}[mesurable function]
An $\mathbb R$-valued function $f$ with domain $X$ is said to be {\bf$\mX$-measurable} if for any real number $\alpha$ the set
\eqn{
	\{x\in\mX|f(x)>\alpha\}
}
belongs to $\mX$.
\end{defi}

\begin{defi}[almost everywhere]
Let $\mu$ be a measure on $\mX$.
A certain proposition is said to hold {\bf$\mu$-almost everywhere} on $X$ if there exists a subset $N\in X$ with $\mu(N)=0$ such that the proposition holds on $X\backslash N$.
\end{defi}

\begin{theo}\label{ae}
Suppose that $f$ is a nonnegative $\mX$-measurable function.
Then, $f(x)=0$ $\mu$-almost everywhere on $X$ iff
\eqn{
	\int f d\mu =0.
} 
\end{theo}

\section{Classification of measures}
\begin{defi}[absolutely continuous]
A measure $\nu$ on $\mX$ is said to be {\bf absolutely continuous} with respect to a measure $\mu$ on $\mX$ if $E\in\mX$ and $\mu(E)=0$ imply $\nu(E)=0$. In this case, we write $\nu \ll \mu$.
\end{defi}
\begin{lemm}
Let $\mu$ and $\nu$ be finite measures on $\mX$.
Then $\nu\ll\mu$ iff for every $\epsilon>0$ there exists a $\delta>0$ such that $E \in \mX$ and $\mu(E)<\delta$ imply that $\nu(E)<\epsilon$.
\end{lemm}

\begin{proof}
If this condition is satisfied and $\mu(E)=0$, then $\nu(E)<\epsilon$ for all $\epsilon>0$, which implies $\nu(E)=0$.

Conversely, suppose that there exist some $\epsilon>0$ and $E_n\in\mX$ with $\mu(E_n)<2^{-n}$ and $\nu(E_n)\ge\epsilon$. 
Let $F_n = \bigcup_{k=n}^\infty E_k$, so that $\mu(F_n) < 2^{-n+1}$ and $\nu(F_n)\ge \epsilon$.
Since $(F_n)$ is a decreasing sequence of measurable sets and $\mu, \nu$ are finite measures, we have
\eqn{
	\mu\left(\bigcap_{n=1}^\infty F_n\right) = \lim_{n\to\infty} \mu(F_n) = 0,\ 
	\nu\left(\bigcap_{n=1}^\infty F_n\right) = \lim_{n\to\infty} \nu(F_n) \ge\epsilon.
}
Therefore, $\nu$ is not absolutely continuous with respect to $\mu$.
\end{proof}

Intuitively, $\nu\ll\mu$ means that a set that has a small $\mu$-measure also has a small $\nu$-measure.

\begin{defi}[singular]
Two measures $\mu$ and $\nu$ on $\mX$ are said to be {\bf mutually singular} if there are sets $A$ and $B\in \mX$ that satisfy $X=A\cup B$, $\emptyset = A\cap B$ and $\mu(A)=\nu(B)=0$. In this case, we write $\mu \perp \nu$.

Despite this symmetric definition, we also say that $\nu$ is {\bf singular} with respect to $\mu$. 
\end{defi}

\begin{lemm}\label{unique}
Let $\alpha$ be a measure such that $\alpha \ll \mu$ and $\alpha \perp \mu$, then $\alpha =0$.
\end{lemm}

\begin{proof}
Since $\alpha\perp\mu$, there exist sets $A$ and $B$ such that
\eqn{
	X=A\cup B,\ \emptyset=A\cap B,\ \alpha(A)=0,\ \mu(B)=0.
}
Since $\alpha \ll \mu$ and $\mu(B)=0$,
$
	\alpha(B) = 0.
$
Then, due to the additivity of $\alpha$, we have
\eqn{
	\alpha(X) = \alpha(A) +\alpha(B) = 0.
}
It follows that for all $E\in\mX$
\eqn{
	0\le \alpha(E) = \alpha(X)-\alpha(X \backslash E)\le 0,
}
which implies $\alpha = 0$. 
\end{proof}

\begin{defi}[discontinuous point]
For $\omega\in X$, if $\mu(\{\omega\}) > 0$, then $\omega$ is said to be a {\bf discontinuous point} of $\mu$.
\end{defi}

\begin{defi}[discrete measure]
Let $C$ denote the set of all the discontinuous points of $\mu$. Then, $\mu$ is said to be {\bf discrete measure} if $\mu(X)=\mu(C)$.
\end{defi}

\begin{defi}[continuous measure]
A measure $\mu$ is said to be a {\bf continuous measure} if $\mu$ has no discontinuous points.
\end{defi}

\section{Radon-Nikod\'ym theorem}
\begin{theo}[Radon-Nikod\'ym theorem]
Let $\mu$ and $\nu$ be $\sigma$-finite measures defined on $\mX$ and suppose that $\nu$ is absolutely continuous with respect to $\mu$. Then, there exists a nonnegative $\mX$-measurable function $f$ such that
\eqn{\label{RN}
	\nu(E) = \int_E f d\mu,\ \ \ ^\forall E\in\mX.
}
Moreover, the function $f$ is uniquely determined $\mu$-almost everywhere.
\end{theo}

The function $f$ is referred to as the Radon-Nikod\'ym derivative, and formally written as
\eqn{
	f = \frac{d\nu}{d\mu}.
}
Moreover, $f$ is the transformation function from $\nu$ to $\mu$.
In fact, for an arbitrary $\mX$-measurable function $g$, we have
\eqn{
	\int_E g d\nu = \int_E g  \frac{d\nu}{d\mu}d\mu = \int_E gfd\mu
	,\ \ \ ^\forall E\in\mX.
}

\section{Lebesgue's decomposition theorem}
\begin{theo}[Lebesgue's decomposition theorem 1]\label{LDT}
Let $\mu$ and $\nu$ be $\sigma$-finite measures defined on a $\sigma$-algebra $\mX$.
Then there exist measures $\nu_{\rm AC}$ and $\nu_{\rm S}$ such that $\nu = \nu_{\rm AC} + \nu_{\rm S}$, $\nu_{\rm AC}\ll\mu$ and $\nu_{\rm S} \perp \mu$.
Moreover, the measures $\nu_{\rm AC}$ and $\nu_{\rm S}$ are unique.
\end{theo}

\begin{proof}
Let $\lambda = \mu + \nu$. Then, $\mu$ and $\nu$ are absolutely continuous with respect to $\lambda$. Therefore, we can apply the Radon-Nikod\'ym theorem to obtain
\eqn{
	\mu(E)=\int_E f d\lambda,\ \ \ \nu(E)=\int_E g d\lambda
}
for all $E\in\mX$, where $f$, $g$ are nonnegative $\mX$-measurable functions.
Let $A=\{x|f(x)>0\}$ and $B=\{x|f(x)=0\}$ so that $A\cap B = \emptyset$ and $X=A\cup B$.

Define $\nu_{\rm AC}$ and $\nu_{\rm S}$ for $E\in\mX$ by
\eqn{
	\nu_{\rm AC}(E) = \nu(E\cap A),\ \ \ \nu_{\rm S}(E) = \nu(E\cap B).
}
Since $\nu$ is additive, $\nu(E)=\nu_{\rm AC}(E)+\nu_{\rm S}(E)$.
To see $\nu_{\rm AC}\ll\mu$, we note that if $\mu(E)=0$, then
\eqn{
	\int_E fd\lambda=0.
}
In accordance with Theorem \ref{ae}, $f(x)=0$ for $\lambda$-almost all $x\in E$, which means $\lambda(E\cap A)=0$.
Since $\nu\ll\lambda$, $\nu(E\cap A)=0$, and then $\nu_{\rm AC}(E)=0$. Thus, $\nu_{\rm AC}$ is absolutely continuous with respect to $\mu$.
On the other hand, since $\nu_{\rm S}(A)=\mu(B)=0$,
$\nu_{\rm S}$ is singular with respect to $\mu$.

The uniqueness of this decomposition can be established by Lemma \ref{unique}.
\end{proof}

\begin{lemm}\label{dis}
Let $\nu$ be a measure defined on a $\sigma$-algebra $\mX$.
Then there exist measures $\nu_{\rm c}$ and $\nu_{\rm d}$ such that $\nu=\nu_{\rm c}+\nu_{\rm d}$, where $\nu_{\rm c}$ is continuous and $\nu_{\rm d}$ is discrete. Moreover, the measures $\nu_{\rm c}$ and $\nu_{\rm d}$ are unique.
\end{lemm}

\begin{proof}
Let $C$ be the set of all the discontinuous points of $\nu$.
Define $\nu_{\rm c}$ and $\nu_{\rm d}$ for $E\in\mX$ by
\eqn{
	\nu_{\rm c}(E) = \nu(E\backslash C),\ \ \ \nu_{\rm d}(E) = \nu(E\cap C).
}
Then, we can show that $\nu_{\rm c}$ is continuous and $\nu_{\rm d}$ is discrete.

Suppose
\eqn{\label{twodecom}
	\nu_{\rm c} + \nu_{\rm d} = \nu'_{\rm c} + \nu'_{\rm d},
}
where $\nu'_{\rm c}$ is continuous and $\nu'_{\rm d}$ is discrete.
If $\nu_{\rm d}\neq\nu'_{\rm d}$, there exists a single point $\omega\in X$ such that $\nu_{\rm d}(\{\omega\})\neq\nu'_{\rm d}(\{\omega\})$. On the other hand, due to the continuity, $\nu_{\rm c}(\{\omega\})=\nu'_{\rm c}(\{\omega\})=0$. These relations lead to
\eqn{
	\nu_{\rm c}(\{\omega\}) + \nu_{\rm d}(\{\omega\})
	 \neq \nu'_{\rm c}(\{\omega\}) + \nu'_{\rm d}(\{\omega\}),
}
which contradicts Eq.~(\ref{twodecom}). Hence, $\nu_{\rm d}=\nu'_{\rm d}$ and therefore the uniqueness of the decomposition is established.
\end{proof}

To introduce a stronger version of Lebesgue's decomposition, we prove the following lemma.

\begin{lemm}\label{cduni}
Let $\mu$ and $\nu$  be measures on a $\sigma$-algebra.
If $\mu$ is continuous and $\nu$ is discrete,
then $\nu$ is singular with respect to $\mu$.
\end{lemm}
\begin{proof}
Let $C$ denote the set of all the discontinuous points of $\nu$.
Then, $\nu(X\backslash C)=0$. On the other hand, since $\mu$ is continuous and $C$ is countable, $\mu(C)=0$. Thus, $\nu$ and $\mu$ are mutually singular.
\end{proof}

\begin{theo}[Lebesgue's decomposition theorem 2]
Let $\mu$ and $\nu$ be $\sigma$-finite measures defined on a $\sigma$-algebra $\mX$ and suppose $\mu$ is continuous.
Then there exist measures $\nu_{\rm ac}$, $\nu_{\rm sc}$ and $\nu_{\rm d}$ such that $\nu = \nu_{\rm ac} + \nu_{\rm sc}+\nu_{\rm d}$, where $\nu_{\rm ac}\ll\mu$; $\nu_{\rm sc} \perp \mu$ and $\nu_{\rm sc}$ is continuous; $\nu_{\rm d}$ is discrete.
Moreover, the measures $\nu_{\rm ac}$, $\nu_{\rm sc}$ and $\nu_{\rm d}$ are unique.
\end{theo}

\begin{proof}
We can apply Theorem \ref{LDT} to uniquely decompose
\eqn{
	\nu = \nu_{\rm ac} + \nu_{\rm s},
}
where $\nu_{\rm ac}$ is absolutely continuous with respect to $\mu$,  and $\nu_{\rm s}$ is singular with respect to $\mu$.
In accordance with Lemma \ref{dis}, we can uniquely decompose $\nu_{\rm s}$ into two parts:
\eqn{
	\nu_{\rm s} = \nu_{\rm sc} + \nu_{\rm d},
}
where $\nu_{\rm sc}$ is continuous and $\nu_{\rm d}$ is discrete. Thus, we obtain the following decomposition:
\eqn{
	\nu = \nu_{\rm ac} + \nu_{\rm sc} + \nu_{\rm d}.
}
The uniqueness of this decomposition follows from Lemma~\ref{cduni}.
\end{proof}

The continuity of $\mu$ is needed to establish the uniqueness of the decomposition.
If $\mu$ has a discontinuous point $\omega\in X$, then the measure $\nu_\omega$ that satisfies $\nu_\omega(\{\omega\})=\nu_\omega(X)$ is absolutely continuous with respect to $\mu$ and discrete at the same time.
\end{appendix}

\chapter*{Acknowledgement}
The studies in this thesis was done when the author was a master-course student in Masahito Ueda group in the University of Tokyo.
This thesis would not be completed without help of a lot of collaborators and colleagues.

First of all, I would like to express my best gratitude to my supervisor, Prof. Masahito Ueda.
He made me cognizant of the topic in this thesis and gave me a lot of  insightful comments throughout discussions.
He also read the manuscript with great attention and gave me enormous suggestions.
I would also thank him for the best research environment that he arranged for me.

I would also like to thank my collaborator, Ken Funo, for fruitful discussions on our studies and for teaching me quantum aspects of nonequilibrium equalities. 

I am thankful for my collaborator, Yuto Ashida, for his constructive ideas from his deep comprehension of our field.

I am grateful to Prof. Takahiro Sagawa for his critical comments and beneficial discussions on our work.

I appreciate critical comments and constructive suggestions by Prof. Shin-ichi Sasa.

I acknowledge comments  from anonymous referees of our article \cite{MFU14}, which I find very useful to clarify physical meanings of our work.

I would like to express my deep sense of gratitude to the members in Masahito Ueda group,
especially to Yui Kuramochi and Tomohiro Shitara for fruitful discussions on mathematical aspects of our work, and to Tatsuhiko N. Ikeda for suggestive comments.

I am also indebted to the members in Masaki Sano group in the University of Tokyo, especially to Kyogo Kawaguchi for lecturing me on theoretical aspects of mesoscopic physics and to Yohei Nakayama, Yuta Hirayama and Daiki Nishiguchi for teaching me experimental techniques of this field. 

Finally, I acknowledge financial support from the Japan Society for the Promotion of Science (JSPS) through the Program for Leading Graduate Schools (MERIT).

\bibliography{reference}
\end{document}